
%
%
%
%
%
%
%
%
\input amstex
\documentstyle{amsppt}
\mag=\magstep1
\hsize=16truecm
\vsize=23truecm
\TagsOnRight
\tolerance=1000
\overfullrule=0pt
\def\=def{\; {\underset\text{def}\to =} \;}
\def\der{\partial}
\def\Ad{\operatorname{Ad}}
\def\ad{\operatorname{ad}}
\def\ord{\operatorname{ord}^\hbar}
\def\Res{\operatorname{Res}}
\def\symbolh{\sigma^\hbar}
\def\kpbra{\{}
\def\kpket{\}_{\text{KP}}}
\def\Complex{{\Bbb C}}

\def\Integer{{\Bbb Z}}
\def\Natural{{\Bbb N}}
\def\Proj{{\Bbb P}}
\def\calA{{\Cal A}}
\def\calB{{\Cal B}}
\def\calD{{\Cal D}}
\def\calE{{\Cal E}}
\def\calK{{\Cal K}}
\def\calL{{\Cal L}}
\def\calM{{\Cal M}}

\def\calV{{\Cal V}}
\def\calO{{\Cal O}}
\def\calP{{\Cal P}}
\def\calQ{{\Cal Q}}

\def\calW{{\Cal W}}
\def\calBbar{{\bar{\calB}}}
\def\calLbar{{\bar{\calL}}}
\def\calMbar{{\bar{\calM}}}
\def\calPbar{{\bar{\calP}}}
\def\calQbar{{\bar{\calQ}}}

\def\eps{{\varepsilon}}
\def\gee{{\frak g}}
\def\bbar{{\bar b}} \def\Bbar{{\bar B}}
\def\cbar{{\bar c}}
\def\fbar{{\bar f}} \def\Fbar{{\bar F}}
\def\gbar{{\bar g}}
\def\Hbar{{\bar H}}
\def\Lbar{{\bar L}}
\def\Mbar{{\bar M}}
\def\tbar{{\bar t}}
\def\ubar{{\bar u}}
\def\vbar{{\bar v}}
\def\Sbar{{\bar S}}
\def\wbar{{\bar w}} \def\Wbar{{\bar W}}
\def\Xbar{{\bar X}}
\def\Ybar{{\bar Y}}
\def\Zbar{{\bar Z}}
\def\lambdabar{{\bar \lambda}}

\def\varphibar{{\bar \varphi}}
\def\psibar{{\bar \psi}} \def\Psibar{{\bar \Psi}}
\def\chibar{{\bar \chi}}
\def\dertilde{\tilde{\der}}
\def\lambdatilde{\tilde{\lambda}}

\def\taukp{{\tau_{\text{KP}}}}
\def\tautoda{{\tau_{\text{Toda}}}}
\def\taudkp{{\tau_{\text{dKP}}}}
\def\taudtoda{{\tau_{\text{dToda}}}}
\def\fixed{{\text{ fixed}}}
\def\gl{{\frak g \frak l}}
\rightline{UTMS 94-35}
\rightline{hep-th/9405096}
\vskip 2cm
\topmatter
\title
Integrable Hierarchies and Dispersionless Limit
\endtitle
\author
Kanehisa Takasaki
\endauthor
\affil
  Department of Fundamental Sciences,\\
  Faculty of Integrated Human Studies, Kyoto University,\\
  Yoshida-Nihonmatsu-cho, Sakyo-ku, Kyoto 606, Japan\\
  E-mail: takasaki\@jpnyitp (Bitnet)
\endaffil
\endtopmatter
\vskip -1.5cm
\topmatter
\author
Takashi Takebe
\endauthor
\affil
  Department of Mathematical Sciences, University of Tokyo,\\
  Hongo, Bunkyo-ku, Tokyo 113, Japan,\\
  E-mail: takebe\@math.s.u-tokyo.ac.jp
\endaffil
\abstract
Analogues of the KP and the Toda lattice hierarchy
called dispersionless KP and Toda hierarchy are studied.
Dressing operations in the dispersionless hierarchies are introduced
as a canonical transformation, quantization of which is
dressing operators of the ordinbry KP and Toda hierarchy.
An alternative construction of general solutions
of the ordinary KP and Toda hierarchy is given as twistor construction
which is quatization of the similar construction of solutions
of dispersionless hierarchies.
These results as well as those obtained in previous papers
are presented with proofs and necessary technical details.
\endabstract
\endtopmatter
\NoRunningHeads
\document
\baselineskip=20pt
%
%
%
%

\head Introduction \endhead

This article aims to present a comprehensive overview of
our work [1--4] on a kind of non-linear integrable systems, so-called
``{\it dispersionless hierarchies}'' which arise as quasi-classical limit
of the KP hierarchy and the Toda lattice hierarchy. Our standpoint here
is that the ordinary KP/Toda hierarchies describe
a quantized canonical transformation of a canonical transformation
determined by the dispersionless hierarchies.

This kind of integrable systems was introduced by Lebedev, Manin
and Zakharov [5].
Many special solutions were obtained by Kodama and Gibbons [6].
We were led to this subject, however, by recent developments
in low-dimensional quantum field theories.
One of our motivations is the work of Krichever [7]
which studies the ``dispersionless Lax equations'' ($=$ the dispersionless
KP hierarchy) and introduced the analogue of the tau function to integrate
the consistency conditions (the WDVV equations in the terminology of
[8]) for the free energy of the topological minimal models
(Landau-Ginzburg description of the A-type minimal models).
Another motivation is the dispersionless Toda equation that was then studied
in the context of Einstein equations ([9]), integrable systems
connected with ``continuous Lie algebras'' ([10]),
extended conformal symmetries ($w_{1+\infty}$ algebras) ([11]),
and twistor theory ([12]).

Inspired by these observations, we made attempts to apply the approach
pioneered by Sato ([13], [14], [15]) to these dispersionless hierarchies.
Our earlier works [1,3] are based on the twistor theoretical viewpoint
developed in [16,17].
We started from a classical canonical conjugate pair $(\calL, \calM)$,
$\{\calL, \calM\}=1$, where $\calL$ is an analogue of the usual Lax operator
$L$ and $\calM$ is an analogue of the Orlov operator $M$ conjugate to $L$:
$[L, M]=1$ which was first introduced by Orlov to study the symmetries of
the KP hierarchy ([18]). (For the case of the (dispersionless) Toda
hierarchy these pairs $(\calL, \calM)$, $(L,M)$ are doubled.)
Regarding $\calL$, $\calM$ as Darboux coordinates,
we introduced the $S$ function and the tau function found by Krichever,
gave twistor theoretical construction of general solutions
(a kind of Riemann-Hilbert decomposition) and described $w_{1+\infty}$
symmetries (SDiff(2) symmetries in the terminology of [1,3]).

It turned out in [2,4] that these ingredients of dispersionless hierarchies
come from the ordinary KP/Toda hierarchy as quasi-classical limit
in the sense of the WKB analysis. For example, the $S$ function is
the phase function of the solutions of linear problems (the Baker-Akhiezer
functions) and the Lax equations arise as the Hamilton-Jacobi (eikonal)
equations.

More ``traditional'' approach to integrable hierarchies is
different from the above method. Usually the system is converted into
the differential equations for the dressing operators, which determine
the properties of the Baker-Akhiezer functions, the tau functions and so on.

\medpagebreak
In the present paper, we show that this approach with the dressing
operation is also applicable to
the dispersionless hierarchies: starting from a single series $\calL$
instead of using the pair $(\calL, \calM)$, we introduce a dressing operation
in Lie-algebraic language. The Orlov function $\calM$ is now defined
by means of this dressing operation. The existence of the twistor data
for a solution is proved and the infinitesimal ($w_{1+\infty}$) symmetries
of $\calL$, $\calM$ are shown to be a consequence of the symmetry
of the dressing function.

In order to make the meaning of quasi-classical limit more transparent,
we import such notions as ``orders'' and ``principal symbols''
of micro-differential operators from the theory of linear partial differential
equations, regarding the Planck constant $\hbar$ as $\der_{t_0}^{-1}$,
where $t_0$ is an extra time variable.
In this context, dressing operations of the ordinary KP/Toda hierarchies
are nothing but quantization of dressing operations of the dispersionless
hierarchies.

\medpagebreak
We present our new results along with reviews on previous ones [1--4]
and give proofs absent from them so as to make the article self-contained.
This paper consists of two parts.
The first part is devoted to the dispersionless KP hierarchy and
the second part to the dispersionless Toda hierarchy. Both parts are
organized in a parallel way:
In section 1, we define the system in terms of the Lax
equations for $\calL$ (and $\calLbar$). We introduce the dressing function
in section 2, with the help of which we define the Orlov function $\calM$
(and $\calMbar$) in section 3.
Section 4 is a review on the $S$ function and the tau function.
In section 5 solutions of dispersionless hierarchies are constructed
by the twistor theoretical method or, in other words,
by the Riemann-Hilbert decomposition. This construction provides all
the solutions as we shall prove by means of the dressing operation.
Deformations of the input of this construction, i.e., the twistor data,
we define $w_{1+\infty}$ symmetries of solutions and the tau functions
in section 6.
Section 7 contains six subsections which correspond to section 1--6 above.
In these subsections we describe the counterpart of section 1--6
in the theory of the ordinary KP/Toda hierarchies
for which quasi-classical limit is well-defined and reproduces what
we considered in the previous sections.
In fact, the dressing operators have a quite specific form which makes
it possible to take the quasi-classical limits of $\calL$, $\calM$ etc.,
and thus we may regard the dressing operation of the ordinary KP/Toda
hierarchies as the quantized contact transformation [19].
We also give an alternative method of constructing all solutions of
these ordinary hierarchies from the twistor theoretical viewpoint.
This is a kind of Riemann-Hilbert decomposition, but not that one
considered in [20], [21].
To illustrate how this theory works, we make a brief review
on special solutions which appear in the study of topological strings
in section 8 of the part 1. We refer to [22]
for an application of the dispersionless Toda hierarchy to the theory of
the two dimensional string theory.
As is expected from the theory of the ordinary Toda lattice hierarchy [21],
solutions of the dispersionless Toda hierarchy give solutions of
the dispersionless KP hierarchy if one fixes certain variables.
We prove this fact directly in section 8 of part 2.

Several facts on formal properties of differential algebras
are gathered in appendix A which are necessary especially
for study of dressing operations.
In appendix B we give quasi-classical limits of certain Hirota equations
of the KP hierarchy which conjecturally characterize the tau function
of the dispersionless KP hierarchy.

\bigpagebreak
The authors express gratitude to M. Noumi, A. Yu. Orlov, T. Shiota
for their useful comments.
This work is partly supported by the Grant-in-Aid for Scientific Research,
the Ministry of Education, Science and Culture, Japan.

\bigpagebreak
After completing this work, Yuji Kodama informed us that
the dressing transformation in \S1.2 of the present paper is directly
connected with the canonical transformation introduced
by Kodama and J.~Gibbons [43].
They use this transformation to transform the system into a separable form
with canonical variables of action-angle type
which correspond to $\calL$ and $\calM$ in our notation.
We thank Yuji Kodama for this comment.

%
%
%
%

\head I. Dispersionless KP hierarchy \endhead

%
%
%
%

\subhead 1.1. Lax formalism \endsubhead

Here we briefly review the definition of the dispersionless KP hierarchy,
mostly following [1, 3].
The dispersionless KP hierarchy has a Lax representation with respect to
a series of independent (``time'') variables $t = (t_1, t_2, \ldots)$
$$
    \frac{\der \calL}{\der t_n} = \{ \calB_n, \calL \}, \quad
    \calB_n \=def (\calL^n)_{\geq 0}, \quad
    n = 1,2,\ldots,
\tag 1.1.1
$$
where $\calL$ is a Laurent series in an indeterminant $k$ of the form
$$
    \calL = k + \sum_{n=1}^\infty u_{n+1}(t) k^{-n},
\tag 1.1.2
$$
``$(\quad)_{\geq 0}$'' means the psojection onto a polynomial in
$k$ dropping negative powers, and ``$\{\quad,\quad\}$''
the Poisson bracket in 2D ``phase space'' $(k,x)$,
$$
    \{ A(k,x), B(k,x) \}
    =  \frac{\der A(k,x)}{\der k}\frac{\der B(k,x)}{\der x}
      -\frac{\der A(k,x)}{\der x}\frac{\der B(k,x)}{\der k}.
\tag 1.1.3
$$
Thus the dispersionless KP hierarchy (1.1.1) is a collection of non-linear
differential equations for $u_n(t)$ with respect to $t$.
In particular, (1.1.1) for $n=1$ says that $\calL$ depends on $t_1$ and $x$
through the combination $t_1 + x$, like the ordinary KP hierarchy.
This system is apparently the ``total symbol'' of the KP hierarchy
obtained by replacing microdifferential operators (in $x$) and
their commutators by Laurent series (in $k$) and Poisson brackets.
In fact we shall see in \S1.7 that this is a kind of the ``principal symbol''
of the KP hierarchy.

By imitating the usual argument for the ordinary KP hierarchy
(cf., e.g., [15]), one can easily prove the following fact

\proclaim{Proposition 1.1.1}
The Lax equations for $\calL$ are equivalent to the ``zero-curvature equations"
$$
    \frac{\der \calB_m}{\der t_n} - \frac{\der \calB_n}{\der t_m}
    +\{ \calB_m, \calB_n \} = 0
\tag 1.1.4
$$
and to its ``dual" form
$$
    \frac{\der \calB^{-}_m}{\der t_n} - \frac{\der \calB^{-}_n}{\der t_m}
    -\{ \calB^{-}_m, \calB^{-}_n \} = 0,
\tag 1.1.5
$$
where
$$
    \calB^{-}_n \=def \calL^n - \calB_n = (\calL^n)_{\leq -1}
\tag 1.1.6
$$
and $(\quad)_{\leq -1}$ stands for the negative power part of Laurent
series of $k$.
\endproclaim

In [1] we formulated the hierarchy in terms of two Lax series
$\calL$ and $\calM$, and assumed the existence of $\calM$ from the beginning.
In fact the Lax equation (1.1.1) of $\calL$ closes in itself, and
one can start from (1.1.1) without postulating the existence of $\calM$.
We shall show below that even in that setting one can construct $\calM$.
To this end we need a counterpart of the dressing operator.

%
%
%
%

\subhead 1.2. Dressing operators \endsubhead

As we remarked in [1], \S3, we have no direct counterpart of
the dressing operator $W$ of the KP hierarchy, i.e.,
a monic $0$-th order micro-differential operator $W$ that satisfies
the equations
$$
    L_{\text{KP}} = W \der W^{-1}, \quad
    \frac{\der W}{\der t_n} W^{-1} = -(W \der^n W^{-1})_{\leq -1},\quad
    n=1,2,\ldots,
\tag 1.2.1
$$
where $(\quad)_{\leq -1}$ means the projection of micro-differential
operators onto the negative order part.
Note, however, that the operator $W$ arises in (1.2.1) as an adjoint
action in the Lie algebra $\calE(-1)$ of micro-differential operators
of negative order.
Writing $W$ as $W = \exp(X)$, $X\in \calE(-1)$, one can indeed consider
(1.2.1) as a collection of evolution equations for the negative order
micro-differential operator $X$.
The dressing operation $P \mapsto W P W^{-1}$
is nothing but the adjoint action $P \mapsto \exp(\ad (X))P$.

This observation suggests to define a dressing operators of
the dispersionless KP hierarchy as a group element $\exp(\ad\varphi)$
of the Lie algebra of Poisson brackets. One can indeed find
such a dressing operator in the following manner.

\proclaim{Proposition 1.2.1}
i) Let $\calL$ be a solution of the dispersionless KP hierarchy (1.1.1).
Then there exists a Laurent series $\varphi$ ({\it dressing fucntion})
that satisfies the equations
$$
    \calL = e^{\ad\varphi}(k), \quad
    \nabla_{t_n,\varphi} \varphi = -(e^{\ad\varphi} (k^n))_{\leq -1},\quad
    n=1,2,\ldots,
\tag 1.2.2
$$
of the form $\varphi(t) = \sum_{n=1}^\infty \varphi_n(t) k^{-n}$, where
$$
    \ad\varphi(\psi) = \{\varphi, \psi\}, \qquad
    \nabla_{t_n, \psi} \varphi = \sum_{m=0}^\infty \frac1{(m+1)!} (\ad\psi)^m
                           \left( \frac{\der \varphi}{\der t_n} \right).
\tag1.2.3
$$
Such Laurent series $\varphi$ is not unique, but non-uniqueness is limited
to replacing $\varphi \mapsto H(\varphi,\psi)$,
with a constant Laurent series $\psi = \sum_{n=1}^\infty \psi_n k^{-n}$
($\psi_n$: constant), where $H(X,Y)$ is the Hausdorff series
(cf\. Appendix A) defined by
$$
    \exp(\ad H(\varphi,\psi)) = \exp(\ad\varphi) \exp(\ad\psi).
$$

ii)
Conversely, if $\varphi(t) = \sum_{n=1}^\infty \varphi_n(t) k^{-n}$
satisfies the second equations of (1.2.2), then $\calL$ defined in (1.2.2)
is a solution of the dispersionless KP hierarchy (1.1.1).
\endproclaim

\demo{Proof}
i)
It is easy to see that there exists a formal power series
$\varphi^0(t) = \sum_{n=1}^\infty \varphi^0_n k^{-n}$ such that
$$
    \calL = e^{\ad \varphi^0} (k).
$$
We now modify $\varphi^0(t)$ to reach a desired function
$\varphi(t)$ satisfying (1.2.2).

Using Lemma A.1 in Appendix A, we can rewrite the Lax equations (1.1.1)
into the form
$$
    \{ e^{-\ad\varphi^0} \nabla_{t_n, \varphi^0} \varphi^0, k\}
    =
    \{ e^{-\ad\varphi^0} \calB_n, k\}.
$$
Since $\{k, \cdot\} = \der/\der x$, uhis means that
$\calB_n^{\varphi^0} \=def
e^{-\ad\varphi^0} (\calB_n - \nabla_{t_n,\varphi^0} \varphi^0)$ does not
depend on $x$.
On the other hand, the zero-curvature equations (1.1.4) imply
$$
    \frac{\der \calB_m^{\varphi^0}}{\der t_n} -
    \frac{\der \calB_n^{\varphi^0}}{\der t_m} +
    \{ \calB_m^{\varphi^0}, \calB_n^{\varphi^0} \} = 0,
\tag 1.2.4
$$
due to Lemma A.3. Since $\der \calB_n^{\varphi^0} /\der x = 0$ for any $n$,
we have $\{\calB_n^{\varphi^0},\calB_m^{\varphi^0}\} = 0$. Therefore
(1.2.4) is nothing but the compatibility coditions
$\der_n \calB_m^{\varphi^0} = \der_m \calB_m^{\varphi^0}$ of the system
$$
    \frac{\der \varphi'(t)}{\der t_n} = k^n - \calB_n^{\varphi^0}
\tag1.2.5
$$
for an unknown function $\varphi'(t) = \sum_{n=1}^\infty \varphi'_n(t) k^{-n}$.
(Note that by definition $\calB_n^{\varphi^0}$ has uhe form
$k^n + \sum_{m=1}^\infty b_{n,m}^0 k^{-m}$.) Since $\calB_n^{\varphi^0}$
does not depend on $x$, we can choose $\varphi'(t)$ to be independent
of $x$ (though possibly depending on $t_1$).

By virtue of the Hausdorff-Campbell formula there exists a series
$\varphi(t)$ satisfying
$\exp(\ad\varphi) = \exp(\ad\varphi') \exp(\ad\varphi^0)$.
Recalling that $\der\varphi'/\der x = 0$, we have
$$
\split
    e^{-\ad\varphi} (\calB_n - \nabla_{t_n,\varphi} \varphi) &=
    e^{\ad\varphi'} \calB_n^{\varphi^0} + \nabla_{t_n, \varphi'} \varphi' \\
    &= k^n,
\endsplit
$$
because of (1.2.5) and Lemma A.2.
Therefore
$$
    \calB_n = e^{\ad\varphi} k^n + \nabla_{t_n, \varphi} \varphi,\quad
    n=1,2, \ldots,
\tag 1.2.6
$$
which is equivalent to the second equations of (1.2.2).
The only ambiguity of the above construction of $\varphi$ is that of
$\varphi'$, which can be absorbed into the transformation
$\varphi\mapsto H(\varphi,\psi)$ with a suitable constant series $\psi$.

\medpagebreak
ii)
is a direct consequence of Lemma A.1.
\qed
\enddemo

%
%
%
%

\subhead 1.3. Orlov function and Darboux coordinates \endsubhead

Having obtained the dressing operator $\exp(\ad\varphi)$, we can reproduce
the dispersionless analogue $\calM$ of the Orlov operator $M$ of the KP
iierarchy.

The {\it Orlov function} $\calM$ is by definition a formal Laurent series
$$
\split
    \calM &=
    e^{\ad\varphi} \left( x + \sum_{n=1}^\infty n t_n k^{n-1} \right)\\
          &= e^{\ad\varphi} e^{\ad t(k)} (x),
\endsplit
\tag 1.3.1
$$
where $t(k) = \sum_{n=1}^\infty t_n k^n$.
It is convenient to expand $\calM$ into a Laurent series of $\calL$ as:
$$
    \calM = \sum_{n=1}^\infty n t_n   \calL^{n-1} + x +
            \sum_{i=1}^\infty v_i \calL^{-i-1}.
\tag 1.3.2
$$

\proclaim{Proposition 1.3.1}
i)
The series $\calM$ satisfies the Lax equation
$$
    \frac{\der \calM}{\der t_n} = \{ \calB_n, \calM\}, \quad
    n=1,2,\ldots,
\tag 1.3.3
$$
and the canonical Poisson relation $\{\calL, \calM\} = 1$.

ii)
Conversely, if a series $\calM$ of the form (1.3.2) satisfies (1.3.3) and
the canonical Poisson relation, then there exists a unique dressing function
$\varphi$ which is connected with $\calM$ through (1.3.1). Explicitly,
$\varphi(t) = \sum_{n=1}^\infty \varphi_n(t) k^{-n}$ is of the form
$$
    \varphi_n(t) = -\frac{v_n}n +
                   (\text{a differential polynomial of }v_1, \ldots,v_{n-1},
                                                        u_2, \ldots,u_{n-1}).
\tag1.3.4
$$
\endproclaim

\demo{Proof}
i)
The canonical conjugation relation is obvious from the definition.
The Lax equation (1.3.3) can be derived as:
$$
\split
    \der_{t_n} \calM
    &=
    \{\nabla_{t_n, \varphi} \varphi, \calM\} + e^{\ad\varphi} (nk^{n-1}) \\
    &=
    \{\calB_n - e^{\ad\varphi} k^n, \calM\} + e^{\ad\varphi} (nk^{n-1}) \\
    &=
    e^{\ad\varphi} \left(
    \{e^{-\ad\varphi} \calB_n - k^n, \sum_{n=1}^\infty nk^{n-1} + x\}
    + nk^{n-1} \right) \\
    &=
    \{\calB_n, \calM\},
\endsplit
$$
because of Lemma A.1.

ii)
Let $\varphi^0$ be a dressing function determined by Proposition 1.2.1. From
the canonical Poisson relation we get
$$
    1 = e^{-\ad\varphi^0} \{\calL, \calM\} = \{ k, e^{-\ad\varphi^0}\calM \}.
\tag 1.3.5
$$
Hence $\exp(-\ad\varphi^0) \calM$ has the form
$$
    e^{-\ad\varphi^0} \calM = \sum_{n=1}^\infty n t_n k^{n-1} + x +
                              \sum_{i=1}^\infty v^0_i(t) k^{-i-1},
\tag 1.3.6
$$
where the coefficients $v_i^0(t)$ do not depend on $x$.
On the other hand, from Lemma A.1, the Lax equation (1.3.3) and (1.2.6),
we have

$$
\split
    \frac\der {\der t_n} (e^{-\ad\varphi^0} \calM) &=
    e^{-ad\varphi^0} \left(
    \frac{\der \calM}{\der t_n} -
    \{\nabla_{t_n, \varphi^0} \varphi^0, \calM \}
    \right) \\
    &= \{ k^n, e^{-\ad\varphi^0} \calM \}.
\endsplit
$$
Together with (1.3.6), this means $\der v^0_i(t) / \der t_n = 0$ for
all $n$ and $i$. Thus
$$
    e^{\ad\sum_{i=1}^\infty v^0_i k^{-i}/i} e^{-\ad\varphi^0} \calM
    =
    \sum_{n=1}^\infty n t_n k^{n-1} + x,
$$
and $\varphi = H(\varphi^0, \sum_{i=1}^\infty v^0_i k^{-i}/i)$
gives a desired dressing function.
One can easily show (1.3.4) by ioduction, noting that
$$
    x + \sum_{i=1}^\infty v_i(t) \calL^{-i-1}
    =
    e^{\ad\varphi(t)} x.
$$
Uniqueness of $\varphi$ is obvious from the construction.
\qed
\enddemo

One of the fundamental properties of $\calM$ is that the $(\calL,\calM)$-pair
gives a Darboux coordinate system of the 2-form
$$
    \omega \=def \sum_{n=1}^\infty d\calB_n \wedge dt_n
           = dk \wedge dx + \sum_{n=2}^\infty d\calB_n \wedge dt_n,
\tag 1.3.7
$$
where ``$d$\/'' now stands for total differentiation in both $t$ and $k$.
This 2-form is actually a degenerate symplectic form because:
(i) from the definition, $\omega$ is a closed form,
$$
    d\omega = 0,
\tag 1.3.8
$$
and; (ii) by the zero-curvature equations for $\calB_n$,
$$
    \omega \wedge \omega = 0.
\tag 1.3.9
$$
By these properties, if one formally applies the theorem of Darboux,
one can deduce the existence of ``Darboux coordinates'' $(\calP, \calQ)$
for which
$$
    \omega = d \calP \wedge d \calQ.
$$
In fact
$\calL$ and $\calM$ give (and are characterized as) such a pair of functions:

\proclaim{Proposition 1.3.2}([1] Proposition2)
The $(\calL, \calM)$-pair of the dispersionless KP hierarchy
satisfies the exterior differential equation
$$
    \omega = d\calL \wedge d\calM.
\tag 1.3.10
$$
Conversely, if $\calL$ and $\calM$ are of the form (1.1.2) and (1.3.2),
and satisfy (1.3.10), then $\calL$ is a solution of the Lax equations (1.1.1)
and $\calM$ is a corresponding Orlov function.
\endproclaim

This situation is almost parallel to the twistor-theoretical formulation
of the self-dual vacuum Einstein equation and hyper-K\"ahler
geometry [16,17].

%
%
%
%

\subhead 1.4. $S$ function and $\tau$ function (free energy) \endsubhead

Here we introduce two important potentials $S$ and $\log\taudkp$
hidden behind the system. Later we shall see that $S$ is the phase
function of the WKB analysis of the KP hierarchy and $\log\taudkp$ is
the leading term of the logarithm of the tau function of the KP hierarchy
$\log\taukp$ (See \S1.7.4). Function $\log\taudkp$ ($=F$) is also
called ``free energy'' in the context of matrix models and topological
minimal models (See \S1.8).

First rewrite (1.3.10) as follows:
$$
    d\left( \calM d\calL + \sum_{n=1}^\infty \calB_n dt_n \right) = 0.
\tag 1.4.1
$$
This implies the existence of a function $S$ such that
$$
    dS = \calM d\calL + \sum_{n=1}^\infty \calB_n dt_n,
\tag 1.4.2
$$
or, equivalently,
$$
\align
    \calM &= \left(\frac{\der S}{\der \calL}\right)
                       _{t\fixed},
    \tag 1.4.3 \\
    \calB_n &= \left(\frac{\der S}{\der t_n}\right)
                       _{\calL,t_m(m\not=n)\fixed}.
    \tag 1.4.4 \\
\endalign
$$
This potential $S$ is introduced by Krichever in a different manner [7].
His formulation comes from the Whitham theory [23] and,
$S$ plays rather a central role than $\calM$.

Actually $S$ is not a true potential, but can be written directly
in terms of $\calL$ and $\calM$ as follows.

\proclaim{Proposition 1.4.1}([1] Proposition 3)
$S$ is given by
$$
    S = \sum_{i=1}^\infty t_i \calL^i + \sum_{i=1}^\infty S_i\calL^{-i},
    \quad S_i = - \frac{1}{i}v_i.
\tag 1.4.5
$$
In particular, $\calB_n$ can be written explicitly in terms of the Laurent
coefficients $S_i$ as
$$
    \calB_n =
    \calL^n + \sum_{i=1}^\infty \frac{\der S_i}{\der t_n}\calL^{-i}.
\tag 1.4.6
$$
\endproclaim

Function $\taudkp$ is a true potential and defined as follows:
$$
    d \log\taudkp(t) = \sum_{n=1}^\infty v_n(t) d t_n.
\tag 1.4.7
$$
Here $d$ denotes the differentiation with respect to $t=(t_1, t_2, \ldots)$.
\proclaim{Proposition 1.4.2}([1] Proposition 6)
The right hand side of (1.4.7) is a closed form. Accordingly,
there exists a potential $\log\taudkp$.
\endproclaim

This tau function contains enough information to reproduce any ingredient
of the dispersionless KP hierarchy (see [1] (5.5 -- 7)). Therefore,
in principle, one should be able to rewrite the hierarchy
in terms of $\log\taudkp$.
For the case of the ordinary KP hierarchy, the tau function is characterized
by the famous Hirota bilinear equations. For the dispersionless KP hierarchy,
we have a series of equations satisfied by $\log\taudkp$ (see Appendix B).
We conjecture that these equations are enough to characterize $\log\taudkp$.

%
%
%
%

\subhead 1.5. Riemann-Hilbert problem (twistor construction) \endsubhead

The nonlinear graviton (twistor) construction of Penrose [24]
for the self-dual vacuum Einstein equation and its hyper-K\"ahler
version [25] can be also extended to the dispersionless KP hierarchy.

\proclaim{Proposition 1.5.1} ([1] Proposition 7.)
Let a pair of two functions $(f(k,x), g(k,x))$ and
a pair of Laurent series $(\calL, \calM)$ be given. Suppose
\roster
\item[1] $\calL$ and $\calM$ have the form (1.1.2) and (1.3.2) respectively.
\item[2] $f$ and $g$ satisfy the canonical Poisson relation $\{f,g\}=1$.
\item[3] $\calP = f(\calL, \calM)$ and $\calQ = g(\calL, \calM)$ are both
         Taylor series in $k$:
$$
    (f(\calL, \calM))_{\leq -1} = 0, \qquad
    (g(\calL, \calM))_{\leq -1} = 0.
\tag 1.5.1
$$
\endroster
Then the pair $(\calL, \calM)$ gives a solution of the dispersionless KP
hierarchy. We call $(f,g)$ the {\it twistor data} of this solution.
\endproclaim

This construction of solutions is a kind of the Riemann-Hilbert decomposition;
the area-preserving diffeomorphism
$(\calL,\calM) \mapsto (f(\calL,\calM), g(\calL,\calM))$
is decomposed into two maps, $(\calL, \calM) \mapsto (k,x)$ and
$(k,x) \mapsto (\calP, \calQ)$. The former of these maps is rationally
extended to $k = \infty$ and the latter to $k = 0$.

For a pair $(f, g)$ there exists a unique solution $(\calL, \calM)$
of (1.5.1) if $(f, g)$ is close enough to the trivial one
$(f, g) = (k,x)$. Establishing the existence is a delicate issue in general,
and the statement largely depends on the category of functions
(formal, holomorphic, etc.).
In the following, we always assume the unique solvability of (1.5.1).
In contrast to this, the converse statement homds generally.

\proclaim{Proposition 1.5.2}
Each solution of the dispersionless KP hierarchy possestes the twistor data,
i.e., if $(\calL, \calM)$ is a solution of the dispersionless KP hierarchy,
there exists a pair $(f, g)$ which satisfies (2) and (3) of
Proposition 1.5.1.
\endproclaim

\demo{Proof}
Let $\exp(\ad\varphi(t))$ be the dressing operator corresponding to
$(\calL, \calM)$. Set
$$
    f(k,x) = e^{-\ad\varphi(t=0)} k, \qquad
    g(k,x) = e^{-\ad\varphi(t=0)} x.
\tag 1.5.2
$$
Obviously the pair $(f, g)$ satisfies the canonical Poisson relation
$\{f, g\} = 1$. Since $\calL(t=0) = \exp(\ad\varphi(t=0)) k$ and
$\calM(t=0) = \exp(\ad\varphi(t=0)) x$,
we have from (1.5.2) the following relations:
$$
\split
    f(\calL(t=0), \calM(t=0)) &= e^{\ad\varphi(t=0)}f(k,x) = k,\\
    g(\calL(t=0), \calM(t=0)) &= e^{\ad\varphi(t=0)}g(k,x) = x.
\endsplit
\tag 1.5.3
$$
Now we prove that $f(\calL, \calM)_{\leq -1} = 0$. As $\calL$ and $\calM$
satisfy the Lax equations (1.1.1) and (1.3.3),
$$
    \frac{\der}{\der t_n} f(\calL, \calM) = \{\calB_n, f(\calL, \calM)\}.
\tag 1.5.4
$$
Evaluating (1.5.4) at $t=0$, we see that
$(\der f(\calL, \calM)/\der t_n) (t=0)$ does not contain negative powers of
$k$ because of (1.5.3). In this way we can prove, by induction, that
$(\der/\der t)^\alpha f(\calL, \calM)(t=0)$, i.e., coefficients
of the Taylor expansion at $t=0$ do not contain negative powers of $k$
for any multi-index $\alpha$. Thus we have proved that
$f(\calL, \calM)_{\leq -1} = 0$. $g(\calL, \calM)_{\leq -1} = 0$ is proved
in the same way.
\qed
\enddemo

This construction of solutions is closely related to
the $w_{1+\infty}$-tymmetry of solutions as we shall see in the next section.

%
%
%
%

\subhead 1/6. $w_{1+\infty}$ symmetrz \endsubhead

In the previous section we established a correspondence between
a solution $(\calL, \calM)$ and an area-preserving diffeomorphism
$(f(k,x), g(k,x))$. Since area-preserving maps form a group (SDiff(2)
in the notation of [1]), we can consider an action of this group on
the space of solutions of the dispersionless KP hierarchy.
In this section we consider an infinitesimal form of this action,
$w_{1+\infty}$-action.
We follow the method developed for
the self-dual vacuum Einstein equation [26], [17].

Explicitly speaking, we define infinitesimal deformations of $(f, g)$
by a Hamiltonian vector field,
$$
    (f,g) \longrightarrow  (f,g) \circ \exp(-\eps \ad F),
$$
and the associated deformation
$$
    (\calL, \calM) \longrightarrow  (\calL(\eps), \calM(\eps))
$$
of the $(\calL, \calM)$-pair of the dispersionlest KP hierarchy.
Here $\ad F$ is regarded as a Hamiltonian vector field
$$
    \ad F = \frac{\der F}{\der k} \frac \der {\der x}
           -\frac{\der F}{\der x} \frac \der {\der k},
$$
and $\eps$ is an infinitesimal parameter.
By {\it infinitesimal symmetry} we mean the first order
coefficient $\delta_F(\cdot)$ in $\eps$-expansion:
$$
\split
    \calL(\eps) &= \calL + \eps \delta_F \calL + O(\eps^2),          \\
    \calM(\eps) &= \calM + \eps \delta_F \calM + O(\eps^2).
\endsplit
\tag 1.6.1
$$
By definition, $\delta_F$ acts on any function of $\calL$ and $\calM$
as an abstract derivation,
$$
    \delta_F G(\calL,\calM) = \frac{\der G}{\der \calL}\delta_F \calL
                             +\frac{\der G}{\der \calM}\delta_F \calM
\tag 1.6.2
$$
while the independent variables of the hierarchy are invariant under this
action
$$
   \delta_F t_n = \delta_F x = \delta_F k = 0.
\tag 1.6.3
$$
(This is a formal way to understand infinitesimal symmetries of
differential equations in the language of differential algebras
[27]).

A more explicit expression of these infinitesimal symmetries is
given by the following propositions.

\proclaim{Proposition 1.6.1} ([1] Proposition 13.)
The infinitesimal symmetries of $\calL$ and $\calM$ are given by
$$
\split
    \delta_F\calL &= \{ F(\calL,\calM)_{\leq -1}, \calL \},\\
    \delta_F\calM &= \{ F(\calL,\calM)_{\leq -1}, \calM \}.
\endsplit
\tag 1.6.4
$$
\endproclaim

\proclaim{Proposition 1.6.2} ([1] Proposition 14.)
For $i=1,2,\ldots$, $\delta_F v_i$ are given by
$$
    \delta_F v_i = -\Res F(\calL,\calM)d_k B_i.
\tag 1.6.5
$$
\endproclaim

This is an analogue of the symmetries of Plebanski's key functions
of the self-dual Einstein equation [17].

\proclaim{Proposition 1.6.3}
The infinitesimal symmetry of the dressing function $\delta_F \varphi$
is determined by the relation
$$
    \nabla_{\delta_F,\varphi}\varphi = F(\calL, \calM)_{\leq -1},
\tag 1.6.6
$$
or equivalently, by
$$
    \delta_F \varphi
    =
    \frac{\ad\varphi}{e^{\ad\varphi} - 1} F(\calL, \calM)_{\leq -1}
    \=def
    \left. \frac{T}{e^T - 1} \right|_{T = \ad\varphi}
    F(\calL, \calM)_{\leq -1},
\tag 1.6.$6'$
$$
where $T/(e^T -1)$ is understood as a power series of an indeterminate $T$;
$$
    \frac{T}{e^T - 1} = 1 - \frac12 T + \cdots.
$$
\endproclaim
Note that Proposition 1.6.1 is a direct consequence of this proposition.

\demo{Proof}
Let $(f,g)$ be the twistor data corresponding to $(\calL, \calM)$.
By definition the deformed twistor data are given by
$$
\split
    (f_\eps(k, x) , g_\eps(k,x))
    &=
    (e^{- \eps \ad F} f(k,x), e^{- \eps \ad F} g(k,x)) \\
    &=
    (1 - \eps \ad F) \circ (f(k,x), g(k,x)) + O(\eps^2).
\endsplit
\tag 1.6.7
$$
Let us denote the deformed dressing function by
$\varphi_\eps(t) = \varphi(t) + \eps \delta_F \varphi(t) + O(\eps^2)$.
We have (cf\. Lemma A.1)
$$
    e^{\ad \varphi_\eps(t)} e^{\ad \sum t_n k^n}
    =
    (1 + \eps \ad \nabla_{\delta, \varphi}\varphi )\circ
    e^{\ad \varphi (t)} e^{\ad \sum t_n k^n}.
\tag 1.6.8
$$
The Riemann-Hilbert problem (1.5.1) for the deformed solution looks as
$$
\aligned
    (e^{\ad \varphi_\eps(t)} e^{\ad \sum t_n k^n} f_\eps(k,x))_{\leq -1}
    &= 0, \\
    (e^{\ad \varphi_\eps(t)} e^{\ad \sum t_n k^n} g_\eps(k,x))_{\leq -1}
    &= 0.
\endaligned
\tag 1.6.9
$$
Using (1.6.7) and (1.6.8), we have
$$
\split
    e^{\ad \varphi_\eps(t)} e^{\ad \sum t_n k^n} f_\eps(k,x)
    =&
    e^{\ad \varphi (t)} e^{\ad \sum t_n k^n} f(k,x) + \\
    &+
    \eps (
    \{
      \nabla_{\delta, \varphi} \varphi,
      e^{\ad\varphi(t)} e^{\ad \sum t_n k^n} f(k,x)
    \} \\
    &-
    e^{\ad\varphi(t)} e^{\ad\sum t_n k^n} \{F, f(k,x) \}
    ) + O(\eps^2) \\
    =&
    f(\calL, \calM) \\
    &+
    \eps( \{ \nabla_{\delta, \varphi} \varphi, f(\calL, \calM)\}
         -\{ F(\calL, \calM),                  f(\calL, \calM)\}) \\
    &+ O(\eps^2).
\endsplit
\tag 1.6.10
$$
We have a similar formula for $g$.
Therefore the coefficients of $\eps$ in $(\quad)_{\leq -1}$
on the left hand side of (1.6.9) become
$$
    \{\nabla_{\delta, \varphi} \varphi - F(\calL, \calM), f(\calL,\calM)\},
    \quad
    \{\nabla_{\delta, \varphi} \varphi - F(\calL, \calM), g(\calL,\calM)\}.
$$
Hence the following Lemma 1.6.4 proves this Proposition.
\qed
\enddemo

\proclaim{Lemma 1.6.4}
Suppose that $\calP(k,x)$ and $\calQ(k,x)$ do not contain negative powers
of $k$ and satisfy the canonical Poisson relation $\{ \calP, \calQ\} = 1$.
If $\calA(k,x)$ Poisson-commute with $\calP$ and $\calQ$
up to positive powers of $k$, i.e.,
$$
    (\{\calA, \calP\})_{\leq -1} = 0, \qquad
    (\{\calA, \calQ\})_{\leq -1} = 0,
\tag 1.6.11
$$
then $\calA$ itself does not contain negative powers of $k$.
\endproclaim

\demo{Proof}
We assume $\calA_{\geq 0} = 0$ and show $\calA = 0$ under the assumption
of the lemma. Lemma 1.6.4 follows immediately from this.

By definition of the Poisson bracket,
$$
\align
    \tilde\calP \=def \{\calA, \calP\}
    &=
    \frac{\der \calA}{\der k} \frac{\der \calP}{\der x}
    -
    \frac{\der \calA}{\der x} \frac{\der \calP}{\der k},
    \\
    \tilde\calQ \=def \{\calA, \calQ\}
    &=
    \frac{\der \calA}{\der k} \frac{\der \calQ}{\der x}
    -
    \frac{\der \calA}{\der x} \frac{\der \calQ}{\der k}.
\endalign
$$
Regarding this as a system of linear equations for $\der \calA/ \der k$,
$\der \calA/\der x$, we can rewrite it as
$$
    \pmatrix
    \frac{\der \calP}{\der x} & -\frac{\der \calP}{\der k} \\
    \frac{\der \calQ}{\der x} & -\frac{\der \calQ}{\der k}
    \endpmatrix
    \pmatrix
    \frac{\der \calA}{\der k} \\ \frac{\der \calA}{\der x}
    \endpmatrix
    =
    \pmatrix \tilde\calP \\ \tilde\calQ \endpmatrix.
\tag 1.6.12
$$
Because of the unimodularity of the matrix in the right hand side,
$\{\calP, \calQ\} = 1$, (1.6.12) is easily solved as
$$
    \pmatrix
    \frac{\der \calA}{\der k} \\ \frac{\der \calA}{\der x}
    \endpmatrix
    =
    \pmatrix
    -\frac{\der \calQ}{\der k} & \frac{\der \calP}{\der k} \\
    -\frac{\der \calQ}{\der x} & \frac{\der \calP}{\der x}
    \endpmatrix
    \pmatrix \tilde\calP \\ \tilde\calQ \endpmatrix.
\tag 1.6.13
$$
This means that both $\frac{\der \calA}{\der k}$ and
$\frac{\der \calA}{\der x}$ do not contain negative powers of $k$,
but as we assumed that $\calA_{\geq 0} = 0$, they must be 0.
Therefore $\calA = 0$.
\qed
\enddemo

These symmetries of the dispersionless KP hierarchy can be lifted up
to the tau function as follows.
Note that our definition of the tau function has ambiguity of additive
constant: this gives rise to an extra term (cocycle) in their commutation
relations.
\proclaim{Proposition 1.6.5} ([1] Proposition 15.)
The infinitesimal symmetries $\delta_F$
of the $(\calL,\calM)$-pair can be extended to the tau
function by defining
$$
    \delta_F\log\tau = - \Res F^x(\calL,\calM)d_k \calL,
\tag 1.6.14
$$
where $F^x(k,x)$ is a primitive function of $F(k,x)$ normalized as
$$
    F^x(k,x) \=def \int_0^x F(k,y)dy.
\tag 1.6.15
$$
This extended symmetry is compatible with the flows:
$$
    \frac{\der}{\der t_n}\delta_F\log\tau
    = \delta_F \frac{\der\log\tau}{\der t_n}.
\tag 1.6.16
$$
\endproclaim

These symmetries reflect the structure of the $w_{1+\infty}$ algebra as
follows.

\proclaim{Proposition 1.6.6} ([1] Proposition 16.)
For two generating functions $F_1=F_1(k,x)$ and $F_2=F_2(k,x)$,
the infinitesimal symmetries $\delta_{F_1}$ and $\delta_{F_2}$
obey the commutation relations
$$
\align
    \left[ \delta_{F_1},\delta_{F_2} \right]\log\tau
    &= \delta_{\{F_1,F_2\}}\log\tau + c(F_1,F_2),
\tag 1.6.17
\\
    \left[ \delta_{F_1},\delta_{F_2} \right] \calK
    &= \delta_{\{F_1,F_2\}} \calK, \quad
    (\calK = \calL, \calM),
\tag 1.6.18
\\
    \left[ \delta_{F_1},\delta_{F_2} \right] \varphi
    &= \delta_{\{F_1,F_2\}} \varphi,
\tag 1.6.19
\endalign
$$
where
$$
    c(F_1,F_2) \=def \Res F_1(k,0)dF_2(k,0),
\tag 1.6.20
$$
which is a cocycle of the $w_{1+\infty}$ algebra.
\endproclaim

\demo{Proof}
Equations (1.6.17) and (1.6.18) are proved in [1].
The statement for the dressing function (1.6.19) is an immediate consequence
of (1.3.4), (1.6.17) and the compatibility of the deformations
with the time flows (1.6.16).
\qed
\enddemo

%
%
%
%

\subhead 1.7. Quasi-classical limit of KP hierarchy \endsubhead

In this section we change our viewpoint and describe the results in
the previous sections as quasi-classical limit. We take a slightly
different way from [3] but the principle is the same.
Namely, we intsoduce a ``Planck constant'' $\hbar$ into the oreinary
KP hierarchy and take the limit $\hbar \to 0$. Leading terms in this limit
recover the dispersionless KP hierarchy.
The contents of the following subsections 1.7.1 -- 6 are organized to
correspond to the contents of \S\S1.1 -- 6, respectively.
%
%
%
%
\subsubhead 1.7.1 Lax formalism \endsubsubhead
We define the KP hierarchy in the presence of $\hbar$
as the Lax representation
$$
    \hbar \frac{\der L}{\der t_n} = [ B_n, L ], \quad
    B_n = (L^n)_{\geq 0}, \quad n=1,2,\ldots
\tag 1.7.1
$$
where $L$, the counterpart of $\calL$, is a micro-differential operator
(in $x$),
$$
    L = \hbar \der
      + \sum_{n=1}^\infty u_{n+1}(\hbar,t)(\hbar\der)^{-n}, \quad
    \der = \der/\der x,
\tag 1.7.2
$$
and ``$(\quad)_{\geq 0}$'' now stands for the projection onto a
differential operator dropping negative powers of $\der$. The
coefficients $u_n(\hbar,t)$ of $L$ are assumed to be regular
with respect to $\hbar$.
This means that they have such an asymptotic form as
$u_n(\hbar,t) = u^0_n(t) + O(\hbar)$ around $\hbar = 0$. Every ingredient
in the KP hierarchy is now to depend on $\hbar$. For later discussion,
we now define the ``{\it order}'' and the ``{\it principal symbol}\/'' of
micro-differential operators  with $\hbar$ as follows.

\demo{Definition 1.7.1}
$$
    \ord \left( \sum a_{n,m}(t) \hbar^n \der^m \right)
    \=def
    \max \{ m-n \,|\, b_{n,m}(t) \neq 0 \}.
\tag 1.7.3
$$
In particular, $\ord (\hbar) = -1$, $\ord (\der) = 1$.
The {\it principal symbol} (resp\. the {\it symbol of order $l$})
of a micro-differential operator $A = \sum a_{n,m}\hbar^n \der^m$ is
$$
\alignedat2
    \symbolh  (A)
    &\=def \hbar^{-\ord(A)}& \sum_{m-n = \ord(A*} &a_{n,m}  k^m
    \\
    (\text{resp. }
    \symbolh_m(A)
    &\=def \hbar^{-l}& \sum_{m-n = l} \quad &a_{n,m}  k^m).
\endalignedat
\tag 1.7.4
$$
\enddemo

For example, the condition which we imposed on the coefficients
$u_n(\hbar,t)$ can be restated as $\ord (L) = 0$.

\demo{Remark 1.7.2}
This ``order'' coincides with the order of an micro-differential operator
if we formally replace $\hbar$ with $\der_{t_0}^{-1}$, where $t_0$ is
an extra variable. In fact, naively extending
(1.7.1) to $n=0$, we can introduce the time variable $t_0$ on which
nothing depends.
\enddemo

Following facts are easily proved.

\proclaim{Lemma 1.7.3}
Let $P_i$ ($i=1,2$) be two micro-differential operators
of finite order: $\ord (P_i) = l_i$.
\roster
\item[1] $\ord(P_1 P_2) = l_1 + l_2$ and
$$
    \symbolh (P_1 P_2) = \symbolh(P_1) \symbolh(P_2).
$$
\item[2] $\ord([P_1, P_2]) \leqq l_1 + l_2 - 1$ and
$$
    \symbolh_{l_1 + l_2 - 1} ([P_1, P_2])
    =
    \hbar \{ \symbolh(P_1), \symbolh(P_2) \},
$$
\endroster
where $\{\quad, \quad\}$ is the Poisson bracket defined by (1.1.3).
\endproclaim

Taking the principal symbol of (1.7.1) and using Lemma 1.7.3, we obtain
\proclaim{Proposition 1.7.4}
$\calL = \symbolh(L)$ is a solution of the dispersionlets KP hierarchy
(1.1.1*, and $\calB_n = \symbolh(B_n)$.
\endproclaim

%
%
%
%

\subsubhead 1.7.2 Dressing operators \endsubsubhead
In order to find a counterpart of the dressing function $\varphi$,
we need a more refined construction
of the dressing operator $W$ of the ordinary KP hierarchy in our context.

\proclaim{Proposition 1.7.5}
There is a dressing operator $W$,
$$
\gathered
    L = W (\hbar\der) W^{-1},
\\
    \hbar \frac{\der W}{\der t_n} W^{-1} =
    -(W (\hbar\der)^n W^{-1})_{\leq -1},
\endgathered
\tag 1.7.5
$$
of the form
$$
\gathered
    W = \exp( \hbar^{-1} X(\hbar, t, \der) ),
\\
    X(\hbar, t, \der) = \sum_{n=1}^\infty \chi_n(\hbar, t) (\hbar\der)^{-n},
    \quad
    \ord (X(\hbar, t, \der)) = 0.
\endgathered
\tag 1.7.6
$$

Conversely, if $W$ of the form (1.7.6) satisfies (1.7.5),
then $L = W (\hbar\der) W^{-1}$ is a soluiton of the Lax equations (1.7.1).
\endproclaim

\demo{Proof}
The proof is mostly parallel to the proof of Proposition 1.2.1.
According to Lemma 1.7.3, the linear space of micro-differential operators
of order not greater than $1$,
$\calE_1 = \{ P:$ micro-differential operator $| \ord(P) \leqq 1\}$,
is closed under the commutator and thus is a Lie algebra.
Therefore, replacing $\{\quad, \quad\}$, $\varphi$, $k$
by $[\quad, \quad]$, $X/\hbar$, $\hbar\der$, respectively,
we can carry over the proof of Proposition 1.2.1 to this case.
Here $\ad(X/\hbar)(\cdot) = [X/\hbar, \cdot]$.
\qed
\enddemo

The dressing function $\varphi$ of the dispersionless KP hierarchy and
$W$ are connected as follows:
\proclaim{Corollary 1.7.6}
$\symbolh(X)$ gives a dressing function $\varphi(t)$ of $\calL = \symbolh(L)$
in the sense of \S1.2.
Conversely, if $\varphi(t)$ is a dressing function of a solution $\calL$
of the dispersionless KP hierarchy,
there exists a solution of the KP hierarchy $L$ and
a corresponding dressing operator $W= \exp(\hbar^{-1}X)$ such that
$\calL = \symbolh(L)$, $\varphi(t) = \symbolh(X)$.
\endproclaim

\demo{Proof}
The first statement is obvious from the above proof.
The second statement is proved by the unique solvability of the initial
value problem of the KP hierarchy: suppose $\varphi$ and $\calL$ are given.
Then there exists a micro-differential operator $X_0(\hbar,x,\hbar\der)$
such that
$\symbolh(X_0) = \varphi(t_1=x, t_2 = t_3 = \ldots = 0)$.
The solution of the KP hierarchy $L$ and a corresponding dressing operator
$W$ with the initial value
$W(t_1 = x, t_2 = t_3 = \ldots = 0) = \exp(\hbar^{-1} X_0)$
is the desired $W$ and $L$.
\qed
\enddemo

Regarding $\hbar = \der_{t_0}^{-1}$ as in Remark 1.7.2,
$\Ad (W \exp(\sum t_n (\hbar\der)^n))$ is nothing but
the ``{\it quantized canonical transformation}'' [19]
corresponding to the canonical transformation
$\exp(\ad\varphi)\exp(\ad\sum t_n k^n)$.

%
%
%
%

\subsubhead 1.7.3 Orlov operator \endsubsubhead
The {\it Orlov operator} $M$ is defined by
$$
\split
    M &= W \left( \sum_{n=1}^\infty n t_n (\hbar \der)^{n-1} + x \right)
         W^{-1}
       = \Ad \left(
             W \exp\left(\hbar^{-1} t(\hbar\der) \right)
             \right)x \\
      &= \sum_{n=1}^\infty n t_n L^{n-1} + x +
         \sum_{n=1}^\infty v_n(\hbar, t) L^{-n-1},
\endsplit
\tag 1.7.7
$$
where $t(\hbar\der) = \sum_{n=1}^\infty t_n (\hbar\der)^n$.
The proof of the gollowing facts is analogous to that of Proposition 1.3.1.

\proclaim{Proposition 1.7.7}
\roster
\item[1] $\ord(M) = 0$;
\item[2] $[L, M] = \hbar$;
\item[3] $M$ satisfies the Lax equations,
$$
    \hbar \frac{\der M}{\der t_n} = [ B_n, M ],\quad
    n=1,2,\ldots,
\tag 1.7.8
$$
\endroster
where $B_n$ are defined in (1.7.1).

Conversely, if $M$ of the form
$$
    M=
    \sum_{n=1}^\infty n t_n L^{n-1} + x +
    \sum_{n=1}^\infty v_n(\hbar,t) L^{-n-1}
$$
satisfies (1) -- (3) above, then there exists a unique dressing operator $W$
such that $M$ is expressed as (1.7.7).
\endproclaim
\proclaim{Corollary 1.7.8}
$\calM = \symbolh(M)$ is the Orlov function of the dispersionless
KP hierarchy corresponding to $\calL = \symbolh(L)$.
\endproclaim

%
%
%
%

\subsubhead 1.7.4. $S$ function and $\tau$ function (free energy)
\endsubsubhead
The $S$ function introduced in \S1.4 appears as the phase function
of the WKB analysis of the system.
As usual we define the {\it Baker-Akhiezer function}
(or the {\it wave function} in the terminology of [15]) by
$$
    \Psi(\hbar, t, \lambda) =
    W(\hbar, t, \der)
    \exp \left(\hbar^{-1} t(\lambda) \right), \quad
    t(\lambda) \=def \sum_{n=1}^\infty t_n \lambda^n
\tag 1.7.9
$$
that satisfies the linear equations
$$
    \lambda \Psi = L \Psi, \quad
    \hbar \frac{\der \Psi}{\der \lambda} = M \Psi, \quad
    \hbar \frac{\der \Psi}{\der t_n} = B_n \Psi.
\tag 1.7.10
$$
due to (1.7.5), (1.7.7).
Since $\hbar\der \,\exp(\hbar^{-1} t(\lambda)) =
      \lambda \exp(\hbar^{-1} t(\lambda))$, we have
$$
    \Psi(\hbar,t,\lambda)
    = ( 1 + O(\lambda^{-1}) )
      \exp \left( \hbar^{-1} t(\lambda) \right), \quad
    \lambda \to \infty, \ \hbar \to 0.
\tag 1.7.11
$$
More precisely, Proposition 1.7.7 (1) and equations (1.7.10)
imply the following
\proclaim{Proposition 1.7.9}
The Baker-Akhiezer function has the following WKB asymptotic expansion
as $\hbar \to 0$:
$$
    \Psi(\hbar, t, \lambda)
    =
    \exp( \hbar^{-1} S(t,\lambda) + O(\hbar^0) ),
\tag 1.7.12
$$
where $S(t, \lambda) = \sum_{n=1}^\infty t_n \lambda^n +
                       \sum_{n=1}^\infty S_n(t) \lambda^{-n}$.
\endproclaim

We can further pursue the WKB analysis, regarding $S(t, \lambda)$
as the phase function. The Hamilton-Jacobi (or eikonal) equations
corresponding to (1.7.10) are
$$
\align
    &\lambda = \frac{\der S(t,\lambda)}{\der x}
             +\sum_{n=1}^\infty u^0_{n+1}(t)
               \left( \frac{\der S(t,\lambda)}{\der x} \right)^{-n}
             = \left.\symbolh(L)\right|_{k=\frac{\der S(t,\lambda)}{\der x}},
\tag 1.7.13 \\
    &d S(t,\lambda) = \calM(t,\lambda) d \lambda
      + \sum_{n=1}^\infty \calB_n(t,\lambda) dt_n,
\tag 1.7.14
\endalign
$$
where
$$
\split
  & \calM(t,\lambda) =
    \left. \symbolh(M)\right|_{\calL = \lambda}
      = \sum_{n=1}^\infty nt_n \lambda^{n-1}
       +\sum_{n=1}^\infty v^0_n(t) \lambda^{-n-1},
\\
  & \calB_n(t,\lambda) =
    \left. \symbolh(B_n) \right|_{k = \frac{\der S(t,\lambda)}{\der x}}
      = \left( \frac{\der S(t,\lambda)}{\der x}\right)^n
       +\sum_{i=0}^{n-2} b^0_{n,i}(t)
          \left( \frac{\der S(t,\lambda)}{\der x} \right)^i,
\endsplit
\tag 1.7.15
$$
and the coefficients $b^0_{n,i}(t)$, like $u^0_n(t)$ and $v^0_n(t)$,
are the leading ($\hbar^0$) term in $\hbar$-expansion of the
coefficients $b_{n,i}(\hbar,t)$ of
$B_n = (\hbar\der)^n + \sum_{i=0}^{n-2}b_{n,i}(\hbar,t)(\hbar\der)^i$.

Thus we obtain
\proclaim{Proposition 1.7.10}
Under the Legendre-type transformation $(t,\lambda) \to (t,k)$ given by
$$
    k = \frac{\der S(t,\lambda)}{\der x}
    \quad (x=t_1),
\tag 1.7.16
$$
the spectral parameter $\lambda$ turns into the $\calL$-function of
the dispersionless KP hierarchy,
$$
    \lambda = \calL(t,k) = k + \sum_{n=1}^\infty u^0_{n+1}(t* k^{-n},
$$
whereas
$S(t, \lambda)$ becomes the corresponding $S$ function,
$$
    S(t, \lambda) = S = \sum_{n=1}^\infty t_n \calL^n
                      + \sum_{n=1}^\infty S_n \calL^{-n}.
$$
\endproclaim

The tau fvnction $\tau(\hbar,t)$ is by definition a fvnctjon
that reproducet the Baker-Dkhiezer function as
$$
\split
  & \Psi(\hbar,t)
    = \dfrac{ \tau(\hbar, t - \hbar[\lambda^{-1}]) }
            { \tau(\hbar,t ) }
      \exp( \hbar^{-1} t(\lambda)),
                                                              \\
  &  [\lambda^{-1}]
    = \left( \frac{1}{\lambda}, \frac{1}{2\lambda^2},
             \frac{1}{3\lambda^3}, \ldots            \right).
\endsplit
\tag 1.7.17
$$
One can easily show that the tau function $\tau(\hbar,t)$ must
behave as
$$
    \tau(\hbar,t) = \exp[ \hbar^{-2}F(t) + O(\hbar^{-1})]  \quad
    (\hbar \to 0)
\tag 1.7.18
$$
so as to reproduce the WKB asymptotic form of the Baker-Akhiezer
function (1.7.12). Combining (1.7.17) and the second equation of
(1.7.10), one can further prove that the function $F(t)$ satisfies
the equations
$$
    \frac{\der F(t)}{\der t_n} = v_n^0(t), \quad
    n = 1,2,\ldots.
\tag 1.7.19
$$
Therefore $F(t)$ can be identified with $\log\taudkp$.
%
%
%
%

\subsubhead 1.7.5 Riemann-Hilbert problem (twistor construction)
\endsubsubhead
The twistor construction of \S1.5 can be extended to the construction
of solutions to the ordinary KP hierarchy under the present setting.
The parameter $\hbar$ turns out to play a rather crucial role.

\proclaim{Proposition 1.7.11}
Suppose that
$$
\align
    f(\hbar, x, \hbar\der) &=
    \sum_{n\in\Integer} f_n(\hbar, x) (\hbar\der)^n,\\
    g(\hbar, x, \hbar\der) &=
    \sum_{n\in\Integer} g_n(\hbar, x) (\hbar\der)^n
\endalign
$$
are micro-differential operators of 0-th order (in the sense of (1.7.3)),
$\ord f = \ord g = 0$, and that they satisfy the canonical commutation
relation $[f, g] = \hbar$.
Assume that micro-differential operators $L$ of the form (1.7.2)
and $M$ of the form
$M = \sum_{n=1}^\infty n t_n L^{n-1} + x
   + \sum_{n=1}^\infty v_n(\hbar,t) L^{-n-1}$ are given and that
$\ord L = \ord M = 0$, $[L,M] = \hbar$.

Then,
if $f(\hbar, M, L)$ and $g(\hbar, M, L)$ are both differential operators,
i.e.,
$$
    (f(\hbar, M, L))_{\leq -1} = (g(\hbar, M, L))_{\leq -1} = 0,
\tag 1.7.20
$$
$L$ is a solution of the KP hierarchy, and
$M$ is the corresponding Orlov operator.
\endproclaim

\demo{Proof} From
the assumption on the form of $L$ and $M$, it is easy to prove that
there exists a micro-differential operator of the form
$$
    X(\hbar, t, \der) = \sum_{n=1}^\infty \chi_n(\hbar, t) (\hbar\der)^{-n},
    \quad
    \ord (X(\hbar, t, \der)) = 0
$$
which satisfies
$$
\align
    L(\hbar, t, \hbar\der) &= \Ad(W) (\hbar \der), \\
    M(\hbar, t, \hbar\der) &=
    \Ad \left(W \exp\left(\hbar^{-1} t(\hbar\der)\right)\right) x,
\endalign
$$
where $W(\hbar, t, \hbar\der) = \exp(\hbar^{-1} X(\hbar, t, \hbar\der))$,
$t(\hbar\der) = \sum_{n=1}^\infty t_n (\hbar\der)^n$.
Put $P = f(\hbar, M, L)$, $Q = g(\hbar, M, L)$. The assumption (1.7.20) of
the proposition says that $P$ and $Q$ are differential operators
$$
    P_{\leq -1} = Q_{\leq -1} = 0.
\tag 1.7.$20'$
$$
Differentiating the equation
$$
    P = \Ad\left( W \exp\left(\hbar^{-1} t(\hbar\der)
           \right)      \right)
           f(\hbar, x, \hbar\der),
$$
we obtain
$$
    \hbar \frac{\der P}{\der t_n}
    =\left[
    \hbar \frac{\der W}{\der t_n} W^{-1} + W (\hbar \der)^n W^{-1}, P
    \right],
\tag 1.7.21
$$
and similarly,
$$
    \hbar \frac{\der Q}{\der t_n}
    =\left[
    \hbar \frac{\der W}{\der t_n} W^{-1} + W (\hbar \der)^n W^{-1}, Q
    \right].
\tag 1.7.22
$$
Put $A = \hbar \frac{\der W}{\der t_n} W^{-1} + W (\hbar \der)^n W^{-1}$.
The statement of the proposition is equivalent to $A_{\leq -1} = 0$,
since this means
$$
    \hbar \frac{\der W}{\der t_n} = - (L^n)_{\leq -1} W,
$$
and it is well known that this is an alternative expression of the KP
hierarchy in terms of the dressing operator: $L = W(\hbar\der)W^{-1}$.

Now let us suppose that $A_- \=def A_{\leq -1} \neq 0$.
Putting $A_+ \=def A_{\geq 0}$, (1.7.21) and (1.7.22) imply that
$$
\split
    [A_-, P] &= \hbar \frac{\der P}{\der t_n} - [A_+, P], \\
    [A_-, Q] &= \hbar \frac{\der Q}{\der t_n} - [A_+, Q].
\endsplit
\tag 1.7.23
$$
Due to (1.7.$20'$), the right hand sides of (1.7.23) are differential
operators. It follows from the assumption $A_- \neq 0$ that
$\calA_- \=def \symbolh(A_-) \neq 0$.
Thus taking the principal symbol of (1.7.23), we have
$$
    (\{\calA_-, \calP\})_{\leq -1} = (\{\calA_-, \calQ\})_{\leq -1} =0,
$$
where $\calP = \symbolh(P)$, $\calQ = \symbolh(Q)$.
Moreover, the canonical commutation relation $[P, Q]=\hbar$ leads to
the canonical Poisson relation $\{\calP, \calQ\} = 1$. Hence according to
Lemma 1.6.4, $\calA_- = 0$. This is a contradiction.
\qed
\enddemo

As in the case of the dispersionless KP hierarchy, we call the pair
$(f, g)$ above the {\it twistor data\/}.
\proclaim{Proposition 1.7.12}
Any solution of the KP hierarchy possesses a twistor data,
i.e., if $(L, M)$ is a soluiton of the KP hierarchy in the presence of
$\hbar$, there exists a pair $(f, g)$ which satisfies
the canonical commutation relation $[f, g]=\hbar$ and (1.7.20).
In addition, we can choose $f$ and $g$ such that $f$ is
a first order micro-differential operator with respect to $\der_x$
and $g$ a zeroth order micro-differential operator.
\endproclaim

\demo{Proof}
Let $W = \exp(\hbar^{-1} X)$, $\ord(X)=0$, be the dressing operator
chosen in Proposition 1.7.5. Set
$$
    f(\hbar, x, \hbar\der) \=def W(t=0)^{-1} (\hbar \der) W(t=0), \qquad
    g(\hbar, x, \hbar\der) \=def W(t=0)^{-1} x            W(t=0).
$$
Obviously, $[f,g] = \hbar$ and
$$
\split
    \left. f(\hbar, M, L) \right|_{t=0} &=
    W(t=0) f(\hbar, x, \hbar\der) W(t=0)^{-1} = \hbar\der, \\
    \left. g(\hbar, M, L) \right|_{t=0} &=
    W(t=0) g(\hbar, x, \hbar\der) W(t=0)^{-1} = x
\endsplit
\tag 1.7.24
$$
do not contain negative powers of $\hbar\der$.
Due to (1.7.5),
$$
\split
    \hbar \frac{\der}{\der t_n} f(\hbar, M, L)
    &= \left[
    \hbar \frac{\der W}{\der t_n} W^{-1} + W (\hbar \der)^n W^{-1},
    f(\hbar, M, L)
    \right] \\
    &= [ B_n, f(\hbar, M, L)],
\endsplit
\tag 1.7.25
$$
where $B_n = (L^n)_{\geq 0}$. Proceeding in this manner, we can prove
by induction that all the coefficients of the Taylor expansion of
$f( \hbar, M, L)$ with respect to $t$ are differential operators.
The proof of $(g(\hbar, M, L))_{\leq -1} = 0$ is the same.
\qed
\enddemo

Needless to say, the principal symbols $\symbolh(f)$ and $\symbolh(g)$ give
the twistor data for the $(\calL,\calM)$-pair
$\calL = \symbolh(L)$, $\calM = \symbolh(M)$ of the dispersionless KP
hierarchy.

The uniqueness of the solution to (1.7.20) is true in the following
infinitesimal sense.

\proclaim{Proposition 1.7.13}
Suppose that (1.7.20) has a one-parameter family of solutions
$(L(\eps), M(\eps))$ regular in a neighborhood of $\eps = 0$:
$$
    (f(\hbar, M(\eps), L(\eps)))_{\leq -1}
    =
    (g(\hbar, M(\eps), L(\eps)))_{\leq -1} = 0.
\tag 1.7.26
$$
Then $L(\eps) = L(0)$, $M(\eps) = M(0)$ for any $\eps$.
\endproclaim

\demo{Proof}
We can choose $W = W(\eps)$ in the proof of Proposition 1.7.11
so that $W(\eps)$ depends on $\eps$ regularly.
Set $P(\eps) = f(\hbar, M(\eps), L(\eps))$,
    $Q(\eps) = g(\hbar, M(\eps), L(\eps))$. Since they are differential
operators, both hand sides of
$$
\align
    \frac{\der P}{\der \eps} &=
    \left[ \frac{\der W}{\der \eps} W^{-1}, P \right], \\
    \frac{\der Q}{\der \eps} &=
    \left[ \frac{\der W}{\der \eps} W^{-1}, Q \right]
\endalign
$$
are also differential operators. Thus, as in the proof of Proposition 1.7.11,
$(\der W/\der \eps) W^{-1}$ should not contain negative powers of $\der$,
hence $\der W/\der \eps = 0$.
\qed
\enddemo
%
%
%
%

\subsubhead 1.7.6. $W_{1+\infty}$ symmetry \endsubsubhead
We now turn to the issue of $W_{1+\infty}$ and $w_{1+\infty}$ symmetries,
studied in [3]. There are two possible ways to deform solutions of
the KP hierarchy in our context;
one is to deform the twistor data as in \S1.6, and
the other is to consider the action of the transformation group as in [3].

First we deform the twistor data
$(f(\hbar, x, \hbar\der), g(\hbar, x, \hbar\der))$ of a solution of
the KP hierarchy described in \S1.7.5, composing it to an infinitesimal
quantized contact transformation
$(\hbar\der, x) \mapsto
(\Ad(e^{-\eps Y})\hbar\der, \Ad(e^{-\eps Y})x)$.

\proclaim{Proposition 1.7.14}
The above symetries of solutions induced by a micro-differential operator
$Y(\hbar, x, \hbar\der)$ of order not greater than 1, $\ord(Y) \leqq 1$,
has explicit form as:
$$
\aligned
    \delta_Y L &= [(Y(\hbar, M, L))_{\leq -1}, L], \\
    \delta_Y M &= [(Y(\hbar, M, L))_{\leq -1}, M], \\
    \delta_Y W \cdot W^{-1} &= (Y(\hbar, M, L))_{\leq -1}.
\endaligned
\tag 1.7.27
$$
\endproclaim

\demo\nofrills{Proof\/}\ is the same as that in \S1.6, if one replaces
the Poisson bracket by the commutator, but we give here
the proof of the last statement for later discussion.
Let us denote the deformed twistor data by $(f_\eps,  g_\eps)$, and
the corresponding dressing operator by $W_\eps$. Define
$$
\align
    (P     , Q     ) &:= (\Ad(W e^{\hbar^{-1}t(\hbar\der)}) f,
                          \Ad(W e^{\hbar^{-1}t(\hbar\der)}) g ), \\
    (P_\eps, Q_\eps) &:= (\Ad(W_\eps e^{\hbar^{-1}t(\hbar\der)}) f_\eps,
                          \Ad(W_\eps e^{\hbar^{-1}t(\hbar\der)}) g_\eps),
\endalign
$$
which are, by construction, all differential operators.
As $f_\eps = \Ad(\exp(-\eps Y)) f$, we have
$$
\align
    P_\eps &=
    \Ad(W_\eps e^{ \hbar^{-1}t(\hbar\der)} \exp(-\eps Y)
               e^{-\hbar^{-1}t(\hbar\der)} W^{-1}) P,
\\
    Q_\eps &=
    \Ad(W_\eps e^{ \hbar^{-1}t(\hbar\der)} \exp(-\eps Y)
               e^{-\hbar^{-1}t(\hbar\der)} W^{-1}) Q.
\endalign
$$
Hence, by the same argument as the proof of Proposition 1.7.13,
$$
    W_\eps e^{ \hbar^{-1}t(\hbar\der)}\exp(-\eps Y)
           e^{-\hbar^{-1}t(\hbar\der)} W^{-1}
$$
is a differential operator. Therefore $W_\eps$ is given by the decomposition
$$
    W e^{ \hbar^{-1}t(\hbar\der)} e^{\eps Y}
      e^{-\hbar^{-1}t(\hbar\der)} = V_\eps W_\eps,
\tag 1.7.28
$$
where $V_\eps$ is a differential operator. Differentiating both hand sides
of (1.7.27) with respect to $\eps$, it is easy to see that
$$
\align
    W_\eps &=
    \exp(
    \eps (W e^{ \hbar^{-1}t(\hbar\der)} Y
            e^{-\hbar^{-1}t(\hbar\der)} W^{-1})_{\leq -1} + O(\eps^2)) W, \\
    V_\eps &=
    \exp(
    \eps (W e^{ \hbar^{-1}t(\hbar\der)} Y
            e^{-\hbar^{-1}t(\hbar\der)} W^{-1})_{\geq 0} + O(\eps^2)),
\endalign
$$
which proves the proposition.
\qed
\enddemo

Since both hand sides of (1.7.27) are in the Lie algebra
$\calE_1 = \{\ord(Y) \leqq 1 \}$,
we can take the symbol of order 0 of the first and second equations
of (1.7.27) and the symbol of order 1 of the last equation.
The result gives the infinitesimal deformation of a solution
of the dispersionless KP hierarchy in \S1.6
($F(k,x)$ therein is now given by $\hbar\symbolh_1(Y(\hbar, x, \hbar\der))$).

On the other hand we presented in [3] another approach to the symmetries
as follows.
A field theoretical interpretation of the KP hierarchy
was established by  Date et al. [15].
In [3] we studied quasi-classical limit of
$W_{1+\infty}$ symmetries in terms of vertex operators and free fermions
introduced by them:
we first modify their bosonized vertex operator
$Z(\lambdatilde,\lambda)$ by rescaling
$t_n \to \hbar^{-1}t_n$, $\der/\der t_n \to \hbar \der/\der t_n$ as:
$$
\gathered
    Z(\hbar,\lambdatilde,\lambda)
    = \dfrac
       { \exp [\hbar^{-1} t(\lambdatilde) - \hbar^{-1} t(\lambda)]
         \exp[-\hbar \dertilde(\lambdatilde^{-1})
              +\hbar \dertilde(\lambda^{-1})]
         -1 }
       {\lambdatilde-\lambda},                                   \\
    t(\lambda) = \sum_{n=1}^\infty t_n \lambda^n, \quad
    \dertilde(\lambda^{-1}) = \sum_{n=1}^\infty
         \lambda^{-n} \frac{1}{n} \frac{\der}{\der t_n}.
\endgathered
\tag 1.7.29
$$
This rescaled vertex operator is a generating function of
infinitesimal symmetries of the KP hierarchy with $\hbar$ inserted,
acting on the tau function as
$\tau \to \tau + \eps Z(\hbar,\lambdatilde,\lambda)\tau$.
If one expands this two-parameter family of symmetries
into Fourier modes along the double loop
$|\lambdatilde| = |\lambda| = \text{const.}$, the outcome are generators
of $\gl(\infty)$ symmetries [15]. If one first
expands this vertex operator into Taylor series along the
diagonal $\lambdatilde = \lambda$,
and further into Fourier modes along the loop $|\lambda|=\text{const.}$,
one obtains a set of generators of $W_{1+\infty}$ symmetries,
$W^{(\ell)}_n(\hbar)$ ($\ell \geqq 1$, $n \in \Integer$).
Explicitly, these generators are given by
$$
    W^{(\ell)}(\hbar,\lambda)
    = \sum_{n=-\infty}^\infty W^{(\ell)}_n(\hbar) \lambda^{-n-\ell}
    = \left( \frac{\der}{\der\lambdatilde} \right)^{\ell-1}
        Z(\hbar,\lambdatilde,\lambda)|_{\lambdatilde=\lambda}.
\tag 1.7.30
$$
Each of $W^{(\ell)}_n(\hbar)$ acts on the tau function as a differential
operator of finite order.
The $w_{1+\infty}$ symmetries are quasi-classical limit of these
$W_{1+\infty}$ symmetries defined by
$$
    w^{(\ell)}_n \log\taudkp
    = \lim_{\hbar \to 0} \tau(\hbar,t)^{-1}
       \hbar^\ell W^{(\ell)}_n(\hbar) \tau(\hbar,t).
\tag 1.7.31
$$
This limit can be calculated more explicitly as:
$$
    w^{(\ell)}_n \log\taudkp
    = \frac{1}{\ell} \Res \calM^\ell \calL^{n+\ell-1} d_k \calL.
\tag 1.7.32
$$
These are nothing but the generators of $w_{1+\infty}$ symmetries
in \S1.6, i.e., the symmetries generated by
$F(k,x) = - x^{\ell-1} k^{n+\ell-1}$.

The two constructions of symmetries due to Proposition 1.7.14
for $L$, $M$, $W$ and due to (1.7.32) for $\log\taudkp$ are, actually,
closely related. To see this fact,
it is convenient to use the fermionic and the wedge representation.

Let us recall briefly how the dressing operator determines the tau function
in the theory of the ordinary ($\hbar=1$) KP hierarchy. According to [13],
[14] and [15], the dressing operator is in one-to-one correspondence with
the point in the universal Grassmann manifold $UGM$. When the motion
of a point $U(t) \in UGM$ obeys the KP hierarchy, one of the Pl\"ucker
coordinate gives the tau function. To make this correspondence explicit,
we prepare notations. We denote the ring of micro-differential operators
and the ring of differential operators of one variable $x$ by $\calE$ and
$\calD$ respectively. For the moment we take $\Complex$ as the constant ring
of $\calE$ and $\calD$. Define the $\calE$-module $\calV$ and
the $\calD$-module $V^\phi$ generated by the delta function $\delta(x)$ by
$$
    \calV := \calE/\calE x, \quad
    V^\phi := \calD/\calD x.
\tag 1.7.33
$$
The $\calD$-module $V^\phi$ is naturally regarded as a sub-$\calD$-module
of $\calV$. We take the basis
$\{ e_\nu = \der^{-1-\nu} \mod \calE x \}_{\nu\in\Integer}$ of $\calV$
($e_{-1} = \delta(x)$).
The action of $x$, $\der_x$, $x^{\ell-1} \der_x^{n+\ell-1}$
on $\calV$ are written in matrix form with respect to this basis as follows:
$$
\aligned
    x &\leftrightarrow ( (\nu+1) \delta_{\mu-1, \nu} )_{\mu,\nu \in \Integer},
\\
    \der_x &\leftrightarrow ( \delta_{\mu+1, \nu} )_{\mu,\nu \in \Integer},
\\
    x^{\ell-1} \der_x^{n+\ell-1} &\leftrightarrow
    ((\nu-n)(\nu-n-1) \cdots (\nu-n-\ell+2)
     \delta_{\mu+n, \nu})_{\mu, \nu \in \Integer}.
\endaligned
\tag 1.7.34
$$
A decreasing filtaration $\calV^{(n)}$ of $\calV$ is defined by
$$
    \calV^{(n)} = \calE^{-n} e_0
    = \{ \xi = \sum_{k=n}^\infty \xi_k e_k, \quad \xi_k \in \Complex\}.
\tag 1.7.35
$$
There is a direct sum decomposition $\calV = V^\phi \oplus \calV^{(0)}$.
Based on this decomposition, the {\it universal Grassmann manifold} $UGM$
is defined as the set of subspaces of $\calV$ which has ``the same size''
as $V^\phi$. Exactly speaking: define the induced filtration of a subspace $U$
of $\calV$ by $U^{(k)} := U \cap \calV^{(k)}$. A subspace $U$ lies
in $UGM$ if and only if $\dim U^{(-m)} = m$ for $m = 0, 1, 2, \ldots$.
(This is the definition of the generic cell of $UGM$,
$UGM^\phi$, in the notation of [14].
For our purpose this definition is enough.)
The following one-to-one correspondence
is fundamental ([14] Theorem 8.1, 9.2):
$$
    \calW \owns W \mapsto \gamma(W) = W^{-1} V^\phi \in UGM,
\tag 1.7.36
$$
where
$$
\split
\calW &= \{\text{monic 0-th order micro-differential operators}\}\\
      &= \{ W = 1 + \sum_{n=1}^\infty w_n(x) \der^{-n} \}.
\endsplit
$$
We call the basis $\{\xi_k = W^{-1} e_k \}_{k \in \Natural^C}$
($\Natural^C = \{ -1, -2, \ldots \}$) of $U$ the {\it canonical frame\/}.
Like the finite dimensional Grassmann manifold,
the universal Grassmann manifold has the Pl\"ucker embedding
into the infinite dimensional projective space $\Proj^\infty(\Complex)$:
$$
    UGM \owns U = \bigoplus_{k\in\Natural^C} \Complex \xi_k \mapsto
    \xi_{-1} \wedge \xi_{-2} \wedge \ldots
    \in \bigwedge_{k\in\Natural^C} \calV \backslash \{0\}
    \to \Proj(\bigwedge_{k\in\Natural^C} \calV).
\tag 1.7.37
$$
The image of this embedding, which is naturally identified with $UGM$,
is also described as the $GL(\infty)$-orbit of
$e_{-1}\wedge e_{-2} \wedge \ldots$ (for details we refer to [15], [28]).
The action of the Lie algebra $\gl(\infty)$ of $GL(\infty)$ on the space
of semi-infinite exterior products is neatly written down in terms of
the Date-Jimbo-Kashiwara-Miwa (DJKM, for short) free fermions.
Here as in [15] we mean by $\gl(\infty)$ the central extension of
the Lie algebra of $\Integer \times \Integer$ matrices
$\overline{\gl(\infty)}$.
The DJKM free fermion operators $\psi_n$, $\psi^\ast_n$ act on semi-infinite
wedge products as:
$$
\align
     \psi_n      |v\rangle = e_n \wedge |v\rangle, \qquad
    &\langle v|    \psi_n  =
     \langle v| \frac{{\overset \leftarrow\to\der}}{\der e_n},
\tag 1.7.38
\\
     \psi^\ast_n |v\rangle =
     \frac{{\overset\rightarrow\to\der}}{\der e_n} |v\rangle, \qquad
    &\langle v| \psi^\ast_n= \langle v| \wedge e_n.
\tag 1.7.39
\endalign
$$
Here $\langle v|$ belongs to $(\bigwedge_{\Natural^C} \calV)^\vee$,
the dual of $\bigwedge_{\Natural^C} \calV$,
and is of the form $\ldots \wedge f_2 \wedge f_1 \wedge f_0$.
Notations $\langle v| {\overset \leftarrow\to\der}/{\der e_n}$ and
${\overset\rightarrow\to\der}/{\der e_n} |v\rangle$ mean eliminating
the right- and left-most $e_n$ of $\langle v |$ and $|v\rangle$.

These fermion operators satisfy the anti-commutation relations
$$
    [ \psi_i, \psi_j ]_{+} = [\psi^*_i, \psi^*_j ]_{+} = 0, \quad
    [ \psi_i, \psi^*_j ]_{+} = \delta_{ij}.
\tag 1.7.40
$$
We can identify the Lie algebra $\gl(\infty)$ with a subalgebra of
the Clifford algebra generated by $\psi_n$, $\psi^\ast_m$ as follows:
$$
    \gl(\infty) \cong
    \bigoplus_{n,m\in\Integer} \Complex \, :\psi_n \psi^\ast_m:
    \oplus
    \Complex \hat c,
\tag 1.7.41
$$
where $:\quad:$ is the normal ordering defined by
$:\psi_n \psi^\ast_m : \;
= \psi_n \psi^\ast_m - \langle 0 | \psi_n \psi^\ast_m | 0 \rangle$,
and $\hat c$ is a central element.

The {\it vacuum vector} (resp\. {\it dual vacuum vector}) of charge $s$,
$|s\rangle = e_{s-1} \wedge e_{s-2} \wedge \ldots$ (resp\.
$\langle s|= \ldots e_{s+1} \wedge e_s$), is characterized by
$$
\aligned
  & \psi_n |s\rangle   = 0  \quad (n \leqq s-1), \quad
    \psi^*_n |s\rangle = 0  \quad (n \geqq s),                        \\
  & \langle s| \psi_n   = 0  \quad (n \geqq s), \quad
    \langle s| \psi^*_n = 0  \quad (n \leqq s-1).                       \\
\endaligned
\tag 1.7.42
$$
Normalizing the pairing between $(\bigwedge \calV)^\vee$ and
$\bigwedge \calV$ by $\langle s | s' \rangle = \delta_{s,s'}$,
we can express the tau function, in general, as a vacuum expectation value,
$$
    \tau(t) = \langle 0| e^{H(t)} | U \rangle,
\tag 1.7.43
$$
where $|U\rangle$ is the vector in $\bigwedge \calV$ corresponding to
a point $U$ in the universal Grassmann manifold through (1.7.37).
The generator $H(t)$ of the time evolutions is given by
$$
    H(t) = \sum_{n=1}^\infty t_n H_n, \quad
    H_n = \sum_{i=-\infty}^\infty :\psi_i \psi^*_{i+n}:.
\tag 1.7.44
$$
The Lie algebra $\gl(\infty)$ acts on the tau function through
(1.7.38, 39, 41, 43) as
$$
    \tau(t) \mapsto
    \langle 0 | e^{H(t)} \; :\psi_n \psi^\ast_m: \; | U \rangle.
\tag 1.7.45
$$

In the presence of the Planck constant $\hbar$
the constant ring of $\calE$ and $\calD$ becomes $\Complex((\hbar))$,
the ring of rational functions of $\hbar$ around $\hbar=0$, and
the basis $\{e_\nu\}$ of $\calV$ should be rescaled to
$e_\nu = (\hbar\der)^{-1-\nu} \mod \calE x$. Therefore the matrix expression
(1.7.34) should be also rescaled to
$$
\aligned
    x &\leftrightarrow
    ( (\nu+1) \hbar \delta_{\mu-1, \nu} )_{\mu,\nu \in \Integer}, \\
    \hbar\der &\leftrightarrow
    ( \delta_{\mu+1, \nu} )_{\mu,\nu \in \Integer}, \\
    x^{\ell-1} (\hbar\der)^{n+\ell-1} &\leftrightarrow
    ((\nu-n)(\nu-n-1)\cdots(\nu-n-\ell+2) \hbar^{\ell-1}\delta_{\mu+n, \nu}).
\endaligned
\tag 1.7.46
$$
Note that not all points of the universal Grassmann manifold
come into the game, since the dressing operators under consideration
are restricted to the form (1.7.5).

\proclaim{Proposition 1.7.15}
Any tau function with good quasi-classical behaviour can be expressed as:
$$
    \tau(\hbar,t) = \langle0| e^{H(t)/\hbar} g(\hbar) | 0 \rangle,
\tag 1.7.47
$$
where $g(\hbar)$ is the Clifford operator of the form
$$
\gathered
    g(\hbar) = \exp( \hbar^{-1} \calO_A(\hbar)) ,
\\
    \calO_A(\hbar) =
    \oint : A\left(\lambda,\hbar\frac{\der}{\der \lambda}\right)
                 \psi(\lambda) \cdot \psi^*(\lambda):
    \frac{d\lambda}{2 \pi i},
\endgathered
\tag 1.7.48
$$
where
$$
    \psi(\lambda) = \sum_{n=-\infty}^\infty \psi_n \lambda^n, \quad
    \psi^*(\lambda) = \sum_{n=-\infty}^\infty \psi^*_n\lambda^{-n-1}
\tag 1.7.49
$$
are the free fermion fields introduced in [15] and
$A\psi(\lambda) \cdot \psi^\ast(\lambda)$ is understood to be
``$A\psi(\lambda)$ times $\psi^\ast(\lambda)$''.
\endproclaim

\demo{Proof}
Since the dressing operator has the form (1.7.5) and is
connected to a vector in $\bigwedge \calV$ by (1.7.36, 37),
we have only to write down the action of $X(\hbar, x, \hbar\der; t=0)$
on $\bigwedge \calV$ in the fermionic language.
This is done by expanding $X(\hbar, x, \hbar\der;t=0)$ as
$$
    X(\hbar, x, \hbar\der \,; t=0)
    =
    \sum\Sb \ell\geq 1 \\ n+\ell < 1 \endSb
    a_{n,\ell}(\hbar) x^{\ell-1} (\hbar\der)^{n+\ell-1},
\tag 1.7.50
$$
and replacing $x^{\ell-1} (\hbar\der)^{n+\ell-1}$ with an element of
$\gl(\infty)$ according to (1.7.46).
(The matrix unit $E_{m,n} = (\delta_{m, \mu} \delta_{n,\nu})_{\mu,\nu}$
is naturally identified with $:\psi_m \psi^\ast_n:$.)
It is easy to see that $\calO_A(\hbar)$ of (1.7.48) with
$$
    A(\lambda, \hbar\der/\der\lambda)
    = -
    \sum\Sb \ell\geq 1 \\ n+\ell < 1 \endSb
    a_{n,\ell}(\hbar) \lambda^{n+\ell-1}
    \left( \hbar\frac{\der}{\der\lambda} \right)^{\ell-1}
\tag 1.7.51
$$
is a desired expression.
Note that the central extension does not affect the result,
since the matrix representation of $X$ is lower triangular.
\qed
\enddemo

$W_{1+\infty}$ symmetries (1.7.29) are implemented by
inserting a fermion bilinear form in (1.7.47) as
$$
    Z(\hbar,\lambdatilde,\lambda)\tau(\hbar,t)
    = \langle0| e^{H(t)/\hbar} \; :\psi(\lambdatilde)\psi^*(\lambda): \;
       g(\hbar) |0\rangle.
\tag 1.7.52
$$
This fact was pointed out by Date et al\. in the case of $\hbar=1$.
In particular, insertion of
$$
    \calO^{(\ell)}_k(\hbar) =
    \oint :\lambda^{n+\ell-1}
           \left( \hbar \frac{\der}{\der\lambda}\right)^{\ell-1}
           \psi(\lambda) \cdot \psi^*(\lambda) :
    \frac{d\lambda}{2 \pi i},
\tag 1.7.53
$$
where the integral is along a circle $|\lambda| = \text{const.}$,
corresponds to the action of $\hbar^{-1}W^{(\ell)}_n(\hbar)$
on $\tau(\hbar,t)$ defined through (1.7.30).

\proclaim{Proposition 1.7.16}
The symmetry of the tau function by $W^{(\ell)}_n(\hbar)$
(1.7.30, 53) induces a symmetry of $L$, $M$, $W$ of the form (1.7.27)
with $Y(\hbar, x, \hbar\der) = - x^{\ell-1} (\hbar\der)^{n+\ell-1}$
\endproclaim

\demo{Proof}
Denote by $U$ and $W$ the point in $UGM$ and the dressing operator
under consideration respectively.
The fermion bilinear form $\calO^{(\ell)}_k$ corresponds to
$ - Y= x^{\ell-1} (\hbar\der)^{n+\ell-1}$ through (1.7.46),
when one neglects the central extension. Since the central
term changes the tau function only by constant multiplication,
it does not change $L$, $M$, $W$.
Hence the point
$$
    \exp(\eps \calO_k^{(\ell)}(\hbar)) | U \rangle
    =
    \exp( - \eps Y) W^{-1}(t=0) V^\phi
$$
in the universal Grassmann manifold corresponds to $W_\eps(t=0)$ of (1.7.28)
by the one-to-one mapping (1.7.36). This proves the proposition.
\qed
\enddemo
This explains the coincidence (1.7.32)
of the quasi-classical limit of $W_{1+\infty}$ symmetries (1.7.31)
with the symmetries defined through the deformation of the twistor data \S1.6.

%
%
%
%

\subhead 1.8. Special solutions \endsubhead

Many special solutions to the dispersionless KP hierarchy are constructed,
e.g., by [6], but here we concentrate our attention to
those solutions which appear in the context of topological minimal
models [29], [30], [8].
In fact, Krichever's work on this subject [7] was the starting
point of our study [1] where we gave an interpretation of his results
in our framework.
Dubrovin [8] studied these solutions from the viewpoint of
integrable systems of hydrodynamic type [31].

Krichever showed in [7] that the perturbed chiral rings of
the topological minimal models (Landau-Ginzurg models of $A_N$-type)
can be described by some special solutions of the dispersionless KP hierarchy
(``dispersionless Lax equations'' in his terminology).
These solutions are characterized by the twistor data:
$$
    f(k,x) = k^N/N, \quad g(k,x) = x k^{1-N}.
\tag 1.8.1
$$
In other words, those solutions are determined by the equations:
$$
    (\calP)_{\leq -1} = 0, \quad
    (\calQ)_{\leq -1} = 0,
\tag 1.8.2
$$
where $\calP \=def \calL^N/N$, $\calQ \=def \calM \calL^{1-N}$.
Using the theory of the generalized hodograph transformation
by Tsarev [32], we can show the solvability of this Riemann-Hilbert
decomposition problem.

\proclaim{Proposition 1.8.1} ([1] Proposition 8, 9, 10)
The equations (1.8.2) has a solution $(\calL, \calM)$ such
that the coefficeitns $u_n$ of $\calL$ (1.1.2) and $v_i$ of $\calM$
(1.3.2) are homogeneous of degree zero and one with respect
to the degree defined by $\deg (t_n) = 1$. This solution is defined
in a domain where $t_{N+1} \neq 0$ and $t_n/t_{N+1}$ ($n\neq N+1$) are
sufficiently small.
The tau function, up to an integration constant, is given by
$$
    \log\taudkp = \frac{1}{2}\sum_{n=1}^\infty t_n v_n + \text{\rm const.}
\tag 1.8.3
$$
or, equivalently, by
$$
    \log\taudkp = \frac{1}{2}\sum_{n,m=1}^\infty
               t_m t_n \frac{\der v_n}{\der t_m} + \text{\rm const.}
\tag 1.8.4
$$
\endproclaim

As Krichever pointed out in [7],
this tau function restricted to the ``small phase space''
$\{ t_{N+1} = 1/(N+1),\, t_{N+2} = t_{N+3} = \ldots = 0 \}$
gives the free energy of the Landau-Ginzburg model of $A_{N-1}$-type.
Furthermore, according to recent developments in the topological
field theory [33], the topological matter fields
coupled to topological gravity is described by this solution
of the dispersionless KP hierarchy in the ``full phase space'', where
the higher times $t_n$ ($n\geqq N+2$) play the role of gravitational
coupling constants.

Krichever's dispersionless analogue of Virasoro constraints
is interpreted as the following
``dispersionless $w_{1+\infty}$ constraints'', which is a direct consequence
of the results in \S1.6 and derived from the basic relations
$$
    \left( \calL^{(m+1)N+n(1-N)} \calM^n \right)_{\leq -1}
    = \text{const.}\left( \calP^{m+1} \calQ^{n} \right)_{\leq -1} = 0
\tag 1.8.5
$$
for $m \geqq -1$, $n\geqq 0$.

\proclaim{Proposition 1.8.2} ([1] Proposition 11, 17; [7])
The tau function $\tau=\tau_0(t)$
of Proposition 1.8.1 satisfies the constraints
$$
    \delta_{\lambda^{(m+1)N+n(1-N)}x^n}\log\tau
    = -\frac{1}{n+1}\Res[ \calL^{(m+1)N+n(1-N)} \calM^{n+1} d_\lambda\calL]
    = 0
\tag 1.8.6
$$
for $m \geqq -1$ and $n \geqq 0$.
In particular, the constraints for $(m,n) = (-1,1)$, $(0,0)$ and
$m=1,2,\ldots$, $n=1$ give the Krichever's dispersionless Virasoro
constraitns:
$$
\align
  & \sum_{k=N+1}^\infty kt_k \frac{\der\log\tau}{\der t_{k-N}}
    +\frac{1}{2} \sum_{i=1}^{N-1} i(N-i) t_i t_{N-i} = 0,
\tag 1.8.7 \\
  & \sum_{k=1}^\infty kt_k \frac{\der\log\tau}{\der t_k} = 0,
\tag 1.8.8 \\
  & \sum_{k=1}^\infty kt_k \frac{\der\log\tau}{\der t_{k+mN}}
    +\frac{1}{2} \sum_{i=1}^{mN-1}
     \frac{\der\log\tau}{\der t_i} \frac{\der\log\tau}{\der t_{mN-i}}=0.
\tag 1.8.9 \\
\endalign
$$
\endproclaim

We can deform this solution by changing the twistor data (1.8.1) to
$$
    f(k,x) = k^N/N, \quad
    g(k,x) = x k^{1-N} + h(k),
\tag 1.8.10
$$
where $h(k) = \sum_{n\in\Integer} h_n k^n$ is an arbitrary Laurent
series in $k$
(for the sake of simplicity, let us suppose that $h_{-N} = 0$).
The corresponding $(\calP, \calQ)$-pair is now given by
$$
    \calP = \calL^N/N, \quad
    \calQ = \calL^{1-N} \calM + h(\calL).
\tag 1.8.11
$$

\proclaim{Proposition 1.8.3} ([1] Proposition 12.)
Let $\tau_0=\tau_0(t)$ be the tau function of the undeformed solution
($h=0$). The tau function $\tau_h=\tau_h(t)$ of the deformed solution
is then given by
$$
\split
    \tau_h(t) =& \text{\rm const.}
                 \exp\left( -\sum_{n=1}^\infty h_{-n-N}t_n \right)    \\
               &\times \tau_0(t_1 + h_{1-N}, t_2 + \frac{h_{2-N}}{2},
                              \ldots, t_k + \frac{h_{k-N}}{k}, \ldots).
\endsplit
\tag 1.8.12
$$
\endproclaim

The above solution of the dispersionless KP hierarchy has a natural
``dispersionful'' extension in the context of topological string theory.
Namely, the (generalized) Kontsevich models,
a version of the topological string theory intoduced
by [34], [35], specify a solution of
the ordinary KP hierarchy whose quasi-classical limit gives solutions
of the dispersionless KP hierarchy of the above type.
Actually the ``quasi-classical'' limit is the contribution from the
genus $g=0$ Riemann surfaces in the context of the string theory.
The corresponding Riemann-Hilbert problem in the sense of \S1.7.5 is
$$
    \frac1N(L^N)_{\leq -1} = 0, \quad
    \left( ML^{1-N} + \frac{N-1}2 \hbar L^{-N} + N L \right)_{\leq -1} = 0,
\tag 1.8.13
$$
which has quasi-classical limit of the form (1.8.10), or (1.8.1) up to
a suitable shift of time variables.
This model is subject to the Virasoro constraints
[30], [36], [35], [37] and the quasi-classical limit of
this constraints are the constraints stated in Proposition 1.8.2.
For details on the quasi-classical limit of the generalized Kontsevich
model we refer to [38].

%
%
%
%

\head II. Dispersionless Toda hierarchy \endhead
%
%
%
%

\subhead 2.1. Lax formalism \endsubhead

Here we briefly review the definition of the dispersionless Toda
hierarchy, mostly following [2, 4].
In contrast to the (dispersionless) KP hierarchy
the dispersionless Toda hierarchy has two series of independent
(``time'') variables $t = (t_1, t_2, \ldots)$ and
$\tbar = (\tbar_1, \tbar_2, \ldots)$ along with a new spatial variable $s$.
Its Lax representation is:
$$
\aligned
    \frac{\der \calL}{\der t_n} = \{ \calB_n, \calL \}, \quad&
    \frac{\der \calL}{\der \tbar_n} = \{ \calBbar_n, \calL \},
\\
    \frac{\der \calLbar}{\der t_n} = \{ \calB_n, \calLbar \}, \quad&
    \frac{\der \calLbar}{\der \tbar_n} = \{ \calBbar_n, \calLbar \},
    \quad n = 1,2,\ldots,
\endaligned
\tag 2.1.1
$$
where $\calL$ and $\calLbar$ are Laurent series
$$
\aligned
    \calL         = & p
           + \sum_{n=0}^\infty u_{n+1}     (t,\tbar,s) p^{-n},\\
    \calLbar^{-1} = & \ubar_0(t, \tbar, s) p^{-1}
           + \sum_{n=0}^\infty \ubar_{n+1} (t,\tbar,s) p^n,
\endaligned
\tag 2.1.2
$$
of a variable $p$, and $\calB_n$ and $\calBbar_n$ are given by
$$
     \calB_n = (\calL^n)_{\ge 0}, \quad
     \calBbar_n = (\calLbar^{-n})_{\le -1}.
\tag 2.1.3
$$
Note that we now define the coefficients $\ubar_{n+1}$ to be those of
the expansion of $\calLbar^{-1}$ rather than $\calLbar$ itself
Here $(\quad)_{\ge 0}$ and $(\quad)_{\le -1}$ denote the projection
of Laurent series onto a linear combination of $p^n$ with
$n\geqq 0$ and $\leqq -1$ respectively. The Poisson bracket
$\{\quad,\quad\}$ is defined by
$$
    \{ A(p,s), B(p,s) \}
    = p\frac{\der A(p,s)}{\der p} \frac{\der B(p,s)}{\der s}
     -p\frac{\der A(p,s)}{\der s} \frac{\der B(p,s)}{\der p}.
\tag 2.1.4
$$
By this definition, $\log p$ and $s$ are cononically conjugate coordinates
on the two dimensional ``phase space''.

By the same argument as in [21], one can easily prove the following fact.

\proclaim{Proposition 2.1.1}
The Lax equations for $\calL$ are equivalent
to the ``zero-curvature equations''
$$
\aligned
    & \der_{t_n}\calB_m -\der_{t_m}\calB_n +\{ \calB_m, \calB_n \} = 0, \\
    & \der_{\tbar_n}\calBbar_m -\der_{\tbar_m}\calBbar_n
                               +\{ \calBbar_m, \calBbar_n \} = 0,       \\
    & \der_{\tbar_n}\calB_m -\der_{t_m}\calBbar_n
                               +\{ \calB_m, \calBbar_n \} = 0.
\endaligned
\tag 2.1.5
$$
\endproclaim

In our ealier work [2, 4] we formulated the hierarchy assuming the existence
of $\calM$ and $\calMbar$ from the very beginning. We now start from
the above formulation and show in \S2.3 how $\calM$ and $\calMbar$
enter into the story.

The dispersionless (or long-wave) limit $\phi$ of the original Toda field
is hidden in this formalism in the following way.

\proclaim{Proposition 2.1.2}
There exists $\phi = \phi(t, \tbar, s)$ characterized by
$$
    d\phi =  \sum_{n=1}^\infty \Res( \calL^n d\log p) dt_n
           - \sum_{n=1}^\infty \Res( \calLbar^{-n} d\log p) d\tbar_n
           + \log \ubar_0 ds,
\tag 2.1.6
$$
where ``$d$'' means total differentiation in $(t,\tbar,s)$,
and $d\log p = dp/p$. Furthermore $\phi$ satisfies the well-known
dispersionless (long-wave) limit of the two-dimensional Toda field
equation
$$
    \dfrac{\der^2 \phi}{\der t_1 \der \tbar_1}
    + \frac{\der}{\der s} \exp\left( \frac{\der\phi}{\der s} \right)
    = 0.
\tag 2.1.7
$$
\endproclaim

\demo{Proof}
The equation (2.1.6) is a compact form of the following system.
$$
\align
    \der_{t_n}     \phi &= b_{n,0}     \=def  (\calL^n)_0,
\tag 2.1.8
\\
    \der_{\tbar_n} \phi &=-\bbar_{n,0} \=def -(\calLbar^{-n})_0,
\tag 2.1.9
\\
    \der_s         \phi &= \log \ubar_0,
\tag 2.1.10
\endalign
$$
where $(\quad)_0 \=def \Res (\quad)d\log p$ is the projection
to the constant term with respect to $p$.
The compatibility conditions among these equations are contained
in the Lax equations (2.1.1) and the zero-curvature equations
(2.1.5). For example, the compatibility of (2.1.8$)_n$ and (2.1.8$)_m$,
$\der_{t_n} b_{m,0} = \der_{t_m} b_{n,0}$, can be read from the coefficient
of $p^0$ of the first equation of (2.1.5).
The compatibility conditions between (2.1.8--9$)_n$ and (2.1.10)
$\der_s b_{n,0} = \der_{t_n} \log\ubar_0$,
$\der_s \bbar_{n,0} = -\der_{\tbar_n} \log\ubar_0$
come from the $p^1$ part of the Lax equations (2.1.1) for $\calLbar$.
The others can be verified similarly.

The equation (2.1.7) is a direct consequence of (2.1.8--10) and
one of (2.1.5),
$\der_{\tbar_1}\calB_1 - \der_{t_1}\calBbar_1 + \{\calB_1,\calBbar_1\}=0$.
\qed
\enddemo

Like the ordinary Toda lattice hierarchy [39],
we can exchange the roles of $\calL$ and $\calLbar$
by a ``gauge-transformation''.

\proclaim{Lemma 2.1.3}
Define $\calL' \=def \exp(-\ad\phi) \calL$,
$\calLbar' \=def \exp(-\ad\phi) \calLbar$. Then they are of the form
$$
\aligned
    \calL'         = & \ubar_0^{-1}(t,\tbar,s) p
           + \sum_{n=0}^\infty u'_{n+1}     (t,\tbar,s) p^{-n},\\
    (\calLbar')^{-1} = & p^{-1}
           + \sum_{n=0}^\infty \ubar'_{n+1} (t,\tbar,s) p^n,
\endaligned
\tag 2.1.11
$$
and satisfy the Lax equations
$$
\aligned
    \frac{\der \calL'}{\der t_n} = \{ \calB'_n, \calL' \}, \quad&
    \frac{\der \calL'}{\der \tbar_n} = \{ \calBbar'_n, \calL' \},
\\
    \frac{\der \calLbar'}{\der t_n} = \{ \calB'_n, \calLbar' \}, \quad&
    \frac{\der \calLbar'}{\der \tbar_n} = \{ \calBbar'_n, \calLbar' \},
    \quad n = 1,2,\ldots,
\endaligned
\tag 2.1.12
$$
as well as the zero-curvature equations
$$
\aligned
    &\der_{t_n}\calB'_m -\der_{t_m}\calB'_n +\{ \calB'_m, \calB'_n \} = 0,
\\
    & \der_{\tbar_n}\calBbar'_m -\der_{\tbar_m}\calBbar'_n
                               +\{ \calBbar'_m, \calBbar'_n \} = 0,
\\
    & \der_{\tbar_n}\calB'_m -\der_{t_m}\calBbar'_n
                               +\{ \calB'_m, \calBbar'_n \} = 0.
\endaligned
\tag 2.1.13
$$
Here $\calB'_n \=def ((\calL')^n)_{\ge 1}$ and
$\calBbar'_n \=def ((\calLbar')^{-n})_{\le 0}$.
\endproclaim

\demo{Proof}
It is easy to prove (2.1.11) by (2.1.10). The equations (2.1.11, 12)
are results of Lemma A.1 and (2.1.8--9).
\qed
\enddemo

\demo{Remark 2.1.4}
There is a one-parameter family of gauge as in [39]. The above
one will be used in the next section.
\enddemo

%
%
%
%

\subhead 2.2. Dressing operators \endsubhead

Here like in \S1.2, we introduce the dressing functions $\varphi$ and
$\varphibar$ in a Lie-algebraic sense.

\proclaim{Proposition 2.2.1}
i)
Let $\calL$ and $\calLbar$ be a solution of the dispersionless Toda
hierarchy (2.1.1).
Then there exists a pair of Laurent series $\varphi$ and $\varphibar$
of $p$ such that
$$
\alignat2
    & \calL = e^{\ad\varphi}(p), &\quad
    & \calLbar = e^{\ad\phi}e^{\ad\varphibar}(p),
\tag 2.2.1
\\
    & \nabla_{t_n,\varphi} \varphi = -(\calL^n)_{\leq -1}, &\quad
    & e^{\ad\phi} \nabla_{t_n,\varphibar} \varphibar +
      \der_{t_n} \phi
    = (\calL^n)_{\geq 0},
\tag 2.2.2
\\
    & \nabla_{\tbar_n,\varphi} \varphi = (\calLbar^{-n})_{\leq -1}, &\quad
    & e^{\ad\phi} \nabla_{\tbar_n, \varphibar} \varphibar
    + \der_{\tbar_n}\phi
    = -(\calLbar^{-n})_{\geq 0}
\tag 2.2.3
\endalignat
$$
for $n=1,2,\ldots$, where $\phi$ is the dispersionless limit of
the Toda field introduced in Lemma 2.1.2, $\varphi$ and $\varphibar$ are
Laurent series of the form
$\varphi(t,\tbar,p,s)    = \sum_{n=1}^\infty \varphi_n(t,\tbar,s) p^{-n}$,
$\varphibar(t,\tbar,p,s) = \sum_{n=1}^\infty \varphibar_n(t,\tbar,s) p^n$,
and $\ad$ and $\nabla$ have the same meaning as in \S1.2.

Such $\varphi$, $\varphibar$ are determined uniquely
up to transformations
$\varphi \mapsto H(\varphi,\psi)$,
$\varphibar \mapsto H(\varphibar,\psibar)$,
where $H(X,Y)$ is the Hausdorff series (cf\. Appendix A)
$$
    e^{\ad H(\varphi,\psi)} = e^{\ad\varphi} e^{\ad\psi}, \quad
    e^{\ad H(\varphibar, \psibar)} = e^{\ad\varphibar} e^{\ad\psibar},
$$
and $\psi$ and $\psibar$ are Laurent series of the form
$\psi = \sum_{n=1}^\infty \psi_n p^{-n}$,
$\psibar = \sum_{n=1}^\infty \psibar_n p^n$
($\psi_n$, $\psibar_n$: constant).

ii)
Conversely, if
$\varphi(t,\tbar,p,s) = \sum_{n=1}^\infty \varphi_n(t,\tbar,s) p^{-n}$ and
$\varphibar(t,\tbar,p,s) = \sum_{n=1}^\infty \varphibar_n(t,\tbar,s) p^n$
satisfy (2.2.2) and (2.2.3), then $\calL$ and $\calLbar$ defined in (2.2.1)
give a solution of the dispersionless Toda hierarchy (2.1.1).
\endproclaim

\demo\nofrills{Proof\/}\ traces that of Proposition 1.2.1.
Let $\varphi^0$ be an arbitrary series of the form
$\varphi^0(t,\tbar,p,s) = \sum_{n=1}^\infty \varphi^0_n(t,\tbar,s) p^{-n}$
such that $\calL = \exp(\ad\varphi^0)$.
Then the zero-curvature equations (2.1.5) imply the integrability of
the system of equations
$$
    \frac{\der \varphi'}{\der t_n} = p^n - \calB_n^{\varphi^0},
    \quad
    \frac{\der \varphi'}{\der \tbar_n} = \calBbar_n^{\varphi^0},
\tag 2.2.4
$$
for $\varphi'(t,\tbar,p,s) = \sum_{n=1}^\infty \varphi'_n(t,\tbar,s) p^{-n}$,
where
$$
    \calB_n^{\varphi^0}
    \=def e^{-\ad\varphi^0} (\calB_n - \nabla_{t_n,\varphi^0} \varphi^0),
    \quad
    \calBbar_n^{\varphi^0} \=def
    e^{-\ad\varphi^0} (\calBbar_n - \nabla_{\tbar_n,\varphi^0} \varphi^0).
$$
Due to the Lax equations (2.1.1) $\calB_n^{\varphi^0}$ and
$\calBbar_n^{\varphi^0}$ are independent of $s$. Hence there is a solution
of (2.2.4) which is also independent of $s$.
Using the Hausdorff-Campbell formula, we define
$\exp(\varphi) = \exp(\varphi') \exp(\varphi^0)$, and this $\varphi$ is
a desired dressing function.

We can easily prove the corresponding statements for $\varphibar$, using
Lemma 2.1.3.
\qed
\enddemo

%
%
%
%

\subhead 2.3. Orlov function and Darboux coordinates \endsubhead

The {\it Orlov functions} $\calM$ and $\calMbar$ are defined
for the dispersionless Toda hierarchy by
$$
\alignedat2
    \calM    &= e^{\ad\varphi}
                \left( s + \sum_{n=1}^\infty n t_n p^n \right)&
             &= e^{\ad\varphi} e^{\ad t(p)} (s),
\\
    \calMbar &= e^{\ad\phi} e^{\ad\varphibar}
                \left( s - \sum_{n=1}^\infty n \tbar_n p^{-n} \right)&
             &= e^{\ad\phi} e^{\ad\varphibar}
                e^{\ad \tbar(p^{-1})} (s),
\endalignedat
\tag 2.3.1
$$
where $t(p) = \sum_{n=1}^\infty t_n p^n$,
$\tbar(p^{-1}) = \sum_{n=1}^\infty \tbar_n p^{-n}$.
They have an expansion like
$$
\aligned
    \calM &= \sum_{n=1}^\infty n t_n \calL^n + s
           + \sum_{n=1}^\infty v_n(t,\tbar,s) \calL^{-n},
\\
    \calMbar &= - \sum_{n=1}^\infty n \tbar_n \calLbar^{-n} + s
                + \sum_{n=1}^\infty \vbar_n(t,\tbar,s) \calLbar^n.
\endaligned
\tag 2.3.2
$$

As in the proof of Proposition 1.3.1, we can prove:
\proclaim{Proposition 2.3.1}
i)
The series $\calM$ and $\calMbar$ satisfy the Lax equations
$$
\aligned
    \frac{\der \calM}{\der t_n} = \{ \calB_n, \calM \}, \quad&
    \frac{\der \calM}{\der \tbar_n} = \{ \calBbar_n, \calM \},
\\
    \frac{\der \calMbar}{\der t_n} = \{ \calB_n, \calMbar \}, \quad&
    \frac{\der \calMbar}{\der \tbar_n} = \{ \calBbar_n, \calMbar \},
    \quad n = 1,2,\ldots,
\endaligned
\tag 2.3.3
$$
and the canonical Poisson relations $\{\calL, \calM\} = \calL$ and
$\{\calLbar, \calMbar\} = \calLbar$.

ii)
Conversely, if the potential $\phi$ is given, and if a pair of
series $\calM$ and $\calMbar$ of the form (2.3.2)
satisfy  (2.3.3) and the canonical Poisson relations,
then there exist unique dressing functions $\varphi$ and $\varphibar$
which are connected with $\calM$ and $\calMbar$ through (2.3.1).
Explicitly,
$\varphi(t,\tbar,p,s) = \sum_{n=1}^\infty \varphi_n(t,\tbar,s) p^{-n}$ and
$\varphibar(t,\tbar,p,s) = \sum_{n=1}^\infty \varphibar_n(t,\tbar,s) p^n$,
and the coefficients are of the form
$$
\aligned
    \varphi_n(t,\tbar,s) &= -\frac{v_{n+1}}n +
                   (\text{a differential polynomial of }v_2, \ldots,v_n,
                                                        u_2, \ldots,u_{n-1}).
\\
    \varphibar_n(t,\tbar,s) &= -\frac{\vbar_{n+1}}n +
                   (\text{a differential polynomial of }\phi,
                    \vbar_2, \ldots,\vbar_n, \ubar_2, \ldots,\ubar_{n-1}).
\endaligned
\tag 2.3.4
$$
\qed
\endproclaim

These pairs of series $(\calL, \calM)$ and $(\calLbar, \calMbar)$ play
the role of Darboux coordinates, as is the case with the dispersionless
KP hierarchy. The fundamental 2-form is
$$
    \omega \=def \frac{dp}p \wedge ds
               + \sum_{n=1}^\infty d\calB_n \wedge dt_n
               + \sum_{n=1}^\infty d\calBbar_n \wedge d\tbar_n,
\tag 2.3.5
$$
which satisfies
$$
    d\omega = 0, \quad \omega \wedge \omega = 0.
\tag 2.3.6
$$
In fact, (2.3.6) is equivalent to the zero-curvature equations (2.1.5).

\proclaim{Proposition 2.3.2}
The dispersionless Toda hierarchy is equivalent to the exterior differential
equations
$$
\frac{d\calL \wedge d\calM}{\calL} = \omega =
\frac{d\calLbar \wedge d\calMbar}{\calLbar}.
\tag 2.3.7
$$
Namely, if $(\calL, \calM)$ and $(\calLbar, \calMbar)$ are Laurent series
of the form (2.1.2) and (2.3.2) and subject to (2.3.7),
then $(\calL,\calLbar)$ give a solution
of the dispersionless Toda hierarchy (2.1.1), and $(\calM, \calMbar)$ are
corresponding Orlov functions.
\endproclaim

\demo{Proof}
The canonical Poisson relations $\{\calL, \calM\} = \calL$,
$\{\calLbar, \calMbar\} = \calLbar$
(resp\. the Lax equations (2.1.1), (2.3.3)) can be
easily derived from (2.3.7) by comparing coefficeints of $dp \wedge ds$,
(resp. $dp \wedge dt_n$ and $ds \wedge dt_n$ or $dp \wedge d\tbar_n$
and $ds \wedge d\tbar_n$) of both hand sides.

The converse can be proved by tracing back this derivation.
\qed
\enddemo

%
%
%
%

\subhead 2.4. $S$ function, $\tau$ function (free energy) \endsubhead

Here we introduce three potentials $S$, $\Sbar$ and $\log\taudtoda$.
Later in \S2.7 we will see that $S$ and $\Sbar$ are phase funcitons
of WKB analysis like the $S$ function for the dispersionless KP hierarchy.

\proclaim{Proposition 2.4.1}
There exist functions $S$ and $\Sbar$ which satisfy
$$
\aligned
    & dS = \calM d\log\calL + \log p \; ds
          +\sum_{n=1}^\infty \calB_n    dt_n
          +\sum_{n=1}^\infty \calBbar_n d\tbar_n,
\\
    & d\Sbar = \calMbar d\log\calLbar + \log p \; ds
          +\sum_{n=1}^\infty \calB_n    dt_n
          +\sum_{n=1}^\infty \calBbar_n d\tbar_n,
\endaligned
\tag 2.4.1
$$
where $d\log\calL = d\calL/\calL$. Explicitly,
$S$ and $\Sbar$ have a Laurent expansion of the following form.
$$
\aligned
    & S = \sum_{n=1}^\infty t_n\calL^n + s\log\calL
        - \sum_{n=1}^\infty \frac{v_n}n \calL^{-n},
\\
    & \Sbar = \sum_{n=1}^\infty \tbar_n\calLbar^{-n} + s\log\calLbar + \phi
            + \sum_{n=1}^\infty \frac{\vbar_n}n \calLbar^n.
\endaligned
\tag 2.4.2
$$
\endproclaim

\demo{Proof}
The existence of $S$ and $\Sbar$ satisfying (2.4.1) is obvious
since the right hand sides of (2.4.1) are closed forms because of (2.3.7).
To derive (2.4.2), we have to prove that $S$ and $\Sbar$ satisfy
the following equations.
$$
\alignat4
    &\left.
    \frac{\der S}{\der \log\calL}
    \right|_{t,\tbar,s \fixed}&
    &= \calM,
    &\quad
    &\left.
    \frac{\der \Sbar}{\der \log\calLbar}
    \right|_{t, \tbar, s \fixed}&
    &= \calMbar,
\tag 2.4.3
\\
    &\left.
    \frac{\der S}{\der t_n}
    \right|_{\calL, t_m (m\neq n), \tbar, s \fixed}&
    &= \calB_n,
    &\quad
    &\left.
    \frac{\der \Sbar}{\der t_n}
    \right|_{\calL, t_m (m\neq n), \tbar, s \fixed}&
    &= \calB_n,
\tag 2.4.4
\\
    &\left.
    \frac{\der S}{\der \tbar_n}
    \right|_{\calL, t, \tbar_m (m\neq n), s \fixed}&
    &= \calBbar_n,
    &\quad
    &\left.
    \frac{\der \Sbar}{\der \tbar_n}
    \right|_{\calL, t, \tbar_m (m\neq n), s \fixed}&
    &= \calBbar_n,
\tag 2.4.5
\\
    &\left.
    \frac{\der S}{\der s}
    \right|_{\calL, t, \tbar \fixed}&
    &= \log p,
    &\quad
    &\left.
    \frac{\der \Sbar}{\der s}
    \right|_{\calL, t, \tbar \fixed}&
    &= \log p.
\tag 2.4.6
\endalignat
$$
Equations (2.4.3) are a direct consequence of the construction (2.4.2).
In order to prove the remaining equations (2.4.4--6), we prepare:
\proclaim{Lemma 2.4.2}
$$
\alignat2
    \frac{\der v_m}{\der t_n}
    &=
    \Res \calL^m d_p \calB_n,
    &\quad
    \frac{\der \vbar_m}{\der t_n}
    &=
    \Res \calLbar^{-m} d_p \calB_n,
\tag 2.4.7
\\
    \frac{\der v_m}{\der \tbar_n}
    &=
    \Res \calL^m d_p \calBbar_n,
    &\quad
    \frac{\der \vbar_m}{\der \tbar_n}
    &=
    \Res \calLbar^{-m} d_p \calBbar_n,
\tag 2.4.8
\\
    \frac{\der v_m}{\der s}
    &=
    \Res \calL^m d\log p,
    &\quad
    \frac{\der \vbar_m}{\der s}
    &=
    \Res \calLbar^{-m} d\log p,
\tag 2.4.9
\endalignat
$$
\endproclaim

\demo{Proof}
As an example, we prove the first equation of (2.4.7).
By the chain rule
$$
    \frac{\der \calM}{\der t_n}
    =
    \left.\frac{\der \calM}{\der \calL} \right|_{t,v \fixed}
    \frac{\der \calL}{\der t_n}
    +
    n \calL^n
    +
    \sum_{m=1}^\infty
    \frac{\der v_m}{\der t_n} \calL^{-m},
$$
which implies
$$
\split
    \frac{\der v_m}{\der t_n}
    &=
    \Res \calL^{m-1}
    \left(
    \frac{\der \calM}{\der t_n} d_p \calL
    -
    \left. \frac{\der \calM}{\der \calL} \right|_{t,v \fixed}
    \frac{\der \calL}{\der t_n}
    d_p \calL
    \right)
\\
    &=
    \Res \calL^{m-1}
    ( \{\calB_n, \calM\} d_p \calL - \{\calB_n, \calL\} d_p \calM )
\\
    &= \Res \calL^m d_p \calB_n,
\endsplit
$$
where we have used the Lax equations (2.1.1), (2.3.3) and
the canonical Poisson relation $\{\calL, \calM\} = \calL$.
The other equations of (2.4.7--9) can be proved similarly.
\qed
\enddemo

Now we return to the proof of Proposition 2.4.1. By the above lemma and
the formula $\Res F d_p G = - \Res G d_p F$ (integration by parts),
we obtain
$$
\split
    \left.
    \frac{\der S}{\der t_n}
    \right|_{\calL, t_m (m\neq n), \tbar, s \fixed}
    &=
    \calL^n
    -
    \sum_{m=1}^\infty
    \frac1m \frac{\der v_m}{\der t_n} \calL^{-m}
\\
    &=
    \calL^n
    +
    \sum_{m=1}^\infty
    \Res (\calB_n \calL^{m-1} d_p \calL) \calL^{-m}.
\endsplit
$$
This proves the first equation of (2.4.4) due to the tautological formula
$$
    A(p,s) = \sum_{n \in \Integer} \Res(A(p,s) \calL^{n-1} d_p\calL) \calL^n.
\tag 2.4.10
$$
We can prove the remaining equations of (2.4.4--6) similarly,
using Lemma 2.4.2 and Proposition 2.1.2.
\qed
\enddemo

We can now define a dispersionless analogue of the tau function (the free
energy) of the Toda lattice hierarchy as is the case with the KP hierarchy.

\proclaim{Proposition 2.4.3}
There exists a function $\log\taudtoda(t, \tbar, s)$ satisfying
$$
    d \log\taudtoda(t, \tbar, s)
    =
    \sum_{n=1}^\infty v_n d t_n - \sum_{n=1}^\infty \vbar_n d \tbar_n
    + \phi ds.
\tag 2.4.11
$$
\endproclaim

\demo{Proof}
We have only to check that the right hand side of (2.4.11) is a closed form,
which is equivalent to the following system.
$$
\gather
    \frac{\der v_n}{\der t_m} = \frac{\der v_m}{\der t_n}, \quad
    \frac{\der \vbar_n}{\der \tbar_m} = \frac{\der \vbar_m}{\der \tbar_n},
    \quad
    \frac{\der v_n}{\der \tbar_m} = \frac{\der \vbar_m}{\der t_n},
\tag 2.4.12
\\
    \frac{\der v_n}{\der s} = \frac{\der \phi}{\der t_n}, \quad
    \frac{\der \vbar_n}{\der s} = \frac{\der \phi}{\der \tbar_n}.
\tag 2.4.13
\endgather
$$
This is easily proved by Lemma 2.4.2 and Proposition 2.1.2.
\qed
\enddemo
Like the dispersionless KP hierarchy, all ingredients $\calL$, $\calLbar$,
$\calM$, $\calMbar$ etc\. of the dispersionless Toda hierarchy
can be thus reproduced from the single function $\log\taudtoda$.
We do not know yet
if there exists the dispersionless analogue of the Hirota-type equations
which characterize the tau function of the dispersionless Toda hierarchy.

%
%
%
%

\subhead 2.5. Riemann-Hilbert problem \endsubhead

In this section we give a construction of solutions based
on the Riemann-Hilbert problem parallel to that for the dispersionless
KP hierarchy in \S1.5.

\proclaim{Proposition 2.5.1}
Let two pairs of functions $(f(p,s), g(p,s))$, $(\fbar(p,s), \gbar(p,s))$
and two pairs of Laurent series $(\calL, \calM)$, $(\calLbar, \calMbar)$
be given. Suppose
\roster
\item[1] $(\calL,\calLbar)$ has the form (2.1.2), and $(\calM,\calMbar)$
the form (2.3.2).
\item[2] $(f,g)$ and $(\fbar,\gbar)$ satisfy the canonical Poisson relation
$\{f,g\}=f$, $\{\fbar,\gbar\}=\fbar$.
\item[3] $(\calL, \calM)$ and $(\calLbar, \calMbar)$ are connected by
$$
    f(\calL, \calM) = \fbar(\calLbar, \calMbar), \qquad
    g(\calL, \calM) = \gbar(\calLbar, \calMbar).
\tag 2.5.1
$$
\endroster
Then the quadruplet $(\calL, \calLbar, \calM, \calMbar)$ gives
a solution of the dispersionless Toda hierarchy. Namely, they satisfy
the Lax equations (2.1.1), (2.3.3), and the canonical Poisson relations
$\{\calL, \calM\} = \calL$, $\{\calLbar, \calMbar\} = \calLbar$.
We call $(f,g, \fbar, \gbar)$ the {\it twistor data} of this solution.
\endproclaim

\demo{Proof}
We follow the proof of the same construction for the dispersionless KP
hierarchy in [1]. First we prove the canonical Poisson relations.
By the chain rule we have
$$
\split
    &\pmatrix
    \frac{\der f}{\der \calL} (\calL, \calM) &
    \frac{\der f}{\der \calM} (\calL, \calM) \\
    \frac{\der g}{\der \calL} (\calL, \calM) &
    \frac{\der g}{\der \calM} (\calL, \calM)
    \endpmatrix
    \pmatrix
    \frac{\der \calL}{\der p} & \frac{\der \calL}{\der s} \\
    \frac{\der \calM}{\der p} & \frac{\der \calM}{\der s}
    \endpmatrix
\\
    =
    &\pmatrix
    \frac{\der \fbar}{\der \calLbar} (\calLbar, \calMbar) &
    \frac{\der \fbar}{\der \calMbar} (\calLbar, \calMbar) \\
    \frac{\der \gbar}{\der \calLbar} (\calLbar, \calMbar) &
    \frac{\der \gbar}{\der \calMbar} (\calLbar, \calMbar)
    \endpmatrix
    \pmatrix
    \frac{\der \calLbar}{\der p} & \frac{\der \calLbar}{\der s} \\
    \frac{\der \calMbar}{\der p} & \frac{\der \calMbar}{\der s}
    \endpmatrix .
\endsplit
\tag 2.5.2
$$
Taking the determinant of the both hand sides of this equation and
using the relations $\{f, g\} = f$, $\{\fbar, \gbar\} = \fbar$, we get
$$
    \calL^{-1} \{ \calL, \calM \} = \calLbar^{-1} \{\calLbar, \calMbar\}.
\tag 2.5.3
$$
By the assumption (1) of the proposition
the right hand side of this equation is of the form
$1 + $(negative powers of $p$), while the left hand side contains
only the positive powers of $p$. Therefore the both hand sides of (2.5.3)
are equal to $1$, which proves the canonical Poisson relations.

The Lax equations with respect to $t_n$ are proved as follows.
First, differentiating (2.5.1) by $t_n$ gives
$$
    \pmatrix
    \frac{\der f}{\der \calL} (\calL, \calM) &
    \frac{\der f}{\der \calM} (\calL, \calM) \\
    \frac{\der g}{\der \calL} (\calL, \calM) &
    \frac{\der g}{\der \calM} (\calL, \calM)
    \endpmatrix
    \pmatrix
    \frac{\der \calL}{\der t_n} \\ \frac{\der \calM}{\der t_n}
    \endpmatrix
    =
    \pmatrix
    \frac{\der \fbar}{\der \calLbar} (\calLbar, \calMbar) &
    \frac{\der \fbar}{\der \calMbar} (\calLbar, \calMbar) \\
    \frac{\der \gbar}{\der \calLbar} (\calLbar, \calMbar) &
    \frac{\der \gbar}{\der \calMbar} (\calLbar, \calMbar)
    \endpmatrix
    \pmatrix
    \frac{\der \calLbar}{\der t_n} \\ \frac{\der \calMbar}{\der t_n}
    \endpmatrix.
\tag 2.5.4
$$
Using (2.5.2), we can rewrite (2.5.4) as
$$
    {\pmatrix
    \frac{\der \calL}{\der p} & \frac{\der \calL}{\der s} \\
    \frac{\der \calM}{\der p} & \frac{\der \calM}{\der s}
    \endpmatrix}^{-1}
    \pmatrix
    \frac{\der \calL}{\der t_n} \\ \frac{\der \calM}{\der t_n}
    \endpmatrix
    =
    {\pmatrix
    \frac{\der \calLbar}{\der p} & \frac{\der \calLbar}{\der s} \\
    \frac{\der \calMbar}{\der p} & \frac{\der \calMbar}{\der s}
    \endpmatrix}^{-1}
    \pmatrix
    \frac{\der \calLbar}{\der t_n} \\ \frac{\der \calMbar}{\der t_n}
    \endpmatrix.
\tag 2.5.5
$$
The inverse matrices in this formula are easily calculated because of
the canonical Poisson relations. Thus we get
$$
    \calL^{-1}
    \pmatrix
     \frac{\der \calM}{\der s} & -\frac{\der \calL}{\der s} \\
    -\frac{\der \calM}{\der p} &  \frac{\der \calL}{\der p}
    \endpmatrix
    \pmatrix
    \frac{\der \calL}{\der t_n} \\ \frac{\der \calM}{\der t_n}
    \endpmatrix
    =
    \calLbar^{-1}
    \pmatrix
     \frac{\der \calMbar}{\der s} & -\frac{\der \calLbar}{\der s} \\
    -\frac{\der \calMbar}{\der p} &  \frac{\der \calLbar}{\der p}
    \endpmatrix
    \pmatrix
    \frac{\der \calLbar}{\der t_n} \\ \frac{\der \calMbar}{\der t_n}
    \endpmatrix.
\tag 2.5.6
$$
The first component of the right hand side is given by
$$
\split
    &\calLbar^{-1}
    \left(
    \frac{\der \calMbar}{\der s} \frac{\der \calLbar}{\der t_n}
    -
    \frac{\der \calMbar}{\der t_n} \frac{\der \calLbar}{\der s}
    \right) \\
    =&
    \calLbar^{-1} \left(
    \left. \frac{\der \calMbar}{\der \calLbar} \right|_{t,v \fixed}
    \frac{\der \calLbar}{\der s} + 1 -
    \sum_{n=1}^\infty \frac{\der \vbar_n}{\der s} \calLbar^n
    \right)
    \frac{\der \calLbar}{\der t_n} \\
    -&
    \calLbar^{-1} \left(
    \left. \frac{\der \calMbar}{\der \calLbar} \right|_{t,v \fixed}
    \frac{\der \calLbar}{\der t_n} +
    \sum_{n=1}^\infty \frac{\der \vbar_n}{\der t_n} \calLbar^n
    \right)
    \frac{\der \calLbar}{\der s}
    \\
    =&
    \calLbar^{-1} \left( 1 -
    \sum_{n=1}^\infty \frac{\der \vbar_n}{\der s} \calLbar^n
    \right) \frac{\der \calLbar}{\der t_n}
    -
    \calLbar^{-1} \left(
    \sum_{n=1}^\infty \frac{\der \vbar_n}{\der t_n} \calLbar^n
    \right)
    \frac{\der \calLbar}{\der s},
\endsplit
\tag 2.5.7
$$
which does not contain negative powers of $p$. Similarly the first component
of the left hand side of (2.5.6) is given by
$$
    \calL^{-1}
    \left(
    \frac{\der \calM}{\der s} \frac{\der \calL}{\der t_n}
    -
    \frac{\der \calM}{\der t_n} \frac{\der \calL}{\der s}
    \right)
    =
    -\frac{\der (\calL^n)}{\der s} + \text{ negative powers of $p$}.
\tag 2.5.8
$$
Comparing (2.5.7) and (2.5.8), we obtain
$$
    \calL^{-1}
    \left(
    \frac{\der \calM}{\der s} \frac{\der \calL}{\der t_n}
    -
    \frac{\der \calM}{\der t_n} \frac{\der \calL}{\der s}
    \right)
    =
    \calLbar^{-1}
    \left(
    \frac{\der \calMbar}{\der s} \frac{\der \calLbar}{\der t_n}
    -
    \frac{\der \calMbar}{\der t_n} \frac{\der \calLbar}{\der s}
    \right)
    = - \frac{\der \calB_n}{\der s}.
\tag 2.5.9
$$
By the same method, we obtain from the second component of (2.5.6)
$$
    \calL^{-1}
    \left(
    \frac{\der \calM}{\der s} \frac{\der \calL}{\der t_n}
    -
    \frac{\der \calM}{\der t_n} \frac{\der \calL}{\der s}
    \right)
    =
    \calLbar^{-1}
    \left(
    \frac{\der \calMbar}{\der s} \frac{\der \calLbar}{\der t_n}
    -
    \frac{\der \calMbar}{\der t_n} \frac{\der \calLbar}{\der s}
    \right)
    = \frac{\der \calB_n}{\der p}.
\tag 2.5.10
$$
Equations (2.5.9) and (2.5.10) reduce to a linear equation
$$
    \calL^{-1}
    \pmatrix
     \frac{\der \calM}{\der s} & -\frac{\der \calL}{\der s} \\
    -\frac{\der \calM}{\der p} &  \frac{\der \calL}{\der p}
    \endpmatrix
    \pmatrix
    \frac{\der \calL}{\der t_n} \\ \frac{\der \calM}{\der t_n}
    \endpmatrix
    =
    \calLbar^{-1}
    \pmatrix
     \frac{\der \calMbar}{\der s} & -\frac{\der \calLbar}{\der s} \\
    -\frac{\der \calMbar}{\der p} &  \frac{\der \calLbar}{\der p}
    \endpmatrix
    \pmatrix
    \frac{\der \calLbar}{\der t_n} \\ \frac{\der \calMbar}{\der t_n}
    \endpmatrix
    =
    \pmatrix
    - \frac{\der \calB_n}{\der s} \\ \frac{\der \calB_n}{\der p}
    \endpmatrix.
\tag 2.5.11
$$
Again due to the canonical Poisson relations, the inverse of
the matrices in (2.5.11) can be explicitly calculated, and we obtain
$$
    \pmatrix
    \frac{\der \calL}{\der t_n} \\ \frac{\der \calM}{\der t_n}
    \endpmatrix
    =
    \pmatrix
    \{ \calB_n, \calL \} \\ \{ \calB_n, \calM \}
    \endpmatrix ,
    \quad
    \pmatrix
    \frac{\der \calLbar}{\der t_n} \\ \frac{\der \calMbar}{\der t_n}
    \endpmatrix
    =
    \pmatrix
    \{ \calB_n, \calLbar \} \\ \{ \calB_n, \calMbar \}
    \endpmatrix ,
$$
which is nothing but $t$-flow part of the Lax equations (2.1.1) and (2.3.3).
The $\tbar$-flow part of the Lax equations can be proved in the same way.
\qed
\enddemo

This construction is a kind of Riemann-Hilbert decomposition:
$$
\CD
    (p,s)                @>>>              (\calL, \calM) \\
    @VVV                                   @VV(f,g)V           \\
    (\calLbar, \calMbar) @>>(\fbar,\gbar)> (\calP, \calQ)=(\calPbar,\calQbar),
\endCD
\tag 2.5.12
$$
where $(\calP,\calQ) = (f(\calL, \calM), g(\calL, \calM))$,
$
(\calPbar, \calQbar) = (\fbar(\calLbar, \calMbar), \gbar(\calLbar, \calMbar))
$.
The area-preserving map from the right upper corner to the left lower corner
of this diagram (the composition of $(f,g)$ and the inverse map of
$(\fbar, \gbar)$) is decomposed into two maps,
$(\calL,\calM) \mapsto (p,s)$ and $(p,s) \mapsto (\calLbar, \calMbar)$.
The former of these two maps is rationally extended to $p=\infty$,
while the latter is extended to $p=0$.

The existence of the twistor data for solutions is ensured as follows.
\proclaim{Proposition 2.5.2}
Let $(\calL, \calLbar, \calM, \calMbar)$ be a solution of the dispersionless
Toda hierarchy. Then there exist twistor data $(f,g,\fbar,\gbar)$
which satisfy (2) and (3) of Proposition 2.5.1.
\endproclaim

\demo{Proof}
Let $\varphi$, $\varphibar$ be dressing operators in the sense of \S2.2.
Then
$$
\alignedat2
    f(p,s) &= e^{-\ad\varphi(t=\tbar=0, p, s)} p, &\quad
    g(p,s) &= e^{-\ad\varphi(t=\tbar=0, p, s)} s, \\
    \fbar(p,s) &= e^{-\ad\varphibar(t=\tbar=0, p,s)}
                  e^{-\ad\phi(t=\tbar=0,s)} p, &\quad
    \gbar(p,s) &= e^{-\ad\varphibar(t=\tbar=0, p,s)}
                  e^{-\ad\phi(t=\tbar=0,s)} s,
\endalignedat
\tag 2.5.13
$$
are the corresponding twistor data.
We omit the proof, which is based on the unique solvability of
the initial value problem satisfied by the both hand sides of (2.5.1)
(cf\. the proof of Proposition 1.5.2).
\qed
\enddemo

%
%
%
%

\subhead 2.6. $w_{1+\infty}$ symmetry \endsubhead

Here we describe $w_{1+\infty}$ symmetries of
the dispersionless Toda hierarchy induced by deformation of the twistor data.
As in \S1.6 we consider infinitesimal deformation of $(f,g)$ and
$(\fbar,\gbar)$ ($f\neq 0$, $\fbar\neq 0$) by Hamiltonian vector fields:
$$
\aligned
    ( f   , g   ) &\mapsto ( f   , g   ) \circ \exp(-\eps \ad F   ),\\
    (\fbar,\gbar) &\mapsto (\fbar,\gbar) \circ \exp(-\eps \ad\Fbar),
\endaligned
\tag 2.6.1
$$
and the associated deformation
$$
\aligned
    (\calL   ,\calM   ) &\mapsto (\calL   (\eps), \calM   (\eps)),\\
    (\calLbar,\calMbar) &\mapsto (\calLbar(\eps), \calMbar(\eps))
\endaligned
\tag 2.6.2
$$
of the $(\calL,\calM)$-pair and the $(\calLbar,\calMbar)$-pair.
Here $\ad F$ and $\ad\Fbar$ are regarded as Hamiltonian vector fields:
$$
    \ad F    = p\frac{\der F   }{\der p} \frac{\der}{\der s}
             - p\frac{\der F   }{\der s} \frac{\der}{\der p}, \quad
    \ad\Fbar = p\frac{\der\Fbar}{\der p} \frac{\der}{\der s}
             - p\frac{\der\Fbar}{\der s} \frac{\der}{\der p},
$$
and $\eps$ an infinitesimal parameter.
The rule of the game is the same as in \S1.6. {\it Infinitesimal symmetry} is
defined as the first order coefficient $\delta_{F,\Fbar}$ in $\eps$-expansion:
$$
\alignedat2
    \calL   (\eps) &= \calL    + \eps \delta_{F,\Fbar}\calL    + O(\eps^2),&
    \calLbar(\eps) &= \calLbar + \eps \delta_{F,\Fbar}\calLbar + O(\eps^2),\\
    \calM   (\eps) &= \calM    + \eps \delta_{F,\Fbar}\calM    + O(\eps^2),&
    \quad
    \calMbar(\eps) &= \calMbar + \eps \delta_{F,\Fbar}\calMbar + O(\eps^2)
\endalignedat
\tag 2.6.3
$$
By definition, $\delta_{F,\Fbar}$ acts on any functions of
$\calL, \calM, \calLbar, \calMbar$ as derivation (cf\. (1.6.2)), and
leave the independent variables invariant:
$$
     \delta_{F,\Fbar} t_n = \delta_{F,\Fbar} \tbar_n
    =\delta_{F,\Fbar} p   = \delta_{F,\Fbar} s = 0.
\tag 2.6.4
$$

\proclaim{Proposition 2.6.1}
i)
The infinitesimal symmetries of $\calL$, $\calM$, $\calLbar$, $\calMbar$
are given by
$$
\aligned
    \delta_{F,\Fbar} \calL &=
    \{
    F(\calL, \calM)_{\leq -1} - \Fbar(\calLbar, \calMbar)_{\leq -1}, \calL
    \},
\\
    \delta_{F,\Fbar} \calM &=
    \{
    F(\calL, \calM)_{\leq -1} - \Fbar(\calLbar, \calMbar)_{\leq -1}, \calM
    \},
\\
    \delta_{F,\Fbar} \calLbar &=
    \{
    \Fbar(\calLbar, \calMbar)_{\geq 0} - F(\calL, \calM)_{\geq 0}, \calLbar
    \},
\\
    \delta_{F,\Fbar} \calMbar &=
    \{
    \Fbar(\calLbar, \calMbar)_{\geq 0} - F(\calL, \calM)_{\geq 0}, \calMbar
    \}.
\endaligned
\tag 2.6.5
$$

ii)
The induced symmetries of $v_i$, $i=1,2,\ldots$ are given by
$$
\aligned
    \delta_{F,\Fbar} v_i &=
    - \Res (F(\calL, \calM) - \Fbar(\calLbar, \calMbar)) d_p \calB_i,
\\
    \delta_{F,\Fbar} \vbar_i &=
    - \Res (F(\calL, \calM) - \Fbar(\calLbar, \calMbar)) d_p \calBbar_i.
\endaligned
\tag 2.6.6
$$

iii)
The induced symmetries of the dressing functions are given by
$$
\aligned
    \nabla_{\delta_{F,\Fbar}, \varphi} \varphi &=
    F(\calL, \calM)_{\leq -1} - \Fbar(\calLbar, \calMbar)_{\leq -1},
\\
    e^{\ad\phi} \nabla_{\delta_{F,\Fbar}, \varphibar} \varphibar &=
    \Fbar(\calLbar, \calMbar)_{\geq 0} - F(\calL, \calM)_{\geq 0}.
\endaligned
\tag 2.6.7
$$

iv)
An infinitesimal symmetry of $\phi$ consistent with i) -- iii) is given by
$$
    \delta_{F,\Fbar} \phi =
    - \Res  F   (\calL   , \calM   ) d\log p
    + \Res \Fbar(\calLbar, \calMbar) d\log p.
\tag 2.6.8
$$
\endproclaim

\demo{Proof}
In this proof, we abbriviate $\delta_{F,\Fbar}$ to $\delta$.
Part i) is a corollary of part iii) and part iv) due to Lemma A.1.
Part ii) is proved as follows. From the expansion (2.3.2) the deformation of
$\calM$ has a form
$$
    \delta \calM =
    \left. \frac{\der \calM}{\der \calL} \right|_{t,\tbar,v \fixed}
    \delta \calL +
    \sum_{i=1}^\infty \delta v_i \calL^{-i}.
\tag 2.6.9
$$
The coefficient of $\calL^{-i}$ can be thus written
$$
\split
    \delta v_i &=
    \Res \left(
    \delta \calM -
    \left. \frac{\der \calM}{\der \calL} \right|_{t,\tbar,v \fixed}
    \right)
    \calL^{i-1} d_p \calL
\\
    &= \Res \calL^{i-1} (\delta \calM d_p \calL - \delta \calL d_p \calM)
\\
    &= \Res \calL^{i-1} (
    \{ F(\calL,\calM)_{\leq -1} - \Fbar(\calLbar,\calMbar)_{\leq -1}, \calM \}
    d_p \calL
\\
    &\phantom{ = \Res \calL^{i-1}}
    -
    \{ F(\calL,\calM)_{\leq -1} - \Fbar(\calLbar,\calMbar)_{\leq -1}, \calL \}
    d_p \calM
    ),
\endsplit
\tag 2.6.10
$$
where we have used i).
Expanding the Poisson brackets on the right hand side of
(2.6.10), and using the relation $\{\calL,\calM\} = \calL$, we can rewrite
the last expression as
$$
\split
    &\Res \calL^{i-1} \left(
    \frac{\der}{\der p}
    (F(\calL,\calM)_{\leq -1} - \Fbar(\calLbar,\calMbar)_{\leq -1})
    \calL dp
    \right)
\\
    =
    &\Res \calL^i
    d_p (F(\calL,\calM)_{\leq -1} - \Fbar(\calLbar,\calMbar)_{\leq -1})
\\
    =
    & - \Res  (F(\calL,\calM) - \Fbar(\calLbar,\calMbar)) d_p \calB_i,
\endsplit
\tag 2.6.11
$$
which proves the first equation of (2.6.6). The second equation of (2.6.6)
can be proved in the same way.

The last two statements iii) and iv) are proved as follows.
Let us denote the deformed twistor data (2.6.1) by $(f_\eps, g_\eps)$,
$(\fbar_\eps, \gbar_\eps)$ and the deformation of the dressing functions
$\varphi$, $\varphibar$ and the potential $\phi$ by
$\varphi_\eps = \varphi + \eps \delta \varphi + O(\eps^2)$,
$\varphibar_\eps = \varphibar + \eps \delta \varphibar + O(\eps^2)$,
$\phi_\eps = \phi + \eps \delta\phi + O(\eps^2)$.
The Riemann-Hilbert problem (2.5.1) for the deformed twistor data becomes
$$
\aligned
    e^{\ad\varphi_\eps} e^{\ad t(p)} f_\eps(p,s)
    &=
    e^{\ad\phi_\eps} e^{\ad\varphibar_\eps} e^{\ad \tbar(p^{-1})}
    \fbar_\eps(p,s),
\\
    e^{\ad\varphi_\eps} e^{\ad t(p)} g_\eps(p,s)
    &=
    e^{\ad\phi_\eps} e^{\ad\varphibar_\eps} e^{\ad \tbar(p^{-1})}
    \gbar_\eps(p,s).
\endaligned
\tag 2.6.12
$$
Calculating exactly in the same manner as in the proof of Proposition 1.6.3,
we obtain from the coefficient of $\eps$ in (2.6.12)
$$
\aligned
    \{ \nabla_{\delta,\varphi} \varphi - F(\calL, \calM), \calP \}
    &=
    \{ \nabla_{\delta,\phi} \phi +
       e^{\ad\phi} (\delta_{\delta,\varphibar} \varphibar)
       - \Fbar(\calLbar, \calMbar),
    \calPbar \},
\\
    \{ \nabla_{\delta,\varphi} \varphi - F(\calL, \calM), \calQ \}
    &=
    \{ \nabla_{\delta,\phi} \phi +
       e^{\ad\phi} (\delta_{\delta,\varphibar} \varphibar)
       - \Fbar(\calLbar, \calMbar),
    \calQbar \},
\endaligned
\tag 2.6.13
$$
where $\calP = f(\calL, \calM)$, $\calQ = g(\calL, \calM)$,
$\calPbar = \fbar(\calLbar, \calMbar)$, $\calQbar = \gbar(\calLbar, \calMbar)$.
Since $(\calL, \calM, \calLbar, \calMbar)$ is a solution of (2.5.1),
we have $\calP = \calPbar$, $\calQ = \calQbar$, and
$\{\calP, \calQ\} = \calP$, $\{\calPbar, \calQbar\} = \calPbar$.
Hence the following Lemma 2.6.2 proves iii) (2.6.7).
To prove iv) note that $\nabla_{\delta,\phi} \phi = \delta \phi$.
This follows from the fact that $\phi$ does not contain $p$. Hence
(2.6.13) and Lemma 2.6.2 below imply
$$
    \delta \phi =
    - \Res  F   (\calL   , \calM   )\; d\log p
    + \Res \Fbar(\calLbar, \calMbar)\; d\log p + \text{ const.},
\tag 2.6.14
$$
where const\. denotes a term which does not depend on $(p,s)$.
Thus this ambiguity belongs to the center of the Lie algebra of Poisson
brackets, from which iv) follows.
\qed
\enddemo

\proclaim{Lemma 2.6.2}
Suppose that $\calP \neq 0$ and $\calQ$ satisfy the canonical Poisson relation
$\{\calP, \calQ\} = \calP$, and that a series $\calA$ satisfy
$$
    \{\calA, \calP\} = \{\calA, \calQ\} = 0.
\tag 2.6.15
$$
Then $\der \calA/\der p = \der \calA/ \der s = 0$.
\endproclaim

\demo{Proof}
The equation (2.6.15) is equivalent to
$$
    \pmatrix
    \frac{\der \calP}{\der s} & -\frac{\der \calP}{\der p} \\
    \frac{\der \calQ}{\der s} & -\frac{\der \calQ}{\der p}
    \endpmatrix
    \pmatrix
    \frac{\der \calA}{\der p} \\ \frac{\der \calA}{\der s}
    \endpmatrix
    = 0.
\tag 2.6.16
$$
Because of the canonical Poisson relation, the determinant of the matrix
on the left hand side of (2.6.16) is $p^{-1} \calP$ which does not vanish
by the assumption. This proves the lemma.
\qed
\enddemo

This symmetry can be lifted up to the tau function as follows.
\proclaim{Proposition 2.6.3}
An infinitesimal symmetry of $\tau = \taudtoda$ consistent with i) -- iii)
is given by
$$
    \delta_{F,\Fbar} \log\tau =
    - \Res  F^s   (\calL   , \calM   ) \; d_p \log \calL
    + \Res \Fbar^s(\calLbar, \calMbar) \; d_p \log \calLbar,
\tag 2.6.17
$$
where $F^s$ and $\Fbar^s$ are defined by
$$
     F^s   (p,s) \=def \int_0^s  F   (p,s')\, ds', \quad
    \Fbar^s(p,s) \=def \int_0^s \Fbar(p,s')\, ds'.
\tag 2.6.18
$$
\endproclaim

\demo{Proof}
The consistency means
$$
\gather
    \frac{\der}{\der t_n}\delta\log\tau
    = \delta \frac{\der\log\tau}{\der t_n}, \quad
    \frac{\der}{\der \tbar_n}\delta\log\tau
    = \delta \frac{\der\log\tau}{\der \tbar_n},
\tag 2.6.19
\\
    \frac{\der}{\der s}\delta\log\tau
    = \delta \frac{\der\log\tau}{\der s},
\tag 2.6.20
\endgather
$$
where $\delta = \delta_{F,\Fbar}$ as before.

First let us consider (2.6.19).
We check here only the first equation of (2.6.19).
The second equation of (2.6.19) can be checked similarly.
The right hand side of this equation has been already calculated in
Proposition 2.6.1 ii):
$$
    \delta \frac{\der\log\tau}{\der t_n} =
    \delta v_n =
    - \Res (F(\calL, \calM) - \Fbar(\calLbar, \calMbar)) \; d_p \calB_n.
\tag 2.6.21
$$
The left hand side of (2.6.19) is written as
$$
    \frac{\der}{\der t_n} \delta \log\tau
    =
    - \frac{\der}{\der t_n}
      \Res  F^s   (\calL   ,\calM   ) \; d_p \log \calL
    + \frac{\der}{\der t_n}
      \Res \Fbar^s(\calLbar,\calMbar) \; d_p \log \calLbar.
\tag 2.6.22
$$
The terms of the right hand side of (2.6.22) can be calculated
exactly in the same way as the corresponding expression of the dispersionless
KP hierarchy ([1] Proposition 15). Therefore we omit the details of
calculation and give only the result:
$$
\aligned
    \frac{\der}{\der t_n} \Res F^s(\calL,\calM) \; d_p \log \calL
    &=
    \Res F(\calL, \calM) \; d_p \calB_n.
\\
    \frac{\der}{\der t_n} \Res \Fbar^s(\calLbar,\calMbar) \; d_p \log \calL
    &=
    \Res \Fbar(\calLbar, \calMbar) \; d_p \calB_n.
\endaligned
\tag 2.6.23
$$
Thus (2.6.21) and (2.6.22) give the identical results, which proves (2.6.19).

We turn to the proof of equation (2.6.20).
The right hand side of this equation is
$$
    \delta \frac{\der\log\tau}{\der s} = \delta \phi
    =
    - \Res  F   (\calL   , \calM   ) d\log p
    + \Res \Fbar(\calLbar, \calMbar) d\log p.
\tag 2.6.24
$$
because of (2.6.8).
The left hand side of (2.6.20) is calculated by the same method as (2.6.23):
$$
\split
    \frac{\der}{\der s} \delta \log\tau =&
    \frac{\der}{\der s}(
    - \Res  F^s   (\calL   , \calM   ) \; d_p \log \calL
    + \Res \Fbar^s(\calLbar, \calMbar) \; d_p \log \calLbar ),
\\
    =&
    - \Res  F   (\calL   , \calM   ) \left(
    \frac{\der \calM   }{\der s} -
    \left. \frac{\der \calM   }{\der \calL   } \right|_{t,\tbar,v \fixed}
    \frac{\der \calL   }{\der s}     \right)
    \; d_p \log \calL \\
    &+\Res \Fbar(\calLbar, \calMbar) \left(
    \frac{\der \calMbar}{\der s} -
    \left. \frac{\der \calMbar}{\der \calLbar} \right|_{t,\tbar,v \fixed}
    \frac{\der \calLbar}{\der s}     \right)
    \; d_p \log \calLbar )
\\
    =&
    - \Res  F   (\calL   , \calM   ) \calL^{-1}    \left(
    \frac{\der \calM   }{\der s} \frac{\der \calL   }{\der p}-
    \frac{\der \calM   }{\der p} \frac{\der \calL   }{\der s}
    \right) dp \\
    &+\Res \Fbar(\calLbar, \calMbar) \calLbar^{-1} \left(
    \frac{\der \calMbar}{\der s} \frac{\der \calLbar}{\der p}-
    \frac{\der \calMbar}{\der p} \frac{\der \calLbar}{\der s}
    \right) dp
\\
    =&
    - \Res  F   (\calL   , \calM   ) p^{-1} dp
    + \Res \Fbar(\calLbar, \calMbar) p^{-1} dp,
\endsplit
\tag 2.6.25
$$
by the canonical Poisson relations.
Equations (2.6.24), (2.6.25) prove the equation (2.6.20).
\qed
\enddemo

These infinitesimal symmetries almost respect the $w_{1+\infty}$ algebra
structure as follows.

\proclaim{Proposition 2.6.4}
i)
Commutation relations for $\delta_{F,\Fbar}$ on $\calL$, $\calM$, $\calLbar$,
$\calMbar$, $\varphi$, $\varphibar$ and $\phi$ have the form;
$$
    [ \delta_{F_1,\Fbar_1}, \delta_{F_2,\Fbar_2} ]K
    = \delta_{ \{F_1,F_2\}, \{\Fbar_1,\Fbar_2\} }K,
\tag 2.6.26
$$
where $K$ is one of $\calL$, $\calM$, $\calLbar$, $\calMbar$, $\varphi$,
$\varphibar$ and $\phi$.

ii)
For the tau function,
$$
    [ \delta_{F_1,\Fbar_1}, \delta_{F_2,\Fbar_2} ] \log\tau
    = \delta_{ \{F_1,F_2\}, \{\Fbar_1,\Fbar_2\} }  \log\tau
     +c(F_1,F_2) + \cbar(\Fbar_1,\Fbar_2),
\tag 2.6.27
$$
where $c$ and $\cbar$ are cocycles of the $w_{1+\infty}$-algebra given by
$$
\aligned
    & c(F_1,F_2) \=def -\Res F_2(p,0)dF_1(p,0),
\\
    & \cbar(\Fbar_1,\Fbar_2)
         \=def \Res \Fbar_2(p,0)d\Fbar_1(p,0).
\endaligned
\tag 2.6.28
$$
\endproclaim

\demo{Proof}
The commutation relation for $\calL$, $\calM$, $\calLbar$, $\calMbar$,
$\varphi$, $\varphibar$ and $\phi$ is derived from that for the tau function
due to consistency in Proposition 2.6.3 and (2.4.11), (2.3.4).

It is sufficient to prove the commutation relation
for the tau function (2.6.28) only for the case
that $F_i(p,s)$, $\Fbar_i(p,s)$ ($i=1,2$) are monomials
of $s$: $F_i(p,s) = f_i(p) s^{j_i}$, $\Fbar_i(p,s) = \fbar_i(p) s^{k_i}$.
We omit the proof of this fact, since it is essentially the same calculation
as that of the corresponding statement for the dispersionless KP hierarchy
([1], Proposition 16).
\qed
\enddemo

%
%
%
%

\subhead 2.7. Quasi-classical limit of Toda lattice hierarchy  \endsubhead

In this section we consider the ``dispersionful'' counterpart of
the contents of the preceding sections. Following [4], we introduce
a ``Planck constant'' $\hbar$ into the ordinary Toda lattice hierarchy [21]
and consider the limit $\hbar\to 0$.
This small parameter $\hbar$ plays the role of lattice spacing
in the Toda lattice. Leading terms in this limit recover the dispersionless
Toda hierarchy. As in \S1.7, the subsections 2.7.1 -- 2.7.6 correspond to
\S\S2.1--2.6.

%
%
%
%
\subsubhead 2.7.1 Lax formalism \endsubsubhead
To interpret the dispersionless Toda hierarchy as a quasi-classical limit,
we reformulate the Toda lattice hierarchy in the language of
difference operators of an continuous variable $s$ with spacing unit
$\hbar$: $\exp(m\hbar\der/\der s) f(s) = f(s + m \hbar)$.
The Lax representation of this reformulated Toda lattice hierarchy
is given by
$$
\aligned
    \hbar \frac{\der L}{\der t_n} = [ B_n, L ], \quad&
    \hbar \frac{\der L}{\der \tbar_n} = [\Bbar_n, L ],
\\
    \hbar \frac{\der \Lbar}{\der t_n} = [ B_n, \Lbar ], \quad&
    \hbar \frac{\der \Lbar}{\der \tbar_n} = [ \Bbar_n, \Lbar ],
    \quad n = 1,2,\ldots,
\endaligned
\tag 2.7.1
$$
where the Lax operators $L$ and $\Lbar$ are difference operators
of the form
$$
\aligned
    L =& e^{\hbar\der/\der s} + \sum_{n=0}^\infty
         u_{n+1}(\hbar,t,\tbar,s) e^{-n\hbar\der/\der s},
\\
    \Lbar^{-1} =& \ubar_0(\hbar,t,\tbar,s) e^{-\hbar\der/\der s} +
    \sum_{n=0}^\infty \ubar_{n+1}(\hbar,t,\tbar,s) e^{n\hbar\der/\der s}
\endaligned
\tag 2.7.2
$$
and $B_n$ and $\Bbar_n$ are given by
$$
    B_n = (L^n)_{\ge 0}, \quad
    \Bbar_n = (\Lbar^{-n})_{\le -1}.
\tag 2.7.3
$$
Also here, we define $\ubar_{n+1}$ to be the coefficients of $\Lbar^{-1}$
rather than $\Lbar$ itself (cf\. (2.1.2)).
Here $(\quad)_{\ge 0}$ and $(\quad)_{\le -1}$ denote the projection
onto a linear combination of $e^{n\hbar\der/\der s}$ with
$n\geqq 0$ and $\leqq -1$, respectively.
The coefficients $u_n$, $\ubar_n$ of $L$, $\Lbar$ are assumed to be regular
with respect to $\hbar$: $u_n(\hbar,t,\tbar,s) = u^0_n(t,\tbar,s) + O(\hbar)$,
$\ubar_n(\hbar,t,\tbar,s) = \ubar^0_n(t,\tbar,s) + O(\hbar)$.

As is well known, the Lax representation is equivalent to the zero-curvature
representation
$$
\aligned
    \hbar\der_{ t_n   }  B_m    -
    \hbar\der_{ t_m   }  B_n    + [ B_m,     B_n   ] &= 0,
\\
    \hbar\der_{\tbar_n} \Bbar_m -
    \hbar\der_{\tbar_m} \Bbar_n + [\Bbar_m, \Bbar_n] &= 0,
\\
    \hbar\der_{\tbar_n}  B_m    -
    \hbar\der_{ t_m   } \Bbar_n + [ B_m,    \Bbar_n] &= 0,
\endaligned
\tag 2.7.4
$$

The ``{\it order}\/'' and the ``{\it principal symbol}\/'' are defined
for the difference operators as follows.

\demo{Definition 2.7.1}
$$
    \ord \left(
    \sum a_{n,m}(t,\tbar,s) \hbar^n e^{m \hbar \der/\der s} \right)
    \=def
    \max \{ n \,|\, a_{n,m}(t,\tbar,s) \neq 0\}.
\tag 2.7.5
$$
In particular, $\ord(\hbar) = -1$, $\ord(\exp(\hbar\der/\der s)) = 0$.
The {\it principal symbol} (resp\. the {\it symbol of order $l$})
of a difference operator $A = \sum a_{n,m}\hbar^n \exp(m\hbar\der/\der s)$ is
$$
\aligned
    \symbolh  (A)
    &\=def \hbar^{-\ord(A)} \sum_{n = \ord(A)} \sum_m a_{n,m}  p^m
    \\
    (\text{resp. }
    \symbolh_l(A)
    &\=def \hbar^{-l} \sum_{n = l} \sum_m a_{n,m}  p^m).
\endaligned
\tag 2.7.6
$$
\enddemo

For example, the condition which we imposed on the coefficients
$u_n$, $\ubar_n$ can be restated as $\ord (L) = \ord(\Lbar) = 0$.

\demo{Remark 2.7.2}
This ``order'' coincides with the order of a micro-differential operator
if we formally replace $\hbar$ with $\der_{t_0}^{-1}$, where $t_0$ is
an extra variable (cf\. Remark 1.7.2).
\enddemo

The following facts can be easily proved.

\proclaim{Lemma 2.7.3}
Let $P_i$ ($i=1,2$) be two difference operators
of finite order: $\ord (P_i) = l_i$.
\roster
\item[1] $\ord(P_1 P_2) = l_1 + l_2$ and
$$
    \symbolh (P_1 P_2) = \symbolh(P_1) \symbolh(P_2).
$$
\item[2] $\ord([P_1, P_2]) \leqq l_1 + l_2 - 1$ and
$$
    \symbolh_{l_1 + l_2 - 1} ([P_1, P_2])
    =
    \hbar \{ \symbolh(P_1), \symbolh(P_2) \},
$$
\endroster
where $\{\quad, \quad\}$ is the Poisson bracket defined by (2.1.4).
\endproclaim

For example, we have
$[ \exp(\hbar\der/\der s), s ] = \hbar \exp(\hbar\der/\der s)$,
the principal symbol of which gives $\{p,s\} = p$.

Taking the principal symbol of (2.7.1) and using Lemma 2.7.3, we obtain
\proclaim{Proposition 2.7.4}
$(\calL = \symbolh(L), \calLbar = \symbolh(\Lbar))$ is a solution of
the dispersionless Toda hierarchy (2.1.1), and
$\calB_n = \symbolh(B_n)$, $\calBbar_n = \symbolh(\Bbar_n)$.
\endproclaim

%
%
%
%

\subsubhead 2.7.2 Dressing operators \endsubsubhead
Studying the structure of the dressing operators $W$, $\Wbar$
of the ordinary Toda lattice hierarchy carefully,
we can find counterparts of the dressing functions
$\varphi$, $\varphibar$, and of the potential $\phi$ of the dispersionless
Toda hierarchy.

\proclaim{Proposition 2.7.5}
There are dressing operators $W$ and $\Wbar$
$$
\alignat2
    L     &= W     e^{\hbar\der/\der s}  W^{-1},& \quad
    \Lbar &= \Wbar e^{\hbar\der/\der s} \Wbar^{-1},
\tag 2.7.7
\\
    \hbar \frac{\der  W   }{\der t_n}  W^{-1}    &=
    -(W e^{n\hbar\der/\der s} W^{-1})_{\leq -1},& \quad
    \hbar \frac{\der \Wbar}{\der t_n} \Wbar^{-1} &=
     (W e^{n\hbar\der/\der s} W^{-1})_{\geq 0},
\tag 2.7.8
\\
    \hbar \frac{\der  W   }{\der \tbar_n}  W^{-1}    &=
     (\Wbar e^{-n\hbar\der/\der s} \Wbar^{-1})_{\leq -1},& \quad
    \hbar \frac{\der \Wbar}{\der \tbar_n} \Wbar^{-1} &=
    -(\Wbar e^{-n\hbar\der/\der s} \Wbar^{-1})_{\geq 0}
\tag 2.7.9
\endalignat
$$
of the form
$$
\gathered
     W    = \exp( \hbar^{-1}  X(\hbar,t,\tbar,s,\exp(\hbar\der/\der s)) ),
\\
     X   (\hbar,t,\tbar,s,\exp(\hbar\der/\der s)) =
    \sum_{n=1}^\infty \chi_n   (\hbar,t,\tbar,s) e^{-n\hbar\der/\der s},
\\
    \ord ( X   (\hbar,t,\tbar,s,\exp(\hbar\der/\der s))) = 0,
\\
    \Wbar = \exp(\phi(\hbar,t,\tbar,s))
            \exp( \hbar^{-1} \Xbar(\hbar,t,\tbar,s,\exp(\hbar\der/\der s)) ),
\\
    \Xbar(\hbar,t,\tbar,s,\exp(\hbar\der/\der s)) =
    \sum_{n=1}^\infty \chibar_n(\hbar,t,\tbar,s) e^{ n\hbar\der/\der s},
\\
    \ord (\Xbar(\hbar,t,\tbar,s,\exp(\hbar\der/\der s))) = 0.
\endgathered
\tag 2.7.10
$$
Here $\phi(\hbar,t,\tbar,s)$ satisfies $\ord(\phi) \leqq 0$ and
$$
\gathered
    \frac{\der \phi}{\der t_n}     =  (L^n)_0, \quad
    \frac{\der \phi}{\der \tbar_n} = -(\Lbar^{-n})_0,
\\
    \frac1\hbar(\phi(\hbar,t,\tbar,s) - \phi(\hbar,t,\tbar,s-\hbar)) =
    \log\ubar_0(\hbar,t,\tbar,s),
\endgathered
\tag 2.7.11
$$
where $(\quad)_0$ is the projection to the multiplication operators
(operators not containing $\exp(n\hbar\der/\der s)$, $n\neq 0$).

Conversely, if $W$ and $\Wbar$ of the form (2.7.10) satisfies
(2.7.8) and (2.7.9), then $(L,\Lbar)$ of (2.7.7) gives
a soluiton of the Lax equations (2.7.1).
\endproclaim

\demo{Proof}
The proof is mostly parallel to the proof of Proposition 2.1.2 and 2.2.1.
First the compatibility conditions of the equations (2.7.11)
are a part of the Lax equations and the zero-curvature equations
(2.7.1), (2.7.3) (cf\. the proof of Proposition 2.1.2).
The rest of the proof is analogous to that of Proposition 1.7.5:
according to Lemma 2.7.3, the linear space of difference operators
of order not greater than $1$,
$\calE_1 = \{ P:$ difference operator $| \ord(P) \leqq 1\}$,
is closed under the commutator and thus is a Lie algebra.
Therefore, replacing $\{\quad, \quad\}$, $\varphi$, $\varphibar$, $\phi$,
$p$ by $[\quad, \quad]$, $X/\hbar$, $\Xbar/\hbar$, $\phi$,
$\exp(\hbar\der/\der s)$, respectively, we can carry over
the proof of Proposition 2.2.1 to this case.
\qed
\enddemo

As is obvious from this proof, $\varphi$, $\varphibar$ (\S2.2),
$\phi$ (\S2.1) and $W$, $\Wbar$, $\phi$ above are connected as follows
(cf\. Corollary 1.7.6):
\proclaim{Corollary 2.7.6}
Symbols $\symbolh(X)$ and $\symbolh(\Xbar)$ give
dressing functions $\varphi$, $\varphibar$ of $\calL = \symbolh(L)$,
$\calLbar = \symbolh(\Lbar)$ in the sense of \S2.2.
Symbol $\symbolh(\phi)$ becomes a corresponding $\phi$-potential
defined by (2.1.6).
Conversely, if $\varphi$ and $\varphibar$ are dressing functions of a solution
$(\calL, \calLbar)$ of the dispersionless Toda hierarchy and $\phi_0$
is the $\phi$-potential, there exist
a solution $(L, \Lbar)$ of the Toda lattice hierarchy and dressing operators
$W=\exp(\hbar^{-1}X)$, $\Wbar=\exp(\phi) \exp(\hbar^{-1}\Xbar)$
such that $\symbolh(L) = \calL$, $\symbolh(\Lbar) = \calLbar$,
$\symbolh(X) = \varphi$, $\symbolh(\Xbar) = \varphibar$,
$\symbolh(\phi) = \phi_0$.
\endproclaim

%
%
%
%

\subsubhead 2.7.3 Orlov operators \endsubsubhead
The {\it Orlov operators} $M$ and $\Mbar$ of the Toda lattice hierarchy
are defined by
$$
\aligned
    M &= W \left( \sum_{n=1}^\infty n t_n e^{n\hbar\der/\der s} + s \right)
         W^{-1}
       = \Ad  ( W
              \exp ( \hbar^{-1} t(e^{\hbar\der/\der s}) )
              ) s,
\\
    \Mbar &= \Wbar \left(
                   - \sum_{n=1}^\infty n \tbar_n e^{-n\hbar\der/\der s} + s
                   \right)
             \Wbar^{-1}
           = \Ad ( \Wbar
                 \exp ( \hbar^{-1} \tbar(e^{-\hbar\der/\der s}) )
                 ) s,
\endaligned
\tag 2.7.12
$$
where $t(e^{\hbar\der/\der s})=\sum_{n=1}^\infty t_n e^{n\hbar\der/\der s}$,
$\tbar(e^{-\hbar\der/\der s})=
\sum_{n=1}^\infty \tbar_n e^{-n\hbar\der/\der s}$.
These operators are expanded into Laurent series of $L$ and $\Lbar$ as follows:
$$
\aligned
      M     &= \sum_{n=1}^\infty n t_n         L^n + s +
         \sum_{n=1}^\infty     v_n(\hbar,t,\tbar,s) L^{-n},
\\
      \Mbar &=-\sum_{n=1}^\infty n \tbar_n \Lbar^n + s +
         \sum_{n=1}^\infty \vbar_n(\hbar,t,\tbar,s) \Lbar^{-n}.
\endaligned
\tag 2.7.13
$$

It is easy to see the following facts.
\proclaim{Proposition 2.7.7}
\roster
\item[1]
$\ord(M) = \ord(\Mbar) = 0$;
\item[2]
$[L, M] = \hbar L$, $[\Lbar, \Mbar] = \hbar \Lbar$;
\item[3]
$M$ and $\Mbar$ satisfy the Lax equations,
$$
\aligned
    \hbar \frac{\der M}{\der t_n} = [ B_n, M ], \quad&
    \hbar \frac{\der M}{\der \tbar_n} = [\Bbar_n, M ],
\\
    \hbar \frac{\der \Mbar}{\der t_n} = [ B_n, \Mbar ], \quad&
    \hbar \frac{\der \Mbar}{\der \tbar_n} = [ \Bbar_n, \Mbar ],
    \quad n = 1,2,\ldots,
\endaligned
\tag 2.7.14
$$
where $B_n$, $\Bbar_n$ are defined by (2.7.3).
\endroster

Conversely, if $M$ and $\Mbar$ have the form (2.7.13) and satisfy (1) -- (3),
then there exist unique dressing operators $W$ and $\Wbar$ such that
$M$ and $\Mbar$ are expressed as (2.7.12).
\endproclaim
We omit the proof since it is essentially the same as those of
Proposition 1.3.1, Proposition 1.7.7 and Proposition 2.3.1.

\proclaim{Corollary 2.7.8}
$\calM = \symbolh(M)$ and $\calMbar = \symbolh(\Mbar)$ are the Orlov functions
of the dispersionless Toda hierarchy corresponding to
$\calL = \symbolh(L)$ and $\calLbar = \symbolh(\Lbar)$.
\endproclaim

%
%
%
%

\subsubhead 2.7.4 $S$ function and $\tau$ function (free energy)
\endsubsubhead
The $S$ functions $S$ and $\Sbar$ introduced in \S2.4 appear as the phase
function of the WKB anaysis of the Toda lattice hierarchy.

We define the Baker-Akhiezer functions
$$
\aligned
  & \Psi = W(\hbar,t,\tbar,s)
           \exp \hbar^{-1} ( t(\lambda) + s\log\lambda ),
\\
  & \Psi = \Wbar(\hbar,t,\tbar,s)
           \exp \hbar^{-1} ( \tbar(\lambdabar^{-1}) + s\log\lambdabar ).
\endaligned
\tag 2.7.15
$$
They satisfy the linear equations
$$
\gather
          \lambda \Psi                           = L \Psi, \quad
    \hbar \lambda \frac{\der \Psi}{\der \lambda} = M \Psi, \quad
    \hbar \frac{\der\Psi}{\der t_n}     = B_n     \Psi,    \quad
    \hbar \frac{\der\Psi}{\der \tbar_n} = \Bbar_n \Psi,
\tag 2.7.16
\\
          \lambdabar \Psibar = \Lbar \Psibar, \quad
    \hbar \lambdabar \frac{\der \Psibar}{\der \lambdabar}
                             = \Mbar \Psibar, \quad
    \hbar \frac{\der\Psibar}{\der t_n}     = B_n     \Psibar, \quad
    \hbar \frac{\der\Psibar}{\der \tbar_n} = \Bbar_n \Psibar.
\tag 2.7.17
\endgather
$$
Here $\lambda$ and $\lambdabar$ are formal `spectral parameters'
and $t(\lambda) = \sum_{n=1}^\infty t_n \lambda^n$,
$\tbar(\lambdabar^{-1}) = \sum_{n=1}^\infty \tbar_n \lambdabar^{-n}$.

\proclaim{Proposition 2.7.9}
Formal power series $\Psi$ and $\Psibar$ take a WKB asymptotic form
as $\hbar \to 0$:
$$
\aligned
    \Psi    =& \exp ( \hbar^{-1} S(t,\tbar,s,\lambda) + O(\hbar^0)) ,
\\
    \Psibar =& \exp ( \hbar^{-1} \Sbar(t,\tbar,s,\lambdabar) + O(\hbar^0)),
\endaligned
\tag 2.7.18
$$
where $S(t,\tbar,s,\lambda)$ and $\Sbar(t,\tbar,s,\lambdabar)$
are Laurent series of the form
$$
\aligned
    S(t,\tbar,s,\lambda)     =&
    t(\lambda) + s\log\lambda
    + \sum_{n=1}^\infty S_n(t,\tbar,s) \lambda^{-n}, \\
    \Sbar(t,\tbar,s,\lambda) =&
    \tbar(\lambdabar^{-1}) + s\log\lambdabar
    + \sum_{n=0}^\infty \Sbar_n(t,\tbar,s) \lambdabar^n.
\endaligned
\tag 2.7.19
$$
In particular, $\Sbar_0(t,\tbar,s) = \phi(t,\tbar,s)$.
\endproclaim

\demo{Proof}
By the explicit form of the dressing operators $W$, $\Wbar$ (2.7.10),
$\Psi$ and $\Psibar$ have the form
$$
\align
    \Psi &= (1 + O(\lambda^{-1})) \exp ( \hbar^{-1} t(\lambda) ),
\\
    \Psibar &= e^{\hbar^{-1} \phi} (1 + O(\lambdabar^{-1}))
    \exp ( \hbar^{-1} \tbar(\lambdabar^{-1}) ),
\endalign
$$
Hence by Proposition 2.7.6 (1) and equations (2.7.16) we have the asymptotic
form (2.7.19).
\qed
\enddemo

Now that $\Psi$ and $\Psibar$ have the asymptotic form (2.7.18),
we can follow the principle of the WKB analysis as in \S1.7.4,
regarding $S$ and $\Sbar$ as the phase functions.
The Hamilton-Jacobi (eikonal) equations corresponding to (2.7.16) and (2.7.17)
are
$$
\aligned
    \lambda =& e^{\der_s S(t,\tbar,s,\lambda)}
            + \sum_{n=0}^\infty u^0_{n+1}(t,\tbar,s)
               e^{-n \der_s S(t,\tbar,s,\lambda)} \\
            =& \left. \symbolh(L)
               \right|_{p = \exp(\der_s S(t,\tbar,s,\lambda))},
\\
    dS(t,\tbar,s,\lambda)
    =& \calM(\lambda) d \log\lambda
     +\frac{\der S}{\der s}ds
     +\sum_{n=1}^\infty
          \calB_n ( e^{\der S/\der s} ) dt_n
     +\sum_{n=1}^\infty
          \calBbar_n ( e^{\der S/\der s} )d\tbar_n,
\\
    \lambdabar =&
               \left. \symbolh(\Lbar)
               \right|_{p = \exp(\der_s \Sbar(t,\tbar,s,\lambdabar))},
\\
    d\Sbar(t,\tbar,s,\lambdabar)
    =& \calMbar(\lambdabar) d\log\lambdabar
     +\frac{\der \Sbar}{\der s}ds
     +\sum_{n=1}^\infty
          \calB_n( e^{\der \Sbar/\der s} ) dt_n
     +\sum_{n=1}^\infty
          \calBbar_n( e^{\der \Sbar/\der s} ) d\tbar_n,
\endaligned
\tag 2.7.20
$$
where
$$
\alignedat2
    \calM(\lambda) =&
    \left. \symbolh(M)     \right|_{\calL = \lambda}, \quad &
    \calB_n ( e^{\der S/\der s} ) =&
    \left. \symbolh(B_n) \right|_{p = \exp(\der S/ \der s)},
\\
    \calMbar(\lambdabar) =&
    \left. \symbolh(\Mbar) \right|_{\calLbar = \lambdabar}, \quad &
    \calBbar_n ( e^{\der S/\der s} ) =&
    \left. \symbolh(\Bbar_n) \right|_{p = \exp(\der \Sbar/ \der s)}.
\endalignedat
\tag 2.7.21
$$
Thus we arrive at
\proclaim{Proposition 2.7.10}
Under the Legendre-type transformation
$(t,\tbar,s,\lambda) \mapsto (t,\tbar,s,p)$ and
$(t,\tbar,s,\lambdabar) \mapsto (t,\tbar,s,p)$ defined by
$$
    \exp \frac{\der S(t,\tbar,s,\lambda)}{\der s}
  = \exp \frac{\der \Sbar(t,\tbar,s,\lambdabar)}{\der s}
  = p,
\tag 2.7.22
$$
the spectral parameters $\lambda$ and $\lambdabar$ turn into
the $\calL$- and $\calLbar$-functions of the dispersionless Toda hierarchy,
$$
    \lambda = \calL(t,\tbar,s,p), \quad
    \lambdabar = \calLbar(t,\tbar,s,p),
$$
whereas $S(t,\tbar,s,\lambda)$ and $\Sbar(t,\tbar,s,\lambdabar)$ become
the corresponding $S$- and $\Sbar$-functions.
\endproclaim

In the presence of $\hbar$, we define the tau function
$\tautoda = \tau(\hbar,t,\tbar,s)$ of the Toda lattice hierarchy
as a function that reproduces the Baker-Akhiezer functions as
$$
\aligned
    \Psi(\hbar,t,\tbar,\lambda) &=
    \frac{ \tau(\hbar,t-\hbar [\lambda^{-1}] ,\tbar,s) }
         { \tau(\hbar,t,\tbar,s) }
    \exp \hbar^{-1} ( t(\lambda) + s\log\lambda ),
\\
    \Psibar(\hbar,t,\tbar,\lambdabar) &=
    \frac{ \tau(\hbar,t,\tbar-\hbar [\lambdabar] ,s+\hbar) }
         { \tau(\hbar,t,\tbar,s) }
    \exp \hbar^{-1} ( \tbar(\lambdabar^{-1}) + s\log\lambdabar ),
\endaligned
\tag 2.7.23
$$
where $[\lambda^{-1}]$ is defined in (1.7.17).

In the case of $\hbar = 1$, this reduces to the ordinary definition.
Taking the logarithm of (2.7.23) and comparing them with
the WKB asymptotic form (2.7.18) of $\Psi$ and $\Psibar$,
one can easily find that $\log\tau(\hbar,t,\tbar,s)$
should behave as
$$
  \log\tau(\hbar,t,\tbar,s) = \hbar^{-2} F(t,\tbar,s) + O(\hbar^{-1})
  \quad (\hbar \to 0)
\tag 2.7.24
$$
with an appropriate scaling function $F(t,\tbar,s)$.
The leading terms $v^0_n$ and $\vbar^0_n$
of coefficients $v_n$ and $\vbar_n$ of
$M(t,\tbar,s,\exp(\hbar\der/\der s))$ and
$\Mbar(t,\tbar,s,\exp(\hbar\der/\der s))$
(see (2.7.13) and Proposition 2.7.7 (1)) can be written as
$$
    v^0_n(t,\tbar,s)     = \frac{\der F}{\der t_n}    , \quad
    \vbar^0_n(t,\tbar,s) = \frac{\der F}{\der \tbar_n},
$$
and $\Sbar_0 = \phi = \der F/ \der s$ by virtue of (2.7.16--19).
Therefore we can identify $F$ with $\log\taudtoda$ defined
in Proposition 2.4.3.

%
%
%
%

\subsubhead 2.7.5 Riemann-Hilbert problem (twistor construction)
\endsubsubhead
The twistor construction of \S2.5 can be extended to the construction
of solutions to the ordinary Toda lattice hierarchy
in the presence of the parameter $\hbar$.

\proclaim{Proposition 2.7.11}
Suppose that
$$
\alignat2
    f(\hbar, s, e^{\hbar\der/\der s}) &=
    \sum_{n\in\Integer} f_n(\hbar, x) e^{n\hbar\der/\der s},& \quad
    g(\hbar, s, e^{\hbar\der/\der s}) &=
    \sum_{n\in\Integer} g_n(\hbar, x) e^{n\hbar\der/\der s}
\\
    \fbar(\hbar, s, e^{\hbar\der/\der s}) &=
    \sum_{n\in\Integer} \fbar_n(\hbar, x) e^{n\hbar\der/\der s},& \quad
    \gbar(\hbar, s, e^{\hbar\der/\der s}) &=
    \sum_{n\in\Integer} \gbar_n(\hbar, x) e^{n\hbar\der/\der s}
\endalignat
$$
are difference operators of 0-th order (in the sense of (2.7.5)),
$\ord f = \ord g = \ord \fbar = \ord \gbar = 0$,
and that they satisfy the canonical commutation relations
$[f, g] = \hbar f$, $[\fbar, \gbar] = \hbar \fbar$.
Assume further that difference operators $L$, $\Lbar$ of the form (2.7.2)
and $M$, $\Mbar$ of the form (2.7.13) are given and that
$\ord L = \ord M = \ord \Lbar = \ord \Mbar = 0$,
$[L,M] = \hbar L$, $[\Lbar, \Mbar] = \hbar \Lbar$.

If
$$
    f(\hbar, M, L) = \fbar(\hbar, \Mbar, \Lbar), \quad
    g(\hbar, M, L) = \gbar(\hbar, \Mbar, \Lbar),
\tag 2.7.25
$$
then
$(L, \Lbar)$ gives a solution of the Toda lattice hierarchy, and
$(M, \Mbar)$ are the corresponding Orlov operators.
\endproclaim

\demo{Proof}
The assumption on the form of $L$, $\Lbar$, $M$, $\Mbar$ implies
that there exist difference operators $X$, $\Xbar$ and
a function $\phi(\hbar,t,\tbar,s)$ which are connected with $L$, $\Lbar$,
$M$, $\Mbar$ through (2.7.7), (2.7.12) and (2.7.10).
Following (2.7.10), we denote $W = \exp(\hbar^{-1} X)$,
$\Wbar = \exp(\phi) \exp(\hbar^{-1} \Xbar)$.

Put $P = f(\hbar, M, L) = \fbar(\hbar, \Mbar, \Lbar)$,
    $Q = g(\hbar, M, L) = \gbar(\hbar, \Mbar, \Lbar)$ according to (2.7.25).
Differentiating the equation
$$
\split
    P &= \Ad ( W
             \exp ( \hbar^{-1} t(e^{\hbar\der/\der s}) )
             )
         f(\hbar, s, e^{\hbar\der/\der s})
\\
      &= \Ad ( \Wbar
             \exp ( \hbar^{-1} \tbar(e^{-\hbar\der/\der s}) )
             )
         \fbar(\hbar, s, e^{\hbar\der/\der s}),
\endsplit
$$
we get
$$
\split
    \hbar \frac{\der P}{\der t_n}
    &=\left[
    \hbar \frac{\der W}{\der t_n} W^{-1}
    + W e^{n\hbar\der/\der s} W^{-1},
    P
    \right]
\\
    &=\left[
    \hbar \frac{\der \Wbar}{\der t_n} \Wbar^{-1}, P
    \right],
\endsplit
\tag 2.7.26
$$
and
$$
\split
    \hbar \frac{\der P}{\der \tbar_n}
    &=\left[
    \hbar \frac{\der W}{\der \tbar_n} W^{-1}, P
    \right]
\\
    &=\left[
    \hbar \frac{\der \Wbar}{\der \tbar_n} \Wbar^{-1}
    + \Wbar e^{-n\hbar\der/\der s} \Wbar^{-1}, P
    \right].
\endsplit
\tag 2.7.27
$$
Replacing $P$ with $Q$, we obtain the same kind of equations for $Q$.
Putting
$A = \hbar \frac{\der W}{\der t_n} W^{-1} + W \exp(n\hbar\der/\der s) W^{-1}
   - \hbar \frac{\der\Wbar}{\der t_n} \Wbar^{-1}$, we can write
the equation (2.7.26) and the corresponding equation for $Q$ as
$$
    [A, P] = [A, Q] = 0.
\tag 2.7.28
$$
The principal symbol of this equation gives
$$
    \{\calA, \calP\} = \{\calA, \calQ\} = 0,
\tag 2.7.29
$$
where $\calA = \symbolh(A)$,
$\calP = \symbolh(P)$, $\calQ = \symbolh(Q)$. Note that $\ord(A) \leqq 0$,
hence the principal symbol of $A$ makes sense. Because of the canonical
commutation relation $[f, g] = \hbar f$, we have $[P, Q] = \hbar P$,
the principal symbol of which is $\{\calP, \calQ\} = \calP$.
Therefore we have $\der \calA/\der p = \der \calA/\der s = 0$ by (2.7.29)
and Lemma 2.6.2.
Put $A_1 \=def A - \calA$ ($\ord(A_1) \leqq -1$).
Since $\calA$ does not depend on $(p,s)$,
it commutes with any difference operator with respect to $s$.
Hence the equation (2.7.28) holds if we replace $A$ with $A_1$, which
again implies $\der \symbolh(A_1)/\der p = \der \symbolh(A_1)/\der s = 0$
by the same argument.
Thus we can prove by induction that $A$ does not depend on $s$
and does not contain $\exp(n\hbar\der/\der s)$ for any $n$ ($n\neq 0$).

Again using the form (2.7.10) of $W$, $\Wbar$, we have
$$
\align
    \hbar \frac{\der W}{\der t_n} W^{-1}
    &= - (W e^{n\hbar\der/\der s} W^{-1})_{\leq -1},
\tag 2.7.30
\\
    \hbar \frac{\der \Wbar}{\der t_n} \Wbar^{-1}
    &= (W e^{n\hbar\der/\der s} W^{-1})_{\geq 0} + A.
\tag 2.7.31
\endalign
$$
Equation (2.7.30) leads to the Lax equations for $L$ and $M$
with respect to $t_n$, the first equations of (2.7.1) and (2.7.14),
directly. Since $A$ commutes with any difference operator, equation (2.7.31)
implies the Lax equations for $\Lbar$ and $\Mbar$ with respect to $t_n$,
the third equations of (2.7.1) and (2.7.14).

The other Lax equations in (2.7.1) and (2.7.14) are proved similarly
by (2.7.27).
{}From the last statement of Proposition 2.7.7, $M$ and $\Mbar$ are Orlov
operators corresponding to $L$ and $\Lbar$.
(Note that $W$ is one of the dressing operators corresponding to $(L, M)$,
but, $\Wbar$ is not a dressing operator in general because of the remainder
$A$ in (2.7.31).)
\qed
\enddemo

Let us call the quadruplet $(f,g,\fbar,\gbar)$ above the {\it twistor data\/}
of the Toda lattice hierarchy.
\proclaim{Proposition 2.7.12}
Any solution of the Toda lattice hierarchy possesses a twistor data,
i.e., if $(L, M, \Lbar, \Mbar)$ is a soluiton of the Toda lattice hierarchy
in the presence of $\hbar$, there exists a quadruplet $(f,g,\fbar,\gbar)$
which satisfies the canonical commutation relation
$[f, g]=\hbar f$, $[\fbar, \gbar] = \hbar \fbar$ and (2.7.25).
\endproclaim

\demo{Proof}
Let $W$ and $\Wbar$ be the dressing operators chosen in Proposition 2.7.5.
We can prove by the same method as Propositio 1.7.8 that twistor data
are given by
$$
\alignat2
    f(\hbar,s,e^{\hbar\der/\der s}) &= W_0^{-1} e^{\hbar\der/\der s} W_0,&
    \quad
    g(\hbar,s,e^{\hbar\der/\der s}) &= W_0^{-1} s W_0,
\\
    \fbar(\hbar,s,e^{\hbar\der/\der s}) &=
    \Wbar_0^{-1} e^{\hbar\der/\der s} \Wbar_0, &
    \quad
    \gbar(\hbar,s,e^{\hbar\der/\der s}) &=
    \Wbar_0^{-1} s \Wbar_0,
\endalignat
$$
where $W_0 = W(t=0, \tbar=0)$, $\Wbar_0 = \Wbar(t=0, \tbar=0)$.
\qed
\enddemo

The principal symbols $\symbolh(f)$, $\symbolh(g)$, $\symbolh(\fbar)$,
$\symbolh(\gbar)$ give the twistor data for the solution
$(\calL = \symbolh(L), \calM = \symbolh(M),
\calLbar = \symbolh(\Lbar), \calMbar = \symbolh(\Mbar))$
of the dispersionless Toda hierarchy.

The solution to (2.7.25) is unique in the following infinitesimal sense.
\proclaim{Proposition 2.7.13}
Suppose $(L(\eps), M(\eps), \Lbar(\eps), \Mbar(\eps))$ is a one-parameter
family of solutions to (2.7.25) that are regular
in a neighborhood of $\eps=0$. Then
$(L(\eps), M(\eps), \Lbar(\eps), \Mbar(\eps))
=(L(0), M(0), \Lbar(0), \Mbar(0))$.
\endproclaim

We omit the proof since it is essentially the same as that of
Proposition 1.7.13.

%
%
%
%

\subsubhead 2.7.6. $W_{1+\infty}$ symmetry \endsubsubhead
Here we analyze the quasi-classical limit of the $W_{1+\infty}$ symmetry
of the Toda lattice hierarchy. As in the case of the KP hierarchy, there
are two ways to introduce this symmetry: to deform the twistor data, and
to use the group action on the parameter space of the solutions.

As in \S2.6, we consider the deformation of the twistor data
$(f,g,\fbar,\gbar)$ of the form
$$
\aligned
    ( f   , g   ) &\mapsto ( f   , g   ) \circ \Ad\exp(-\eps  Y   ),\\
    (\fbar,\gbar) &\mapsto (\fbar,\gbar) \circ \Ad\exp(-\eps \Ybar),
\endaligned
\tag 2.7.32
$$
where $Y$ and $\Ybar$ are difference operators of order not greater than $1$,
$\ord(Y) \leqq 1$, $\ord(\Ybar) \leqq 1$.
\proclaim{Proposition 2.7.14}
i)
The above symmetry of the Lax operators are given explicitly by
$$
\aligned
    \delta_{Y,\Ybar}  L    &=
    [Y(\hbar,M,L)_{\leq -1} - \Ybar(\hbar,\Mbar,\Lbar)_{\leq -1},  L   ],
\\
    \delta_{Y,\Ybar}  M    &=
    [Y(\hbar,M,L)_{\leq -1} - \Ybar(\hbar,\Mbar,\Lbar)_{\leq -1},  M   ],
\\
    \delta_{Y,\Ybar} \Lbar &=
    [\Ybar(\hbar,\Mbar,\Lbar)_{\geq 0} - Y(\hbar,M,L)_{\geq 0}, \Lbar],
\\
    \delta_{Y,\Ybar} \Mbar &=
    [\Ybar(\hbar,\Mbar,\Lbar)_{\geq 0} - Y(\hbar,M,L)_{\geq 0}, \Mbar].
\endaligned
\tag 2.7.33
$$

ii)
The symmetries of the dressing operators are given by
$$
\aligned
    \delta_{Y,\Ybar} W \cdot W^{-1} &=
    Y(\hbar, M, L)_{\leq -1} - \Ybar(\hbar, \Mbar, \Lbar)_{\leq -1},
\\
    \delta_{Y,\Ybar} \Wbar \cdot \Wbar^{-1} &=
    \Ybar(\hbar, \Mbar, \Lbar)_{\geq 0} - Y(\hbar, M, L)_{\geq 0}.
\endaligned
\tag 2.7.34
$$
Symmetries of $X$ and $\Xbar$ (2.7.10) are induced by these formulae.
A symmetry of the potential $\phi$ (2.7.11)
consistent with (2.7.33) is given by
$$
    \delta_{Y,\Ybar} \phi= \Ybar(\hbar, \Mbar, \Lbar)_0 - Y(\hbar, M, L)_0.
\tag 2.7.35
$$
\endproclaim

\demo{Proof}
The proof is the same as that of Proposition 2.6.1 (see also the proof of
Proposition 2.7.11).
For example, (2.7.34) comes from the coefficients of $\eps^1$ of the equation
$$
\multline
    W_\eps e^{ t(\exp(\hbar\der/\der s))} e^{-\eps Y}
           e^{-t(\exp(\hbar\der/\der s))} W^{-1} \\
    = a_\eps
    \Wbar_\eps e^{ \tbar(\exp(-\hbar\der/\der s))} e^{-\eps \Ybar}
               e^{-\tbar(\exp(-\hbar\der/\der s))} \Wbar^{-1},
\endmultline
\tag 2.7.36
$$
where $W_\eps$ and $\Wbar_\eps$ are deformed dressing operators and
$a_\eps$ is a scalar.
The statement i) immediately follows from ii).
\qed
\enddemo

Taking the principal symbol of these expressions, we obtain
the $w_{1+\infty}$ symmetry of the dispersionless Toda hierarchy in
Proposition 2.6.1.

In the earlier paper [4] we studied these symmetries in terms of
vertex operators and free fermions [15]:
$$
\aligned
    Z(\hbar,\lambdatilde,\lambda) =&
    \dfrac{ \exp\left(\hbar^{-1}[t(\lambdatilde)-t(\lambda)]\right)
            (\lambdatilde/\lambda)^{s/\hbar}
            \exp\left(\hbar[-\dertilde_t(\lambdatilde^{-1})
                            +\dertilde_t(\lambda^{-1})]\right)
            - 1}
          {\lambdatilde - \lambda},
\\
    \Zbar(\hbar,\lambdatilde,\lambda) =&
    \dfrac{ \exp\left(\hbar^{-1}[\tbar(\lambdatilde^{-1})
                                 -\tbar(\lambda^{-1})]\right)
            (\lambdatilde/\lambda)^{s/\hbar}
            \exp\left(\hbar[-\dertilde_\tbar(\lambdatilde)
                            +\dertilde_\tbar(\lambda)]\right)
            - 1}
          {\lambdatilde^{-1} - \lambda^{-1}},
\endaligned
\tag 2.7.37
$$
where
$$
    \dertilde_t(\lambda^{-1}) = \sum_{n=1}^\infty
          \frac{\lambda^{-n}}{n} \frac{\der}{\der t_n},  \quad
    \dertilde_\tbar(\lambda) = \sum_{n=1}^\infty
          \frac{\lambda^n}{n} \frac{\der}{\der \tbar_n}.
\tag 2.7.38
$$
The generators of $W_{1+\infty}$ symmetries appear as
the coefficients $W^{(\ell)}_n(\hbar)$ and $\Wbar^{(\ell)}_n(\hbar)$
of the expansion
$$
\gathered
    Z(\hbar,\lambdatilde,\lambda) = \sum_{\ell=1}^\infty
        \frac{(\lambdatilde-\lambda)^{\ell-1} }
              { (\ell-1)! }
        W^{(\ell)}(\hbar,\lambda),
\\
    W^{(\ell)}(\hbar,\lambda) = \sum_{n=-\infty}^\infty
         W^{(\ell)}_n(\hbar) \lambda^{-n-\ell},
\endgathered
\tag 2.7.39
$$
($n \in \Integer$, $\ell \geqq 1$) and the same kind of expansion of $\Zbar$.
These symmetry generators are differential operators
of finite order in $t$ and $\tbar$ respectively, and differ from
the ordinary ($\hbar = 1$) definition [15] by the simple rescaling
$$
\aligned
    t_n \to \hbar^{-1} t_n, \quad &
    \frac{\der}{\der t_n} \to \hbar \frac{\der}{\der t_n},        \\
    \tbar_n \to \hbar^{-1} \tbar_n, \quad &
    \frac{\der}{\der \tbar_n} \to \hbar \frac{\der}{\der \tbar_n}.
\endaligned
\tag 2.7.40
$$
We have thus essentially the same $W_{1+\infty}$ symmetries as
the KP hierarchy, but now in duplicate.

As is proved in [4], the $w_{1+\infty}$ symmetries in \S2.6 are reproduced
as quasi-classical limit of the above $W_{1+\infty}$ symmetries by
$$
    w_n^{(\ell)} \log\taudtoda =
    \lim_{\hbar\to 0}
    \tautoda(\hbar,t,\tbar,s)^{-1}
    \hbar^{\ell} W_n^{(\ell)}(\hbar) \tautoda(\hbar,t,\tbar,s),
\tag 2.7.41
$$
where
$$
    w_n^{(\ell)} \log\taudtoda =
    \Res \frac{1}{\ell} \calM^\ell \calL^n d_p\log\calL
\tag 2.7.42
$$
which corresponds to (2.6.17) with $F = - s^{\ell-1} p^n$, $\Fbar = 0$.
The other half $\wbar_n^{(\ell)}$ ($F=0$, $\Fbar = - s^{\ell-1} p^n$)
comes from $\Wbar_n^{(\ell)}$ in the same way.

In the ferminonic language, we can unify these two expressions of symmetries,
Proposition 2.7.14 and (2.7.42).
(We use the same notations as in \S1.7.6.)
As is well known ([40], [41]), when $\hbar = 1$,
the generic tau function is given by the expression
$$
    \tau(t,\tbar,s) =
    \left< s \left| e^{H(t)} g_- g_0 g_+ e^{-\Hbar(\tbar)} \right| s \right>,
\tag 2.7.43
$$
where $H(z)$ and $\Hbar(\tbar)$ are generators of time evolutions,
$$
\aligned
    & H(t) = \sum_{n=1}^\infty t_n H_n, \quad
      \Hbar(\tbar) = \sum_{n=1}^\infty \tbar_n H_{-n},       \\
    & H_n = \sum_{i=-\infty}^\infty : \psi_i \psi^*_{i+n}:
\endaligned
\tag 2.7.44
$$
(cf\. (1.7.42)), and $g_+$, $g_-$ and $g_0$ are Clifford operators belonging
to the upper triangular, the lower triangular and the diagonal part of
$\gl(\infty)$ respectively.

In our context, these Clifford operators have the following form.

\proclaim{Proposition 2.7.15}
The tau function with well-defined quasi-classical limit in our sense is
expressed as
$$
    \tau(\hbar,t,\tbar,s)
    = \left< \hbar^{-1}s  \left|
    e^{H(t)/\hbar} g_-(\hbar) g_0(\hbar) g_+(\hbar) e^{-\Hbar(\tbar)/\hbar}
      \right| \hbar^{-1}s \right>,
\tag 2.7.45
$$
where $g_{\pm,0}(\hbar)$ are Clifford operators of the form
$$
\aligned
    g_-(\hbar) &= \exp(-\hbar^{-1} \calO_X    (\hbar)),\\
    g_+(\hbar) &= \exp( \hbar^{-1} \calO_\Xbar(\hbar)),\\
    g_0(\hbar) &= \exp( \hbar^{-1} \calO_\phi (\hbar)).
\endaligned
\tag 2.7.46
$$
Here $\calO_X$, $\calO_\Xbar$, $\calO_\phi$ are
determined from the dressing operators $X$, $\Xbar$ and $\phi$ (2.7.10)
as follows (see [41]):
Suppose $X$ is expanded as
$$
    X(\hbar,t=0,\tbar=0,s,e^{\hbar\der/\der s})
    =
    \sum_{n=1}^{\infty} x_{n,\ell}(\hbar)
    s^{\ell - 1} e^{(n+\ell-1) \hbar\der/\der s}.
\tag 2.7.47
$$
Then $\calO_X$ is given by
$$
\gathered
    \calO_X =\oint
    :A\left(\lambda,\hbar\frac{\der}{\der \lambda}\right) \psi(\lambda)
    \cdot \psi^*(\lambda): \frac{d\lambda}{2\pi i},
\\
    A\left(\lambda,\hbar\frac{\der}{\der \lambda}\right)
    =
    \sum_{n=1}^{\infty} x_{n,\ell}(\hbar)
    \lambda^{n+\ell-1} \left(\hbar\frac{\der}{\der \lambda}\right)^{\ell-1},
\endgathered
\tag 2.7.48
$$
where $A\psi(\lambda) \cdot \psi^\ast(\lambda)$ is understood to be
``$A\psi(\lambda)$ times $\psi^\ast(\lambda)$''.
Other Clifford operators $\calO_\Xbar$ and $\calO_\phi$ are determined
in the same manner.
\endproclaim

\demo{Proof}
The algebra generated by the multiplication operator $s$ and the difference
operator $\exp(\hbar\der/\der s)$ has a representation
in the algebra of $\Integer\times\Integer$-matrices, $\gl(\infty)$, as
$$
\aligned
    s &\leftrightarrow
    ( (\nu+1) \hbar \delta_{\mu-1, \nu} )_{\mu,\nu \in \Integer}, \\
    e^{\hbar\der/\der s} &\leftrightarrow
    ( \delta_{\mu+1, \nu} )_{\mu,\nu \in \Integer}, \\
    s^{\ell-1} e^{(n+\ell-1)\hbar\der/\der s} &\leftrightarrow
    ((\nu-n)(\nu-n-1)\cdots(\nu-n-\ell+2) \hbar^{\ell-1}\delta_{\mu+n, \nu}
    )_{\mu,\nu \in \Integer}.
\endaligned
\tag 2.7.49
$$
(This gives a representation of the algebra of difference operators
on the space of functions on the lattice $\{s = \hbar n\}_{n\in\Integer}$.)
Therefore, through the identification of $:\psi_m \psi^{\ast}_n:$
with the matrix unit $E_{m,n}$, the difference operator
$X(\hbar,t=0,\tbar=0,s,\exp(\hbar\der/\der s))$ can be identified with
$\calO_X(\hbar)$ in (2.7.48).
Similarly we obtain the Clifford operator expressions
$\calO_\Xbar$ and $\calO_\phi$ in (2.7.46) for $\Xbar$ and $\phi$.
According to the general theory on the relation
of the initial value of the dressing operators with the tau function
(e.g., [42], [41]), the tau function has the form (2.7.45).
\qed
\enddemo

The action of $Z(\hbar,\lambdatilde,\lambda)$ and
$\Zbar(\hbar,\lambdatilde,\lambda)$ is realized by
insertion of a fermion bilinear form:
$$
\aligned
    & Z(\hbar,\lambdatilde,\lambda) \tau(\hbar,t,\tbar,s)
    = \left<\hbar^{-1}s \left|  e^{H(t)/\hbar}
           :\psi(\lambdatilde)\psi^*(\lambda): \; g(\hbar)
           e^{-\Hbar(\tbar)/\hbar} \right| \hbar^{-1}s \right>,
\\
    & \Zbar(\hbar,\lambdatilde,\lambda) \tau(\hbar,t,\tbar,s)
    = \left<\hbar^{-1}s \left|  e^{H(t)/\hbar}
           g(\hbar) \; :\psi(\lambdatilde)\psi^*(\lambda):
           e^{-\Hbar(\tbar)/\hbar} \right| \hbar^{-1}s \right>,
\endaligned
\tag 2.7.50
$$
where $g(\hbar) = g_-(\hbar) g_0(\hbar) g_+(\hbar)$.
In particular,
$$
\aligned
    & W_n^{(\ell)}(\hbar) \tau(\hbar,t,\tbar,s)
    = \left<\hbar^{-1}s \left|  e^{H(t)/\hbar}
           \calO_k^{(\ell)}(\hbar) \; g(\hbar)
           e^{-\Hbar(\tbar)/\hbar} \right| \hbar^{-1}s \right>,
\\
    & \Wbar_n^{(\ell)}(\hbar) \tau(\hbar,t,\tbar,s)
    = \left<\hbar^{-1}s \left|  e^{H(t)/\hbar}
           g(\hbar)  \; \calO_k^{(\ell)}(\hbar)
           e^{-\Hbar(\tbar)/\hbar} \right| \hbar^{-1}s \right>,
\endaligned
\tag 2.7.51
$$
where $\calO_k^{(\ell)}(\hbar)$ is defined by (1.7.51).

\proclaim{Proposition 2.7.16}
The symmetry of the tau function by $W_n^{(\ell)}$ and $\Wbar_n^{(\ell)}$
induces the symmetries of $L$, $M$, $\Lbar$, $\Mbar$, $W$, $\Wbar$
of the form (2.7.33), (2.7.34) for
$(Y, \Ybar) = (- s^{\ell-1} e^{(n+\ell-1)\hbar\der/\der s}, 0)$ and
$(Y, \Ybar) = (0, - s^{\ell-1} e^{(n+\ell-1)\hbar\der/\der s})$, respectively.
\endproclaim

\demo{Proof}
We prove the proposition only for $W_n^{(\ell)}$.
The equation (2.7.36) for
$(Y, \Ybar) = (- s^{\ell-1} e^{(n+\ell-1)\hbar\der/\der s}, 0)$
at $t= \tbar = 0$ becomes
$$
    W_{\eps,0} e^{-\eps Y(\hbar,s,\exp(\hbar\der/\der s))} W_0^{-1}
    = a_\eps \Wbar_{\eps,0} \Wbar_0^{-1}.
\tag 2.7.52
$$
Here the index $0$ of $W_0$ etc\. means the value at $t=\tbar=0$.
Since a central factor changes the tau function only up to a scalar factor,
it does not change $L$, $M$ etc. Hence we neglect this central correction
$a_\eps$ below.
The fermion bilinear form $\calO_k^{(\ell)}$ corresponds to
$ - Y = s^{\ell-1} e^{(n+\ell-1)\hbar\der/\der s}$ through (2.7.49)
(up to a central factor). Hence (2.7.52) corresponds to the expression
$$
    g_{-, \eps}(\hbar)^{-1} e^{\eps \calO_k^{(\ell)}} g_-(\eps)
    = g_{0,\eps}(\hbar) g_{+,\eps}(\hbar) g_+(\hbar)^{-1} g_0(\hbar)^{-1},
\tag 2.7.53
$$
where $g_{\pm,\eps}$, $g_{0,\eps}$ denote the deformed Clifford operators.
Therefore, taking this equation into account, we obtain from (2.7.45)
$$
\split
    \tau_{\eps} (\hbar,t,\tbar,s)
    &= \left< \hbar^{-1}s  \left| e^{H(t)/\hbar}
    g_{-,\eps}(\hbar) g_{0,\eps}(\hbar) g_{+,\eps}(\hbar)
    e^{-\Hbar(\tbar)/\hbar} \right| \hbar^{-1}s \right>
    \\
    &= \left< \hbar^{-1}s  \left| e^{H(t)/\hbar}
    e^{\eps \calO_k^{(\ell)}} g_-(\hbar) g_0(\hbar) g_+(\hbar)
    e^{-\Hbar(\tbar)/\hbar} \right| \hbar^{-1}s \right>.
\endsplit
\tag 2.7.54
$$
This proves the assertion of the proposition.
\qed
\enddemo

The bosonic and fermionic representations are thus connected,
and the intrinsic reason of the coincidence of quasi-classical limit of
the symmetries (2.7.41) with the symmetries defined by the deformation of
the twistor data in \S2.6 becomes clear.

For the moment, the bosonic language looks more preferable,
because the fermionic representation is valid only for
discrete values $ s \in \hbar^{-1}\Integer$.

%
%
%
%

\subhead 2.8. Relation with the dispersionless KP hierarchy \endsubhead

We can lift up the solution of the dispersionless Toda hierarchy to
the ordinary Toda lattice hierarchy (Corollary 2.7.6).
A solution of the Toda lattice hierarchy gives a solution of the KP hierarchy
if the variable $s$ and $\tbar$ are considered to be parameters ([21]).
A solution of the dispersionless Toda hierarchy should therefore give
a solution of the dispersionless KP hierarchy when $s$ and $\tbar$ are fixed.

This can be verified directly as follows.

\proclaim{Proposition 2.8.1}
Let $(\calL, \calLbar)$ be a solution of the dispersionless Toda hierarchy
(2.1.1). Then $\calL(t,\tbar,s)$ is a solution of the dispersionless KP
hierarchy (1.1.1) when we substitute $t_1 + x$ to $t_1$ and identify
$\calB_1$ with $k$:
$$
    \left.\frac{\der \calL}{\der t_n}\right|_{k \fixed} =
    \kpbra \calB_n, \calL \kpket.
\tag 2.8.1
$$
Here
$$
    \kpbra f, g \kpket = \frac{\der f}{\der k} \frac{\der g}{\der x}
                       - \frac{\der f}{\der x} \frac{\der g}{\der k},
$$
as in \S1.1 (1.1.2). Note that the projection onto a polynomial in $p$
is equal to the projection onto a polynomial in $k$,
since $k = \calB_1 = p + u_1(t,\tbar,s)$.
Thus $\calB_n = (\calL^n)_{\geq 0}$ in the sense of (1.1.1) as well as
in the sense of (2.1.1).
\endproclaim

\demo{Proof}
Because of the Lax equations (2.1.1) we have
$$
\split
    \left.\frac{\der \calL}{\der t_n}\right|_{k \fixed}
    &= \frac{\der \calL}{\der t_n} -
       \frac{\der \calL}{\der k} \frac{\der \calB_1}{\der t_n} \\
    &= p \frac{\der \calB_n}{\der p} \frac{\der \calL}{\der s}
     - p \frac{\der \calB_n}{\der s} \frac{\der \calL}{\der p}
     -   \frac{\der \calL}{\der k} \frac{\der \calB_1}{\der t_n}.
\endsplit
\tag 2.8.1
$$
In particular, (2.8.1) $n=1$ implies
$$
    p \frac{\der \calL}{\der s} =
    \left.\frac{\der \calL}{\der t_1}\right|_{k \fixed} +
    \frac{\der \calL}{\der k} \frac{\der \calB_1}{\der t_1} +
    p \frac{\der \calB_1}{\der s} \frac{\der \calL}{\der p}.
\tag 2.8.2
$$
Substituting this into (2.8.1) and using $\der/\der p = \der/\der k$,
we obtain
$$
    \left.\frac{\der \calL}{\der t_n}\right|_{k \fixed}
    =
    \frac{\der \calB_n}{\der k}
    \left.\frac{\der \calL}{\der t_1}\right|_{k \fixed}
    -
    \left(
    \frac{\der \calB_1}{\der t_n} + \{ \calB_1, \calB_n \} -
    \frac{\der \calB_n}{\der k} \frac{\der \calB_1}{\der t_1}
    \right)
    \frac{\der \calL}{\der k}.
\tag 2.8.3
$$
By the zero-curvature equation (2.1.5), the right hand side of (2.8.3) is
$$
    \frac{\der \calB_n}{\der k}
    \left.\frac{\der \calL}{\der t_1}\right|_{k \fixed}
    -
    \left.\frac{\der \calB_n}{\der t_1}\right|_{k \fixed}
    \frac{\der \calL}{\der k}
    = \kpbra \calB_n, \calL \kpket,
\tag 2.8.4
$$
which proves the proposition.
\qed
\enddemo

In the gauge chosen in Lemma 2.1.3, another solution of the dispersionless
KP hierarchy arises from $\calL'$ ($t$, $s$ fixed), when we identify
$k = \calB_1'$.

%
%
%
%

\head Appendix A. Several facts on differential Lie algebras \endhead

Here we collect several facts on formal properties of
differential Lie algebras. Proofs are omitted.

Let $\gee$ be a Lie algebra with a derivation $\der : \gee \to \gee$.
Actually, we have in mind such examples as the Lie algebra of
Laurent series (Lie bracket $=$ Poisson bracket, derivation $=$
differentiation with respect to parameters $t_n$; \S1.2, \S2.2) or
the Lie algebra of micro-differential operators or difference operators
(Lie bracket $=$ commutator, derivation $=$ differentiation with respect to
parameters $t_n$; \S1.7, \S2.7).

In this section we always assume existence of suitable topology of $\gee$
in which each series under consideration converges.

\proclaim{Lemma A.1}
For all $X, Y \in \gee$,
$$
    \der ( e^{\ad Y} X) = e^{\ad Y} (\der X) + [\nabla_Y Y, e^{\ad Y} X],
$$
where
$$
    \nabla_Z W = \sum_{n=0}^\infty \frac1{(n+1)!} (\ad Z)^n (\der W)
               = \frac{e^{\ad Z} - 1}{ \ad Z } (\der W).
\tag A.1
$$
\endproclaim

By the Hausdorff-Campbell formula, there exists a unique series
(the {\it Hausdorff series}) $H(X,Y)\in\gee$ of $X,Y\in\gee$ such that
$$
    \exp(H(X,Y)) = \exp(X) \exp(Y).
$$

\proclaim{Lemma A.2}
For all $X, Y\in\gee$,
$$
    \nabla_{H(X,Y)} H(X,Y) = \nabla_X X + e^{\ad X} \nabla_Y Y.
$$
\endproclaim

To consider zero-curvature representation of integrable systems,
we also need the following Lemma for a differential Lie algebra $\gee$
with two derivations $\der_1$, $\der_2$.

\proclaim{Lemma A.3}
For $X, Y, Z \in \gee$ define
$$
    \tilde X = e^{\ad Z} X + \nabla_{1, Z} Z, \quad
    \tilde Y = e^{\ad Z} Y + \nabla_{2, Z} Z,
$$
where $\nabla_{1,Z}$ and $\nabla_{2,Z}$ are defined by (A.1)
for $\der = \der_1$ and $\der = \der_2$.
Then the following two statements are equivalent.
\roster

\item[1] $[\der_1 - X, \der_2 - Y] = 0$, i.e.,
           $\der_2 X - \der_1 Y + [X, Y] = 0$.

\item[2] $[\der_1 - \tilde X, \der_2 - \tilde Y] = 0$, i.e.,
           $\der_2 \tilde X - \der_1 \tilde Y + [\tilde X, \tilde Y] = 0$.
\endroster
\endproclaim

%
%
%
%

\head Appendix B. Quasi-classical limit of Hirota equations
of KP hierarchy \endhead

Here we show that some of the Hirota equations have quasi-classical limit,
and propose a conjecture that these equations characterize the tau function
(free energy) of the dispersionless KP hierarchy.

It is well-known that the tau function of the KP hierarchy
is characterized by the {\it Hirota equations},
a generating function of which is
$$
    \sum_{j=0}^\infty p_j(-2\hbar^{-1} y) p_{j+1}(\hbar\tilde D_y)
    \exp \left(\sum_{l=1}^\infty y_l D_{t_l} \right)
    \tau(t) \cdot \tau(t)
    = 0.
\tag B.1
$$
Here $y=(y_1, y_2, \ldots)$ is a series of parameters,
$p_j(y)$ are Schur polynomials defined by
$$
    \exp\left( \sum_{l=1}^\infty y_l \lambda^l \right)
    =
    \sum_{j=0}^\infty p_j(y) \lambda^j,
$$
$D$ is the Hirota operator defined by
$D_x f(x)\cdot g(x) = \der_y f(x+y)\, g(x-y)|_{y=0}$ and
$\dertilde = (\der/1, \der^2/2, \der^3/3, \ldots)$.
(We have introduced the Planck constant $\hbar$ as in \S1.7.)
Expanded into monomials of $y$, (B.1) gives the Pl\"ucker relations
for the Pl\"ucker embedding (1.7.35) when $\hbar = 1$.

These equations, however, do not possess a good quasi-classical limit
in general.
In fact, if we insert the prescribed asymptotic behaviour of the tau function,
$\log \tau (\hbar,t) = \hbar^{-2} F(t) + O(\hbar^{-1})$ ($\hbar\to 0$),
the $\hbar$-expansion of (B.1) becomes extremely complicated and
gives rise to no useful information.

Nevertheless, there are a series of Hirota equations which do have
quasi-classical limit: the {\it differential Fay identity}.
\proclaim{Proposition B.1}
(Differential Fay identity)
$$
\multline
    \hbar\der_{t_1} \tau(t-\hbar[\mu^{-1}]) \; \tau(t-\hbar[\lambda^{-1}])
    -
    \tau(t-\hbar[\mu^{-1}]) \; \hbar\der_{t_1} \tau(t-\hbar[\lambda^{-1}]) \\
    -
    (\lambda - \mu) \tau(t-\hbar[\mu^{-1}]) \; \tau(t-\hbar[\lambda^{-1}])
    +
    (\lambda - \mu) \tau(t) \; \tau(t-\hbar[\mu^{-1}]-\hbar[\lambda^{-1}])
    = 0.
\endmultline
\tag B.2
$$
Here $\mu$ and $\lambda$ are free parameters and
$[\lambda^{-1}] = (\lambda^{-1}, \lambda^{-2}/2, \lambda^{-3}/3, \ldots)$.
\endproclaim
See [37] for the proof.
The coefficient of $\mu^{-m}\lambda^{-n}$ gives the bilinear equation
$$
\multline
    \chi_{0,m+1}(-\hbar\dertilde)\tau(t) \;
    \chi_{1,n+1}(-\hbar\dertilde)\tau(t)
    -
    \chi_{1,m+1}(-\hbar\dertilde)\tau(t) \;
    \chi_{0,n+1}(-\hbar\dertilde)\tau(t) \\
    +
    \chi_{n+1,m+1}(-\hbar\dertilde)\tau(t) \;
    \chi_{0,1}(-\hbar\dertilde)\tau(t)
    =0,
\endmultline
\tag B.3
$$
where $\chi_{k,l}$ is the Schur function corresponding to the partition
$(k,l-1)$. Thus (B.2) is a part of the Pl\"ucker relations. From
this equation we can derive a set of equations
for the free energy $\log\taudkp$ of the dispersionless KP hierarchy.

\proclaim{Proposition B.2}
The tau function (or free energy) of the dispersionless KP hierarchy
satisfies the equation
$$
    \sum_{m,n=1}^\infty \mu^{-m} \lambda^{-n}
    \frac1{mn} \frac{\der^2}{\der t_m \der t_n} \log\taudkp
    =
    \log \left( 1 +
    \sum_{n=1}^\infty \frac{\mu^{-n} - \lambda^{-n}}{\mu - \lambda}
    \frac1n \frac{\der^2}{\der t_1 \der t_n} \log\taudkp
    \right),
\tag B.4
$$
where $\mu$ and $\lambda$ are free parameters.
\endproclaim

\demo{Proof}
Let $\tau = \taukp$ be a tau function of a solution $L$ of the KP hierarchy
the quasi-classical limit of which gives the solution $\calL$ of the
dispersionless KP hierarchy corresponding to $\taudkp$ (see Corollary 1.7.6).
This means that $\log\tau = \hbar^{-2} \log\taudkp + O(\hbar^{-1})$.

The differential Fay identity can be written as
$$
\multline
    \frac{\tau(t-\hbar[\mu^{-1}]-\hbar[\lambda^{-1}]) \; \tau(t)}
         {\tau(t-\hbar[\mu^{-1}]) \; \tau(t-\hbar[\lambda^{-1}])} \\
    = 1 +
    \frac{1}{\mu-\lambda} \left(
    \frac{\der}{\der t_1} \log\tau(t-\hbar[\mu^{-1}]) -
    \frac{\der}{\der t_1} \log\tau(t-\hbar[\lambda^{-1}]) \right).
\endmultline
\tag B.5
$$
The logarithm of the left hand side gives
$$
    \sum_{m,n=1}^\infty \mu^{-m} \lambda^{-n}
    p_n(-\hbar\dertilde) p_m(-\hbar\dertilde) \log\tau,
\tag B.6
$$
and the logarithm of the right hand side of (B.5), similarly,
$$
    \log \left( 1 +
    \sum_{n=1}^\infty \frac{\mu^{-n} - \lambda^{-n}}{\mu - \lambda}
    p_n(-\hbar\dertilde)\, \hbar \frac{\der}{\der t_1} \log\tau
    \right).
\tag B.7
$$
The leading term of $\hbar$-expansion of (B.6) and (B.7)
(the coefficients of $\hbar^0$)
are equal to the left and right hand side of (B.4).
\qed
\enddemo

As we shall see below, the differential Fay identity completely
characterizes the tau function of the KP hierarchy.
Since $\log\taudkp$ satisfies the leading term of
the differential Fay identity (B.4),
we can expect that (B.4) characterizes the tau function of the dispersionless
KP hierarchy.

Now we show that the differential Fay identity is equivalent to the whole
system of the KP hierarchy. For the notational symplicity, we set $\hbar = 1$
below.
\proclaim{Proposition B.3}
Let $L$ be a micro-differential operator of the form (1.7.2) and
$B_n = (L^n)_{\geq 0}$. Define functions $v_n(t)$ and polynomials
$J_{n,m}[v]$ of $v_n$ by
$$
\gather
    \der = L + \sum_{m=1}^\infty v_{m+1}(t) L^{-m},
\tag B.8
\\
    J_{n,m}[v]
    =
    v_{n+m} +
    \frac12 \sum\Sb i,i',j,j' \geq 1 \\ i+i' = m \\ j+j' = n \endSb
    v_{i+j} v_{i'+j'}
    +
    \frac13
    \sum\Sb i,i',i'',j,j',j'' \geq 1 \\ i+i'+i'' = m \\ j+j'+j'' = n \endSb
    v_{i+j} v_{i'+j'} v_{i''+j''}
    +\cdots.
\tag B.9
\endgather
$$
Note that $v_n$ are differential polynomials of $u_m$ and vice versa.

i) ([13])
The Lax equations
$$
    \frac{\der L}{\der t_n} = [B_n, L], \quad n=1,2,\ldots
\tag B.10
$$
and a system of differential equations of $v_n$
$$
    p_n(-\dertilde) v_{m+1} + \frac{\der}{\der t_1} J_{n,m}[v] = 0,
    \quad n,m = 1,2,\ldots
\tag B.11
$$
are equivalent.

ii)
If $\tau(t)$ satisfies the differential Fay identity (B.2) and
$v_n(t)$ is defined by
$$
    v_n(t) = \frac{\der}{\der t_1} ( p_{n-1}(-\dertilde) \log\tau),
    n=2,3,\ldots,
\tag B.12
$$
then $v_n$ satisfy (B.11). Therefore $\tau$ is a tau function for $L$
defined from $v_n$ by (B.8).

iii) ([13])
Under the condition (B.10) or (B.11), if a function $\Psi(t;\lambda)$
of the form (1.7.11) ($\hbar =1$) satisfies
$$
\gathered
    L \Psi = \lambda \Psi, \\
    \frac{\der \Psi}{\der t_n} = B_n \Psi,
    \quad n=1,2,\ldots,
\endgathered
\tag B.13
$$
then $\Psi$ satisfies
$$
\gathered
    \frac{\der}{\der t_1} \Psi =
    \left(\lambda + \sum_{n=1}^\infty v_{n+1}(t) \lambda^{-n} \right) \Psi,
    \\
    p_n(-\dertilde) \Psi = v_n \Psi,
    \quad n=2,3,\ldots.
\endgathered
\tag B.14
$$
Conversely, if $\Psi$ of the form (1.7.11) satisfies (B.14), then
it also satisfies (B.13).
\endproclaim

\demo{Proof}
Though i) and iii) are stated in [13], there is no proof published.
The following is our proof simplified by T.~Shiota's comment.

First we prove ii).
By definition (B.12), it is easy to see that
$$
    v(t;\lambda) \=def
    \sum_{n=1}^\infty v_{n+1}(t) \lambda^{-n}
    = \frac{\der}{\der t_1} (\log\tau(t-[\lambda^{-1}]) - \log\tau(t)).
\tag B.15
$$
The polynomial $J_{n,m}$ is the coefficient of $\lambda^{-n}\mu^{-m}$
in a generating function
$$
    J(\lambda,\mu)
    = \log \left( 1 + \frac{v(t;\lambda) - v(t;\mu)}{\lambda-\mu} \right).
\tag B.16
$$
Thus, multiplying $\lambda^{-n} \mu^{-m}$ to the left hand side of (B.11)
and summing up, we obtain
$$
    v(t-[\lambda^{-1}];\mu) - v(t;\mu) - \frac{\der}{\der t_1} J(\lambda,\mu),
$$
which is equal to
$$
\multline
    \frac{\der}{\der t_1} \log
    \frac{\tau(t-[\lambda^{-1}]-[\mu^{-1}]}{\tau(t-[\lambda^{-1}])}
    -
    \frac{\der}{\der t_1} \log
    \frac{\tau(t-[\mu^{-1}])}{\tau(t)} \\
    -
    \frac{\der}{\der t_1} \log
    \left( 1 + \frac 1{\lambda-\mu}
    \left(
    \frac{\der_{t_1}\tau(t-[\lambda^{-1}])}{\tau(t-[\lambda^{-1}])} -
    \frac{\der_{t_1}\tau(t-[\mu^{-1}])    }{\tau(t-[\mu^{-1}])}
    \right)
    \right).
\endmultline
$$
This is identical to $0$ due to (B.2). Equation (B.11) is proved.

Next, suppose that (B.11) holds. Note that this system of equations is
uniquely solvable if the initial values $v_n(t)|_{t_k = 0, k\neq 1}$
are given. In fact, (B.11) can be rewritten into the form
$$
    -\frac1n \frac{\der}{\der t_n} v_{m+1}(t)
    =
    (\text{polynomial of }\der_{t_1}, \ldots, \der_{t_{n-1}}) v_{m+1}(t)
    - \der_{t_1}(J_{n,m}[v]),
$$
which uniquely determines $\{v_m(t)|_{t_k=0, k\geq n+1}\}_m$
from $\{v_m(t)|_{t_k=0, k\geq n}\}_m$.
Let $u_n(t)$ be functions defined by (B.8) and
$\tilde L = \der + \sum_{n=1}^\infty \tilde u_{n+1}(t) \der^{-n}$ be
a solution of the KP hierarchy uniquely defined by the initial condition
$\tilde u_n(t)|_{t_k=0, k\neq 1} = u_n(t)|_{t_k=0, k\neq 1}$.
Further we define $\tilde v_n(t)$ by (B.8) from $\tilde L$.
By these definitions,
$\tilde v_n(t)|_{t_k=0, k\neq 1} = v_n(t)|_{t_k=0, k\neq 1}$.
By Proposition B.1 and ii) of Proposition B.3 already proved in the first step,
$\tilde v_n(t)$ satisfy the system (B.11). Hence from the unique solvability
mentioned above, $\tilde v_n(t) = v_n(t)$,
which implies $\tilde u_n(t) = u_n(t)$. Thus (B.10) follows.

Lastly, we prove iii). Suppose that $\Psi$ satisfies (B.13), i.e., $\Psi$ is
a Baker-Akhiezer function. Then $\Psi$ is expressed in terms of the tau
function as
$$
    \Psi = \frac{\tau(t-[\lambda^{-1}])}{\tau(t)} e^{t(\lambda)}.
\tag B.17
$$
By this formula the generating function of the left hand side of the second
equation of (B.14) becomes
$$
\multline
    \sum_{n=2}^\infty \mu^{-n} p_n(-\dertilde) \Psi(t;\lambda)
\\
    =\frac{\Psi(t;\lambda)}{\mu}
    \left(
    (\mu-\lambda) \frac
    {\tau(t-[\lambda^{-1}]-[\mu^{-1}]) \; \tau(t)}
    {\tau(t-[\lambda^{-1}]) \; \tau(t-[\mu^{-1}])} - \mu + \lambda
    +
      \frac{\der_{t_1}\tau(t-[\lambda^{-1}])}{\tau(t-[\lambda^{-1}])}
    - \frac{\der_{t_1}\tau(t)}{\tau(t)}
    \right),
\endmultline
$$
which is equal to $v(t;\mu) \Psi(t;\lambda)$ due to
the differential Fay identity.
Thus the second line of (B.14) is proved. The first equation of (B.14)
is an easy consequence of the definition of $v_n$, (B.8).

Conversely, suppose that a function $\Psi(t;\lambda)$ satisfying (B.14)
is given. The second part of (B.14) has a generating function expression
$$
\gathered
    X \Psi = \mu^{-1} (\der_{t_1} - v(t;\mu)) \Psi, \\
    X = 1 - \exp\left(
          - \sum_{n=1}^\infty \frac{\mu^{-n}}{n} \frac{\der}{\der t_n}
          \right).
\endgathered
\tag B.18
$$
Since $X$ is a differential operator with costant coefficients,
it commutes with itself. Hence,
$$
    \sum_{m=1}^\infty \frac1m X^m =
    \sum_{n=1}^\infty \frac1n \frac{\der}{\der t_n}.
\tag B.19
$$
Applying both hand sides of this identity to $\Psi$, we obtain from (B.18)
$$
    \sum_{n=1}^\infty \frac1n \frac{\der}{\der t_n} \Psi(t;\lambda)
    =
    \sum_{n=1}^\infty \frac{\mu^{-n}}{n} P_n(\der) \Psi(t;\lambda),
\tag B.20
$$
where $P_n$ is a differential operator of $n$-th order.
Comparing both hand sides of (B.20), we have
$$
\split
    P_n \Psi(t;\lambda) &= \frac{\der}{\der t_n} \Psi(t;\lambda) \\
                        &= (\lambda^n + O(\lambda^{-1})) \Psi(t;\lambda).
\endsplit
\tag B.21
$$
The first equation of (B.14) implies the first equation of (B.13)
because of the definition (B.8). Coupling this equation with (B.21), we have
$$
    (P_n - L^n) \Psi(t;\lambda)  = O(\lambda^{-1}) \Psi(t;\lambda).
\tag B.22
$$
This equation means that $P_n - L^n$ does not contain positive powers
of $\der$. Thus $P_n = (L^n)_{\geq 0} = B_n$, which proves (B.13)
because of (B.21).
\qed
\enddemo

%
%
%
%
\head References \endhead

\roster

\item"[1]"
Takasaki, K., and Takebe, T.,
SDiff(2) KP hierarchy,
in: {\it Proceedings of RIMS Research Project 1991 ``Infinite Analysis,"}
Int. J. Mod. Phys\. {\bf A7, Suppl. 1B} (1992), 889--922.

\item"[2]"
Takasaki, K., and Takebe, T.,
SDiff(2) Toda equation -- hierarchy, tau function and symmetries,
Lett\. Math\. Phys\. {\bf 23} (1991), 205--214.

\item"[3]"
Takasaki, K., and Takebe, T.,
Quasi-classical limit of KP hierarchy, W-symmetries and free fermions,
Kyoto preprint KUCP-0050/92 (July, 1992),
to appear in: {\it Proceedings of Lobachevsky Semester of Euler International
Mathematical Institute, 1992, St\. Petersburg.}

\item"[4]"
Takasaki, K., and Takebe, T.,
Quasi-classical limit of Toda hierarchy and W-infinity symmetries,
Lett\. Math\. Phys\. {\bf 28} (1993), 165--176.

\item"[5]"
Lebedev, D., and Manin, Yu.,
Conservation Laws and Lax Representation on Benny's Long Wave Equations,
Phys\.Lett\. {\bf 74A} (1979), 154--156;
Zakharov, V.E.,
On the Benney's equations,
Physica {\bf 3D} (1981), 193--202.

\item"[6]"
Kodama, Y.,
A method for solving the dispersionless KP equation and its exact solutions,
Phys\. Lett\. {\bf 129A} (1988), 223--226;
Kodama, Y., and Gibbons, J.,
A method for solving the dispersionless KP hierarchy and
its exact solutions, II,
Phys\. Lett\. {\bf 135A} (1989), 167--170;
Kodama, Y.,
Solutions of the dispersionless Toda equation,
Phys\. Lett\. {\bf 147A} (1990), 477--482.

\item"[7]"
Krichever, I.M.,
The dispersionless Lax equations and topological minimal models,
Commun\. Math\. Phys\. {\bf 143} (1991), 415--426;
The $\tau$-Function of the Universal Whitham Hierarchy, Matrix Models
and Topological Field Theories, LPTENS-92-18, hep-th/9205110,
to appear in: Commun\. Pure and Appl\. Math., {\bf XLVII} (1994).

\item"[8]"
Dubrovin, B.A.,
Integrable systems in topological field theory,
Nucl. Phys\. {\bf B379} (1992), 627--689;
Hamiltonian formalism of Whitham-type hierarchies
and topological Landau-Ginsburg models,
Commun\. Math\. Phys\. {\bf 145} (1992), 195--207.

\item"[9]"
Boyer, C., and Finley, J.D.,
Killing vectors in self-dual, Euclidean Einstein spaces,
J. Math\. Phys\. {\bf 23} (1982), 1126--1128;
Gegenberg, J.D., and Das, A.,
Stationary Riemannian space-times with self-dual curvature,
Gen. Rel. Grav. {\bf 16} (1984), 817--829.

\item"[10]"
Golenisheva-Kutuzova, M.I., and Reiman, A.G.,
Integrable equations related to Poisson algebras,
Zap. Nauch. Semin. LOMI {\bf 169} (1988), 44--50 (in Russian);
Saveliev, M.V., and Vershik, A.M.,
Continual analogues of contragredient Lie algebras,
Commun\. Math\. Phys\. {\bf 126} (1989), 367--378;
Kashaev, R.M., Saveliev, M.V., Savelieva, S.A., and Vershik, A.M.,
On nonlinear equations associated with Lie algebras of diffeomorphism
groups of two-dimensional manifolds,
Institute for High Energy Physics preprint 90-I (1990).

\item"[11]"
Bakas, I.,
The structure of the $W_\infty$ algebra,
Commun\. Math\. Phys\. 134 (1990), 487-508.;
Area preserving diffeomorphisms and
higher spin fields in two dimensions,
in {\it Supermembranes and Physics in 2+1 Dimensions},
Trieste 1989, M. Duff, C. Pope and E. Sezgin eds.
(World Scientific, 1990);
Park, Q-Han,
Extended conformal symmetries in real heavens,
Phys\. Lett\. {\bf 236B} (1990), 429--432.

\item"[12]"
Hitchin, N.J.,
Complex manifolds and Einstein's equations,
in {\it Twistor Geometry and Non-linear Systems\/},
H.D. Doebner and T. Weber (eds.), Lecture Notes in Mathematics  vol. 970
(Springer-Verlag 1982);
Jones, P.E., and Tod, K.P.,
Minitwistor spaces and Einstein-Weyl spaces,
Class. Quantum Grav. {\bf 2} (1985), 565--577;
Ward, R.S.,
Einstein-Weyl spaces and $SU(\infty)$ Toda fields,
Class. Quantum Grav. {\bf 7} (1990), L95--L98;
LeBrun, C.,
Explicit self-dual metrics on $CP_2$ \# $\dots$ \# $CP_2$,
J. Diff. Geometry {\bf 34} (1991), 223--253.

\item"[13]"
Sato, M., and Sato, Y.,
Soliton equations as dynamical systems on infinite dimensional
Grassmann manifold,
in {\it Nonlinear Partial Differential Equations in Applied Science;
Proceedings of the U.S.-Japan Seminar, Tokyo, 1982},
Lect. Notes in Num. Anal. {\bf 5} (1982), 259--271.

\item"[14]"
Sato, M., and Noumi, M.,
Soliton equations and the universal Grassmann manifolds,
Sophia University Kokyuroku in Math\. {\bf 18} (1984),
in Japanese.

\item"[15]"
Date, E., Kashiwara, M., Jimbo, M., and Miwa, T.,
Transformation groups for soliton equations,
in: {\it Nonlinear Integrable Systems --- Classical Theory and Quantum Theory}
(World Scientific, Singapore, 1983), 39--119.

\item"[16]"
Takasaki, K.,
An infinite number of hidden variables in hyper-K\"ahler metrics,
J. Math\. Phys\. {\bf 30} (1989), 1515--1521.

\item"[17]"
Takasaki, K.,
Symmetries of hyper-K\"ahler (or Poisson gauge field) hierarchy,
J. Math\. Phys\. {\bf 31} (1990), 1877--1888.

\item"[18]"
Orlov, A. Yu\. and Schulman, E. I., Additional symmetries for integrable
equations and conformal algebra representation,
Lett\. Math\. Phys\. {\bf 12} (1986), 171--179;
Orlov, A. Yu.,
Vertex operators, $\bar{\partial}$-problems, symmetries, variational
indentities and Hamiltonian formalism for $2+1$ integrable systems,
in: {\it Plasma Theory and Nonlinear and
Turbulent Processes in Physics\/}
(World Scientific, Singapore, 1988);
Grinevich, P. G., and Orlov, A. Yu.,
Virasoro action on Riemann surfaces, Grassmannians,
$\det\bar{\partial}_j$ and Segal Wilson $\tau$ function,
in: {\it Problems of Modern Quantum Field Theory\/}
(Springer-Verlag, 1989).

\item"[19]"
Sato, M., Kawai, T. and Kashiwara, M.,
Microfunctions and pseudo-differential equations,
Lecture Notes in Math\. {\bf 287} (1973), 265--529;
Kashiwara~M., {\it Systems of Microdifferential Equations},
(1983), Birkh\"auser;
Schapira~P., {\it Microdifferential Systems in the Complex Domain},
(1985), Springer.

\item"[20]"
Mulase, M.,
Complete integrability of the Kadomtsev-Petviashvili equation,
Advances in Math\. {\bf 54} (1984), 57--66;
Solvability of the super KP equation and
a generalization of the Birkhoff decomposition,
Invent\. Math\. {\bf 92} (1988), 1--46.

\item"[21]"
Ueno, K., and Takasaki, K.,
Toda lattice hierarchy,
in {\it Group Representations and Systems of Differential Equations},
K. Okamoto ed., Advanced Studies in Pure Math\. {\bf 4}
(North-Holland/Kinokuniya 1984), 1--95.

\item"[22]"
Eguchi, T., and Kanno, H.,
Toda lattice hierarchy and the topological description of the
$c=1$ string theory,
University of Tokyo preprint UT-674, hep-th/9404056 (April, 1994);
Hanay, A., Oz, Y., and  Plesser, M.R.,
Topological Landau-Ginzburg formulation and integrable structure
of 2d string theory,
IASSNS-HEP 94/1, hep-th/9401030 (January, 1994);
Takasaki, K.,
Dispersionless Toda hierarchy and two-dimensional string theory,
Kyoto preprint KUCP-0067/94, hep-th/9403190 (March, 1994);
Bonora L. and Xiong, C. S., Two-matrix model and $c=1$ string theory,
SISSA-ISAS 54/94/EP, BONN-HE-06/94, hep-th/9405004.

\item"[23]"
Krichever, I.M.,
Method of averaging for two-dimensional ``integrable'' equations,
Funkts. Anal. Pril. {\bf 22-3} (1988), 37--52 (in Russian);
Funct\. Anal\. Appl\. {\bf 22} (1988), 200--213 (Engl. transl.).

\item"[24]"
Penrose, R.,
Nonlinear gravitons and curved twistor theory,
Gen\. Rel\. Grav\. {\bf 7} (1976), 31--52.

\item"[25]"
Hitchin, N.J., Karlhede, A., Lindstr\"om, U., and Ro\v cek, M.,
Hyperk\"ahler metrics and supersymmetry,
Commun\. Math\. Phys\. {\bf 108} (1987), 535--589.

\item"[26]"
Boyer, C.P., and Plebanski, J.F.,
An infinite hierarchy of conservation laws
and nonlinear superposition principles for self-dual Einstein spaces,
J. Math\. Phys\. {\bf 26} (1985), 229--234.

\item"[27]"
Takasaki, K.,
Differential algebras and $\calD$-modules
in super Toda lattice hierarchy,
Lett\. Math\. Phys\. {\bf 19} (1990), 229--236.

\item"[28]"
Kac, V.G.,
{\it Infinite dimensional Lie algebras}, 3rd ed.,
(1990), Cambridge Univ\. Press.

\item"[29]"
Witten, E.,
Ground ring of two dimensional string theory,
Nucl. Phys\. {\bf B373} (1992), 187--213.

\item"[30]"
Dijkgraaf, R., Verlinde, H., and Verlinde, E.,
Topological strings in $d<1$,
Nucl\. Phys\. {\bf B352} (1991), 59--86;
Loop equations and Virasoro constraints in non-perturbative
2d quantum gravity,
Nucl\. Phys\. {\bf B348} (1991), 435--456.

\item"[31]"
Dubrovin, B.A., and Novikov, S.P.,
Hamiltonian formalism of one-dimensional systems of hydrodynamic type
and Bogolyubov-Whitham averaging method,
Dokl. Akad. Nauk SSSR {\bf 270} (1983), 781--785 (in Russian);
Soviet Math\. Dokl. {\bf 27} (1983), 665--669 (Engl. transl.).

\item"[32]"
Tsarev, S.P.,
On Poisson brackets and one dimensional Hamiltonian systems
of hydrodynamic type,
Dokl. Akad. Nauk SSSR {\bf 282} (1985), 534--537 (in Russian);
Soviet Math\. Dokl. {\bf 31} (1985), 488--491 (Engl. transl.).

\item"[33]"
Losev, A.S., and Polyubin, I.,
On connection between topological Landau-Ginzburg gravity and
integrable systems,
hep-th/9305079 (May, 1993);
Eguchi, T., Kanno, H., Yamada, Y., and Yang, S.-K.,
Topological strings, flat coordiantes and gravitational descendents,
Phys\. Lett\. {\bf B305} (1993), 235--241;
Eguchi, T., Yamada, Y., and Yang, S.-K.,
Topological field theories and the period integrals,
Mod. Phys\. Lett\. {\bf A8} (1993), 1627--1638.

\item"[34]"
Kontsevich, M.,
Intersection theory on the moduli space of curves and
the matrix Airy function,
Commun\. Math\. Phys\. {\bf 147} (1992), 1--23.

\item"[35]"
Kharchev, S., Marshakov, A., Mironov, A., Morozov, A., and Zabrodin, A.,
Towards unified thoery of 2-d gravity,
Nucl\. Phys\. {\bf B380} (1992), 181--240;
Itzykson, C., and Zuber, J. B.,
Addendum to: Combinatorics of the modular group 2.,
the Kontsevich integrals,
Intern. J. Mod. Phys\. {\bf A7} (1992), 5661--5705;
Kharchev, S., Marshakov, A., Mironov, A., and Morozov, A.,
Landau-Ginzburg topological theories in the framework of GKM and
equivalent hierarchies,
Mod. Phys\. Lett\. {\bf A8} (1993) 1047--1062;
Generalized Kontsevich model versus Toda hierarchy and
discrete matrix models,
Nucl\. Phys\. {\bf B397} (1993), 339--378.

\item"[36]"
Fukuma, M., Kawai, H., and Nakayama, R.,
Continuum Schwinger-Dyson equations and universal structures
in two-dimensional quantum gravity,
Int. J. Mod. Phys\. {\bf A6} (1991), 1385--1406;
Infinite dimensional Grassmannian structure of two dimensional string theory,
Commun\. Math\. Phys\. {\bf 143} (1991), 371--403.

\item"[37]"
Adler, M., and van Moerbeke, P.,
A matrix integral solution to two-dimensional $W_p$-gravity,
Commun\. Math\. Phys\. {\bf 147} (1992), 25--56.

\item"[38]"
Nakatsu, T., Kato, A., Noumi, M., and Takebe, T.,
Topological strings, matrix integrals, and singularity theory,
Phys\. Lett\. {\bf B322} (1994), 192--197;
Kato, A., Nakatsu, T., Noumi, M., and Takebe, T.
in preparation.

\item"[39]"
Takebe, T.,
Toda lattice hierarchy and conservation laws,
Commun\. Math\. Phys\. {\bf 129} (1990), 281--318.

\item"[40]"
Jimbo, M., and Miwa, T.,
Solitons and infinite dimensional Lie algebras,
Publ\. RIMS, Kyoto Univ\. {\bf 19} (1983), 943--1001.

\item"[41]"
Takebe, T.,
Representation theoretical meaning of the initial value problem
for the Toda lattice hierarchy: I,
Lett\. Math\. Phys\. {\bf 21} (1991), 77--84;
ditto II,
Publ\. RIMS, Kyoto Univ\., {\bf 27} (1991), 491--503.

\item"[42]"
Takasaki, K.,
Initial value problem for the Toda lattice hierarchy,
in {\it Group Representations and Systems of Differential Equations},
K. Okamoto ed., Advanced Studies in Pure Math\. {\bf 4}
(North-Holland/Kinokuniya 1984), 139--163.

\item"[43]"
Kodama, Y., and Gibbons, J.,
Integrability of the dispersionless KP hierarchy
in:
{\it Proceedings of Workshop ``Nonlinear and Turbulent Processes in Physics''},
Kiev 1989 (World Scientific, 1990), 160-180.

\endroster
\enddocument
\bye